\documentclass[amssymb,aps]{revtex4-2}
\usepackage{graphicx}
\usepackage{enumerate}
\usepackage{amssymb}
\usepackage{amsmath}
\usepackage{amsfonts}
\usepackage{color}
\usepackage{mathrsfs}

\begin{document}
\title{Direct derivation of the modified Langevin noise formalism from the canonical quantization of macroscopic electromagnetism}

\author{A. Ciattoni$^1$}
\email{alessandro.ciattoni@spin.cnr.it}
\affiliation{$^1$CNR-SPIN, c/o Dip.to di Scienze Fisiche e Chimiche, Via Vetoio, 67100 Coppito (L'Aquila), Italy}
\date{\today}

\begin{abstract}
The modified Langevin noise formalism (MLNF) models the interaction of the quantized electromagnetic field with an arbitrary lossy magneto-dielectric object placed in vacuum using three types of non-interacting bosonic polaritons: scattering, electric, and magnetic. These respectively represent free-space photons scattered by the object, and photons radiated by quantized electric and magnetic dipolar sources embedded within its volume. Recently [A. Ciattoni, Phys. Rev. A \textbf{110}, 013707 (2024)], this formalism was justified from the canonical quantization of macroscopic electromagnetism (CQME) [Philbin, New J. Phys. \textbf{12}, 123008 (2010)] in the Heisenberg picture. This was achieved by identifying the polariton operators within the formal solution of the macroscopic Maxwell equations, assuming they obey bosonic commutation relations to retrieve the canonical ones, and showing they diagonalize the CQME Hamiltonian. However, the explicit functional dependence of these polaritons on the underlying canonical field operators remained undetermined. In this paper, we derive the exact analytical expressions for the polariton operators in terms of the canonical CQME field operators. Using these mappings, we provide a direct and rigorous derivation of the MLNF from the canonical theory in the Schr\"odinger picture. Our derivation is structured in three foundational steps: 1) adopting the derived analytical expressions as the constitutive definitions of the polariton operators; 2) mathematically proving that these operators are strictly bosonic as a direct consequence of the canonical commutation relations; and 3) demonstrating that they exactly diagonalize the macroscopic CQME Hamiltonian.
\end{abstract}
\maketitle

\section{Introduction}
The ability to model the interaction of the quantized electromagnetic field with arbitrarily dispersive macroscopic matter is essential in quantum optics, enabling predictions of how realistic macroscopic devices can be used to manipulate the quantum properties of photons. While in ideally transparent dielectric media the quantum field description is easily achieved by exploiting the orthogonality of photonic modes \cite{Glaub1,Dalto1}, field quantization in realistic lossy media is considerably more complex, since matter degrees of freedom must be included in the formalism. Matter is routinely modeled using a continuum of reservoir oscillator fields in all theoretical schemes incorporating damping \cite{Fanooo,Hopfi1,Huttn1,Huttn2,Sutto1,Sutto2,Kheir1,Matloo,Kheir2,Bhattt,Amoos1,Sutto3,Amoos2}, including the canonical quantization of macroscopic electromagnetism (CQME) proposed by Philbin in Ref.\cite{Philbi}. As far as we are aware, CQME is the most general approach for handling causal magneto-dielectric inhomogeneous isotropic media. CQME has a robust theoretical basis: following the introduction of a Lagrangian whose Euler-Lagrange equations yield the macroscopic Maxwell equations, quantization is performed by imposing canonical commutation relations between the electromagnetic and reservoir field operators and their conjugate momenta. The subsequent diagonalization of the Hamiltonian, performed in Ref.\cite{Philbi} through the Fano method, casts the CQME into the familiar language of second quantization via the introduction of bosonic electric and magnetic polaritons---i.e., quanta of dipolar sources embedded within the medium, which fully determine the overall quantum field through the classical dyadic Green’s function. 

It is worth emphasizing that the second-quantized form of CQME derived in Ref.\cite{Philbi} coincides with macroscopic quantum electrodynamics (MQED) or the Langevin noise formalism (LNF) \cite{Grune2,Schee3,Dungg1,Schee4,Buhma1}, a phenomenological scheme introduced prior to the development of CQME in which, guided by the fluctuation-dissipation theorem, quantization is non-canonically performed by imposing the standard field commutation relations of vacuum quantum electrodynamics. A crucial observation is that the quantum electromagnetic field considered in the LNF contains no contributions exhibiting an ingoing far-field behavior (which characterizes incoming free radiation). This clearly indicates that the LNF is strictly valid only for unbounded lossy media, where no free radiation (either ingoing or outgoing) is allowed. However, for finite-size lossy objects, the LNF can still be applied indirectly by resorting to a singular (and generally involved) limiting procedure, where the limits ${\rm Im} (\varepsilon) \rightarrow 0^+$ and ${\rm Im} (\mu) \rightarrow 0^+$, corresponding to the transparent regions, are taken at the end of the calculations \cite{Hanso1}. To avoid such a singular limiting procedure, bosonic operators of free radiation modes (scattering modes) must be explicitly included in the theory. This constitutes the core observation of the modified Langevin noise formalism (MLNF), which has been formulated for one-dimensional planar structures \cite{DiStef}, non-magnetic objects \cite{Dreze1,Dorie1,Naaaa2}, and general magneto-dielectric objects \cite{Ciatt1}. Owing to its flexibility in handling quantized scattering modes, the MLNF has recently been exploited to model the interaction of quantum emitters with dispersive objects \cite{Miano1,Miano2,Miano3} and to develop a general approach to quantum optical scattering by finite-size lossy objects in vacuum \cite{Ciatt2,Ciatt3}.

The variant of the MLNF discussed in Ref.\cite{Ciatt1} is, to the best of our knowledge, the most general one (also due to the inclusion of magnetic polaritons); therefore, from now on, we will simply refer to it as the MLNF. In Ref.\cite{Ciatt1}, the MLNF was justified from the first-quantized form of the CQME by means of an indirect strategy, which we briefly summarize here. First, the Heisenberg equations for the field operators are recognized to be the quantum macroscopic Maxwell equations. For a general finite-size object in vacuum, their formal solution has two contributions: the scattering field (a superposition of the classical scattering modes produced by any vacuum plane wave, containing the scattering polariton operators) and the medium-assisted field (a superposition of the fields radiated by all dipolar sources in the medium, containing the electric and magnetic polariton operators). Second, a fundamental integral relation involving the dyadic Green’s function---and containing a surface term for finite-size objects \cite{Dreze2,Frank1}---made it possible to prove that the canonical commutation relations of the CQME are self-consistently recovered by assuming that the polariton operators satisfy bosonic commutation relations. Third, the CQME Hamiltonian was shown to be diagonalized by these bosonic polariton operators. 

This indirect justification of the MLNF might raise two significant concerns regarding its validity: 1) the polariton operators are not explicitly expressed in terms of the canonical field operators, and thus their mathematical existence could be questioned; 2) both the MLNF and the LNF appear to be the second-quantized form of CQME, an apparent contradiction that we hereafter refer to as the double second-quantization paradox. To dispel any doubts concerning the theoretical foundation of the MLNF, in this paper we first deduce the exact analytical mapping between the canonical field operators and the polariton operators and, subsequently, use this mapping to provide a direct derivation of the MLNF from the CQME. Our derivation follows a logical path analogous to the quantization of a simple harmonic oscillator: 1) adopting the newly derived expressions for the polariton operators in terms of the canonical field operators as constitutive definitions; 2) proving that these polariton operators satisfy bosonic commutation relations as a direct consequence of the canonical commutation relations of the CQME; and 3) showing that such bosonic polariton operators exactly diagonalize the CQME Hamiltonian. Although executing the second and third steps requires an extensive and highly involved algebraic derivation, this analytical procedure is strictly necessary. It provides the definitive proof that the MLNF is structurally exact and firmly grounded in a rigorous canonical framework.

Furthermore, we conclude the paper by demonstrating that the resolution of the double second-quantization paradox lies in the fact that, in Ref.\cite{Philbi}, the CQME Hamiltonian is diagonalized using the Fano method. This method begins with the premise that the canonical operators are entirely expandable as integrals of the diagonalizing (polariton) operators over the medium volume---or, from a physical perspective, that the field is uniquely generated by the dipolar sources embedded within the lossy medium. While this assumption is strictly correct if the lossy medium is unbounded, it fails to account for field sources originating at infinity if the lossy object has a finite size. Therefore, the diagonalization of the CQME Hamiltonian in Ref.\cite{Philbi} yields, as expected, the standard LNF for an unbounded medium but it fails to retrieve the more general MLNF for a finite-size object, since its initial Fano ansatz for the canonical operators artificially omits the angular integrals over the scattering modes, thereby failing to span the full electromagnetic Fock space.

The paper is structured as follows. In Sec.II, we briefly review the MLNF, the first-quantized form of the CQME, and their previously established indirect connection. In Sec.III, we derive the exact analytical mapping between the canonical field operators and the polariton operators. SectionIV is devoted to proving that these polaritons rigorously obey macroscopic bosonic commutation relations. In Sec.V, we perform the exact diagonalization of the CQME Hamiltonian, completing the direct derivation. In Sec.VI, we physically resolve the double second-quantization paradox by examining the foundational assumptions of the standard Fano diagonalization. Finally, our conclusions are drawn in Sec.VII. The most mathematically demanding proofs, dyadic properties, and generalized dispersion relations are deferred to the Appendices.

\section{Modified Langevin noise formalism and canonical quantization of macroscopic electromagnetism}
To frame our results within an established theoretical context and to fix the notation, in this section we summarize the MLNF \cite{Ciatt1} and the formulation of the CQME \cite{Philbi}, respectively. Furthermore, we briefly outline the indirect derivation of the former from the latter discussed in Ref.\cite{Ciatt1}. 

We hereafter consider an arbitrary finite-size lossy object in vacuum, filling the spatial region $V$, whose inhomogeneous isotropic magneto-dielectric optical response is characterized, in the frequency domain ($e^{-i \omega t}$), by the dielectric permittivity $\varepsilon _\omega ^{V} \left( {\bf{r}} \right)$ and the magnetic permeability $\mu _\omega ^{V} \left( {\bf{r}} \right)$ whose only constraint is that they be holomorphic functions of the complex frequency $\omega$ in the upper half-plane ${\mathop{\rm Im}\nolimits} \left( \omega  \right) > 0$ as required by causality. To account for the object being situated in vacuum, it is convenient to define the global permittivity ${\varepsilon _\omega  \left( {\bf{r}} \right)}$ and permeability ${\mu _\omega  \left( {\bf{r}} \right)}$ according to
\begin{align} \label{eps_mu}
\varepsilon _\omega  \left( {\bf{r}} \right) =& \left\{ {\begin{array}{*{20}l}
   {\varepsilon _\omega ^V \left( {\bf{r}} \right),} & {{\bf{r}} \in V}  \\
   {1,} & {{\bf{r}} \notin V}  \\
\end{array} } \right. &
\mu _\omega  \left( {\bf{r}} \right) =& \left\{ {\begin{array}{*{20}l}
   {\mu _\omega ^V \left( {\bf{r}} \right),} & {{\bf{r}} \in V}  \\
   {1,} & {{\bf{r}} \notin V}  \\
\end{array}} \right.
\end{align}
For any Hermitian operator $\hat O$ in the Schr\"odinger picture, we denote by $\hat O_\omega$  ($\omega >0$) the positive frequency spectrum of the associated Heisenberg operator, i.e. $e^{\frac{i}{\hbar }\hat Ht} \hat O e^{ - \frac{i}{\hbar }\hat Ht}  = \int {d\omega e^{ - i\omega t} } \hat O_\omega  + {\rm H.c.}$, where $\int {d\omega }  \equiv \int_0^{ + \infty } {d\omega }$, from which the spectral operator decomposition follows
\begin{equation} \label{SpOpDe}
\hat O = \int {d\omega } \hat O_\omega  + {\rm H.c.}. 
\end{equation}
For convenience, in Appendix A we report the dyadic definitions and  relations that we use throughout the paper.

\subsection{MLNF}
MLNF describes the field-matter system by means of the scattering ($s$) polariton operators ${\bf{\hat g}}_{\omega s} \left( {\bf{n}} \right)$, defined for any unit vector $\bf{n}$ and transverse to it (${{\bf{n}} \cdot {\bf{\hat g}}_{\omega s} \left( {\bf{n}} \right) = 0}$), and the electric ($e$) and magnetic ($m$) polariton operators ${\bf{\hat f}}_{\omega e } \left( {\bf{r}} \right)$ and ${\bf{\hat f}}_{\omega m } \left( {\bf{r}} \right)$, defined strictly within the object's volume (${{\bf{r}} \in V}$) and zero elsewhere. The polariton operators are defined for positive frequencies $\omega >0$ and they satisfy the boson commutation relations
\begin{align} \label{MLNF_Com_Rel}
\left[ {{\bf{\hat g}}_{\omega s} \left( {\bf{n}} \right),{\bf{\hat g}}_{\omega 's}^\dag  \left( {{\bf{n}}'} \right)} \right] &= \delta \left( {\omega  - \omega '} \right)\delta \left( {o_{\bf{n}}  - o_{{\bf{n}}'} } \right){\cal I}_{\bf{n}}, & \left[ {{\bf{\hat f}}_{\omega \nu } \left( {\bf{r}} \right),{\bf{\hat f}}_{\omega '\nu '}^\dag  \left( {{\bf{r}}'} \right)} \right] &= \delta \left( {\omega  - \omega '} \right)\delta _{\nu \nu '} \delta \left( {{\bf{r}} - {\bf{r}}'} \right){\cal I},
\end{align}
with all other commutators vanishing, where ${\cal I}_{\bf{n}}$ is the dyadic projector onto the plane orthogonal to the direction ${\bf{n}} = \sin \theta _{\bf{n}} \left( {\cos \varphi _{\bf{n}} {\bf{u}}_x  + \sin \varphi _{\bf{n}} {\bf{u}}_y } \right) + \cos \theta _{\bf{n}} {\bf{u}}_z$,  ${\cal I}$  is the dyadic identity and $\delta \left( {o_{\bf{n}}  - o_{{\bf{n}}'} } \right) = \delta \left( {\theta _{\bf{n}}  - \theta '_{\bf{n}} } \right)\delta \left( {\varphi _{\bf{n}}  - \varphi '_{\bf{n}} } \right)/\sin \theta _{\bf{n}}$ is the angular delta function. The Hamiltonian and electric field operators are
\begin{eqnarray} \label{MLNF_HE}
\hat H &=& \int {d\omega } \;\hbar \omega \left[ {\int {do_{\bf{n}} } {\bf{\hat g}}_{\omega s}^\dag  \left( {\bf{n}} \right) \cdot {\bf{\hat g}}_{\omega s} \left( {\bf{n}} \right) + \sum\limits_{\nu} {\int {d^3 {\bf{r}}\,} {\bf{\hat f}}_{\omega \nu }^\dag  \left( {\bf{r}} \right) \cdot {\bf{\hat f}}_{\omega \nu } \left( {\bf{r}} \right)} } \right]  , \nonumber \\ 
  {\bf{\hat E}}\left( {\bf{r}} \right) &=& \int{d\omega } \left[ {\int {do_{\bf{n}} } {\cal F}_{\omega s} \left( {\left. {\bf{r}} \right|{\bf{n}}} \right) \cdot {\bf{\hat g}}_{\omega s} \left( {\bf{n}} \right) + \sum\limits_{\nu} {\int {d^3 {\bf{r}}'\,} {\cal G}_{\omega \nu } \left( {\left. {\bf{r}} \right|{\bf{r}}'} \right) \cdot {\bf{\hat f}}_{\omega \nu } \left( {{\bf{r}}'} \right)} } \right] + {\rm H.c.},
\end{eqnarray}
where $do_{\bf{n}}  = \sin \theta _{\bf{n}} d\theta _{\bf{n}} d\varphi _{\bf{n}}$ is the solid angle element around the direction $\bf n$, whereas the scattering $s$, electric $e$ and magnetic $m$ dyadic  kernels ($\nu = e,m$) are
\begin{equation} \label{Kernels}
{\cal F}_{\omega s} \left( {\left. {\bf{r}} \right|{\bf{n}}} \right) = {\cal F}_\omega  \left( {\left. {\bf{r}} \right|{\bf{n}}} \right)\sqrt {\frac{{\hbar k_\omega ^3 }}{{16\pi ^3 \varepsilon _0 }}} ,\quad \quad \begin{array}{*{20}l}
  {\cal G}_{\omega e} \left( {\left. {\bf{r}} \right|{\bf{r}}'} \right) &=& {\cal G}_\omega  \left( {\left. {\bf{r}} \right|{\bf{r}}'} \right) i\sqrt { \displaystyle \frac{{\hbar k_\omega ^4 }}{{\pi \varepsilon _0 }}{\mathop{\rm Im}\nolimits} \left[ {\varepsilon _\omega  \left( {{\bf{r}}'} \right)} \right]},  \\
  {\cal G}_{\omega m} \left( {\left. {\bf{r}} \right|{\bf{r}}'} \right) &=& {\cal G}_\omega  \left( {\left. {\bf{r}} \right|{\bf{r}}'} \right) \displaystyle \frac{{ \times \mathord{\buildrel{\lower3pt\hbox{$\scriptscriptstyle\leftarrow$}} 
\over \nabla } _{{\bf{r}}'} }}{{ik_\omega  }}\sqrt {\frac{{\hbar k_\omega ^4 }}{{\pi \varepsilon _0 }}{\mathop{\rm Im}\nolimits} \left[ {\frac{{ - 1}}{{\mu _\omega  \left( {{\bf{r}}'} \right)}}} \right]} ,
\end{array}
\end{equation}
where $k_\omega   = \omega /c$ is the vacuum wavenumber whereas the modal dyadic ${\cal F}_\omega  \left( {\left. {\bf{r}} \right|{\bf{n}}} \right)$ and the dyadic Green's function ${{\cal G}_\omega  \left( {\left. {\bf{r}} \right|{\bf{r}}'} \right)}$ are defined by the boundary-value problems
\begin{align} \label{GS}
\left[ {\left( {\nabla _{\bf{r}}  \times \frac{1}{{\mu _\omega  \left( {\bf{r}} \right)}}\nabla _{\bf{r}}  \times } \right) - k_\omega ^2 \varepsilon _\omega  \left( {\bf{r}} \right)} \right]{\cal F}_\omega  \left( {\left. {\bf{r}} \right|{\bf{n}}} \right) &=  0,   & 
{\cal F}_\omega  \left( {\left. {r{\bf{m}}} \right|{\bf{n}}} \right) & \mathop  \approx \limits_{r \to + \infty } e^{i\left( {k_\omega  {\bf{n}}} \right) \cdot \left( {r{\bf{m}}} \right)} {\cal I}_{\bf{n}}  + \frac{{e^{ik_\omega  r} }}{r}{\cal S}_\omega  \left( {{\bf{m}}\left| {\bf{n}} \right.} \right),  \nonumber \\
\left[ {\left( {\nabla _{\bf{r}}  \times \frac{1}{{\mu _\omega  \left( {\bf{r}} \right)}}\nabla _{\bf{r}}  \times } \right) - k_\omega ^2 \varepsilon _\omega  \left( {\bf{r}} \right)} \right]{\cal G}_\omega  \left( {\left. {\bf{r}} \right|{\bf{r}}'} \right) & = \delta \left( {{\bf{r}} - {\bf{r}}'} \right){\cal I}, &
{\cal G}_\omega  \left( {\left. {r{\bf{m}}} \right|{\bf{r}}'} \right) & \mathop  \approx \limits_{r \to + \infty } \frac{{e^{ik_\omega  r} }}{r}{\cal W}_\omega  \left( {\left. {\bf{m}} \right|{\bf{r}}'} \right), 
\end{align}
where $\bf m$ is a unit vectors, the symbol $\mathop  \approx \limits_{r \to + \infty }$ stands for the leading-order term of the asymptotic expansion for $r \to  + \infty$ whereas ${\cal S}_\omega  \left( {{\bf{m}}\left| {\bf{n}} \right.} \right)$ is the scattering dyadic \cite{Krist1} and ${\cal W}_\omega  \left( {{\bf{m}}\left| {\bf{r}} \right.} \right)$ is the asymptotic amplitude of the dyadic Green's function (see Appendix B). Note that the electric field operator in the second of Eqs.(\ref{MLNF_HE}) is consistent with the polariton definitions because, from Eqs.(\ref{Kernels}) and Appendix B, ${\cal F}_{\omega s} \left( {\left. {\bf{r}} \right|{\bf{n}}} \right) \cdot {\bf{n}} = 0$ and ${\cal G}_{\omega \nu } \left( {\left. {\bf{r}} \right|{\bf{r}}'} \right) = 0$ for ${\bf{r}}' \notin V$.

\subsection{CQME}
As shown in Ref.\cite{Philbi}, CQME has a solid Lagragian theoretical basis and, in its first-quantized form, it describes the field-matter system by means of the vector potential operator ${\bf{\hat A}}\left( {\bf{r}} \right)$ in the Coulomb gauge ($\nabla  \cdot {\bf{\hat A}} = 0$) and the electric and magnetic reservoir field operators ${\bf{\hat X}}^\Omega  \left( {\bf{r}} \right)$ and ${\bf{\hat Y}}^\Omega  \left( {\bf{r}} \right)$ of all the proper frequencies $\Omega >0$ and defined strictly within the object's volume (${{\bf{r}} \in V}$) and zero elsewhere, together with their conjugate canonical momenta ${\bf{\hat \Pi }}_A \left( {\bf{r}} \right)$, ${\bf{\hat \Pi }}_X^\Omega  \left( {\bf{r}} \right)$ and ${\bf{\hat \Pi }}_Y^\Omega  \left( {\bf{r}} \right)$, the latter two being defined strictly within the object's volume (${{\bf{r}} \in V}$) and zero elsewhere. The six canonical field operators satisfy the commutation relations
\begin{align} \label{CQME_Com_Rel}
 \left[ {{\bf{\hat A}}\left( {\bf{r}} \right),{\bf{\hat \Pi }}_A \left( {{\bf{r}}'} \right)} \right] &= i\hbar {\cal D}^ \bot  \left( {{\bf{r}} - {\bf{r}}'} \right), &
 \left[ {{\bf{\hat X}}^\Omega  \left( {\bf{r}} \right),{\bf{\hat \Pi }}_X^{\Omega '} \left( {{\bf{r}}'} \right)} \right] &= 
 \left[ {{\bf{\hat Y}}^\Omega  \left( {\bf{r}} \right),{\bf{\hat \Pi }}_Y^{\Omega '} \left( {{\bf{r}}'} \right)} \right] = i\hbar \delta \left( {\Omega  - \Omega '} \right)\delta \left( {{\bf{r}} - {\bf{r}}'} \right){\cal I},
\end{align}
together with the vanishing of all the other possible commutation relations, where  ${\cal D}^ \bot  \left( {\bf{r}} \right)$ is the dyadic transverse delta function. The Hamiltonian operator is
\begin{eqnarray} \label{CQME_H}
&& \hat H = \int {d^3 {\bf{r}}} \left\{ {\frac{1}{{2\varepsilon _0 }}\left( {{\bf{\hat \Pi }}_A  + \int {d\Omega } \;\alpha ^\Omega  {\bf{\hat X}}^\Omega  } \right)^2  + \left( {\nabla  \times {\bf{\hat A}}} \right) \cdot \left( {\frac{1}{{2\mu _0 }}\nabla  \times {\bf{\hat A}} - \int {d\Omega } \;\beta ^\Omega  {\bf{\hat Y}}^\Omega  } \right) + \frac{1}{2}\int {d\Omega } \left( {{\bf{\hat \Pi }}_X^{\Omega 2}  + \Omega ^2 {\bf{\hat X}}^{\Omega 2} } \right)} \right. \nonumber \\ 
&& \quad \quad \left. { + \frac{1}{2}\int {d\Omega } \left( {{\bf{\hat \Pi }}_Y^{\Omega 2}  + \Omega ^2 {\bf{\hat Y}}^{\Omega 2} } \right)} \right\},
\end{eqnarray} 
where
\begin{align} \label{alf_bet}
\alpha ^\omega  \left( {\bf{r}} \right) &= \sqrt {\frac{{2\varepsilon _0 }}{\pi }\omega {\mathop{\rm Im}\nolimits} \left[ {\varepsilon _\omega  \left( {\bf{r}} \right)} \right]}, &
\beta ^\omega  \left( {\bf{r}} \right) &= \sqrt {\frac{2}{{\pi \mu _0 }}\omega {\mathop{\rm Im}\nolimits} \left[ {\frac{{ - 1}}{{\mu _\omega  \left( {\bf{r}} \right)}}} \right]} 
\end{align}
are real coupling coefficients fully accounting for the medium optical response and playing a central role in the entire theory presented hereafter, whereas ${\bf{\hat V}}^2  = {\bf{\hat V}} \cdot {\bf{\hat V}}$ is the square modulus of any field operator ${\bf{\hat V}}$. Eventually, the electric and  induction magnetic field operators are

\begin{align} \label{CQME_EB}
{\bf{\hat E}} &=  - \frac{1}{{\varepsilon _0 }}\left( {{\bf{\hat \Pi }}_A  + \int {d\Omega } \;\alpha ^\Omega  {\bf{\hat X}}^\Omega  } \right), & {\bf{\hat B}} = \nabla  \times {\bf{\hat A}}.
\end{align}

\subsection{Indirect justification of MLNF from CQME}
The starting point of the indirect justification of MLNF from CQME discussed in Ref.\cite{Ciatt1} is that the Heisenberg equations for the six canonical field operators ${\bf{\hat A}}$, ${\bf{\hat X}}^\Omega$, ${\bf{\hat Y}}^\Omega$, ${\bf{\hat \Pi }}_A$, ${\bf{\hat \Pi }}_X^\Omega$ and ${\bf{\hat \Pi }}_Y^\Omega$, combined with the expressions of the electric and induction magnetic field operators in Eqs.(\ref{CQME_EB}), lead to the quantum Maxwell equations for the electromagnetic field and to the forced harmonic oscillator equations for the reservoir fields, as shown in Ref.\cite{Philbi}. The frequency domain analysis of such equations shows that the spectra of the canonical field operators are
\begin{align} \label{CQME_Spectra}
{\bf{\hat A}}_\omega   &= \frac{1}{{i\omega }}{\bf{\hat E}}_\omega ^ \bot  , &
{\bf{\hat \Pi }}_{A\omega }  &=  - \varepsilon _0 \varepsilon _\omega  {\bf{\hat E}}_\omega   - i\alpha ^\omega  \sqrt {\frac{\hbar }{{2\omega }}} {\bf{\hat f}}_{\omega e} , \nonumber \\
{\bf{\hat X}}_\omega ^\Omega   & = \frac{{\alpha ^\Omega  }}{{\Omega ^2  - \left( {\omega  + i\eta } \right)^2 }}{\bf{\hat E}}_\omega   + \delta \left( {\omega  - \Omega } \right)i\sqrt {\frac{\hbar }{{2\omega }}} {\bf{\hat f}}_{\omega e}, & 
{\bf{\hat \Pi }}_{X\omega }^\Omega   &=  - i\omega {\bf{\hat X}}_\omega ^\Omega, \nonumber  \\
{\bf{\hat Y}}_\omega ^\Omega   &= \frac{{\beta ^\Omega  }}{{\Omega ^2  - \left( {\omega  + i\eta } \right)^2 }}\frac{{\nabla  \times {\bf{\hat E}}_\omega  }}{{i\omega }} + \delta \left( {\omega  - \Omega } \right)\sqrt {\frac{\hbar }{{2\omega }}} {\bf{\hat f}}_{\omega m}, & 
{\bf{\hat \Pi }}_{Y\omega }^\Omega   &=  - i\omega {\bf{\hat Y}}_\omega ^\Omega   
\end{align}
where ${\bf{V}}^ \bot  \left( {\bf{r}} \right) = \int {d^3 {\bf{r}}'{\cal D}^ \bot  \left( {{\bf{r}} - {\bf{r}}'} \right) \cdot {\bf{V}}\left( {{\bf{r}}'} \right)}$ is the transverse  part of the vector field $
{\bf{V}}$,  the prescription $\eta  \to 0^ +$ accounts for the causality of the electromagnetic response, ${\bf{\hat f}}_{\omega e}$ and ${\bf{\hat f}}_{\omega m}$ are operators pertaining to the freely evolving contributions of the electric and magnetic reservoir fields ${{\bf{\hat X}}^\Omega  }$ and ${{\bf{\hat Y}}^\Omega  }$, whereas ${\bf{\hat E}}_\omega$ is the spectrum of the electric field ${\bf{\hat E}}$ in the first of Eqs.(\ref{CQME_EB}) in turn satisfying the inhomogeneous Helmholtz equation
\begin{equation} \label{CQME_InhHel}
\left[ {\left( {\nabla  \times \frac{1}{{\mu _\omega  }}\nabla  \times } \right) - k_\omega ^2 \varepsilon _\omega  } \right]{\bf{\hat E}}_\omega   = i\omega \mu _0 \left[ {\sqrt {\frac{{\hbar \omega }}{2}} \alpha ^\omega  {\bf{\hat f}}_{\omega e}  + \sqrt {\frac{\hbar }{{2\omega }}} \nabla  \times \left( {\beta ^\omega  {\bf{\hat f}}_{\omega m} } \right)} \right].
\end{equation}
Resorting to classical electrodynamics, the complete solution of Eq.(\ref{CQME_InhHel}) is readily provided by Eq.(\ref{Clas_Sol}) after the replacements ${\bf{\tilde E}}_\omega ^{\left( {in} \right)} \left( {\bf{n}} \right) \to \sqrt {\frac{{\hbar k_\omega ^3 }}{{16\pi ^3 \varepsilon _0 }}} {\bf{\hat g}}_{\omega s} \left( {\bf{n}} \right)$ and ${\bf{J}}_\omega  \left( {\bf{r}} \right) \to \sqrt {\frac{{\hbar \omega }}{2}} \alpha ^\omega  {\bf{\hat f}}_{\omega e}  + \sqrt {\frac{\hbar }{{2\omega }}} \nabla  \times ( {\beta ^\omega  {\bf{\hat f}}_{\omega m} } )$, i.e.
\begin{equation} \label{Eom}
{\bf{\hat E}}_\omega  \left( {\bf{r}} \right) = \int {do_{\bf{n}} } {\cal F}_{\omega s} \left( {\left. {\bf{r}} \right|{\bf{n}}} \right) \cdot {\bf{\hat g}}_{\omega s} \left( {\bf{n}} \right) + \sum\limits_\nu  {\int {d^3 {\bf{r}}'\,} {\cal G}_{\omega \nu } \left( {\left. {\bf{r}} \right|{\bf{r}}'} \right) \cdot {\bf{\hat f}}_{\omega \nu } \left( {{\bf{r}}'} \right)},
\end{equation}
where the dyadic kernels ${\cal F}_{\omega s} \left( {\left. {\bf{r}} \right|{\bf{n}}} \right)$, ${\cal G}_{\omega e} \left( {\left. {\bf{r}} \right|{\bf{r}}'} \right)$ and ${\cal G}_{\omega m} \left( {\left. {\bf{r}} \right|{\bf{r}}'} \right)$ are defined in Eq.(\ref{Kernels}) and an integration by parts has been performed in the magnetic contribution using the second of Eqs.(\ref{IntPart}) combined with the relation $\mu _\omega  \left( \infty  \right) = 1$. Note that the field ${\bf{\hat E}}_\omega$ in Eq.(\ref{Eom}) is precisely the spectrum of the MLNF field  ${\bf{\hat E}}$ in the second of Eqs.(\ref{MLNF_HE}). It is worth stressing that the operators ${\bf{\hat g}}_{\omega s}$, ${\bf{\hat f}}_{\omega e}$ and ${\bf{\hat f}}_{\omega m}$ in Eq.(\ref{Eom}) emerge as integration constants appearing when formally solving the  Heisenberg equations of CQME and; as such, they lack an explicit definition in terms of the six canonical field operators. 

The second step of the indirect justification of MLNF from CQME consists of showing that the canonical commutation relations in Eqs.(\ref{CQME_Com_Rel}) are self-consistently recovered by assuming that the operators ${\bf{\hat g}}_{\omega s}$, ${\bf{\hat f}}_{\omega e}$ and ${\bf{\hat f}}_{\omega m}$ satisfy the boson commutation relations in Eqs.(\ref{MLNF_Com_Rel}). The core of the proof is that such assumption together with Eq.(\ref{Eom}) leads to the commutation relations
\begin{equation} \label{[E,f+]}
\left[ {{\bf{\hat E}}_\omega  \left( {\bf{r}} \right),{\bf{\hat f}}_{\omega '\nu }^\dag  \left( {{\bf{r}}'} \right)} \right] = \delta \left( {\omega  - \omega '} \right){\cal G}_{\omega \nu } \left( {\left. {\bf{r}} \right|{\bf{r}}'} \right)
\end{equation}
and
\begin{equation}
\left[ {{\bf{\hat E}}_\omega  \left( {\bf{r}} \right),{\bf{\hat E}}_{\omega '}^\dag  \left( {{\bf{r}}'} \right)} \right] = \delta \left( {\omega  - \omega '} \right)\left[ {\int {do_{\bf{n}} } {\cal F}_{\omega s} \left( {\left. {\bf{r}} \right|{\bf{n}}} \right) \cdot {\cal F}_{\omega s}^{T*} \left( {\left. {{\bf{r}}'} \right|{\bf{n}}} \right) + \sum\limits_\nu  {\int {d^3 {\bf{s}}\,} } {\cal G}_{\omega \nu } \left( {\left. {\bf{r}} \right|{\bf{s}}} \right) \cdot {\cal G}_{\omega \nu }^{T*} \left( {\left. {{\bf{r}}'} \right|{\bf{s}}} \right)} \right]
\end{equation}
which, combined with the integral relation derived in Ref.\cite{Ciatt1} (see Appendix B)
\begin{equation} 
\int {do_{\bf{n}} } {\cal F}_{\omega s} \left( {\left. {\bf{r}} \right|{\bf{n}}} \right) \cdot {\cal F}_{\omega s}^{T*} \left( {\left. {{\bf{r}}'} \right|{\bf{n}}} \right) + \sum\limits_\nu  {\int {d^3 {\bf{s}}} \,{\cal G}_{\omega \nu } \left( {\left. {\bf{r}} \right|{\bf{s}}} \right) \cdot {\cal G}_{\omega \nu }^{T*} \left( {\left. {{\bf{r}}'} \right|{\bf{s}}} \right)}  = \frac{{\hbar k_\omega ^2 }}{{\pi \varepsilon _0 }}{\mathop{\rm Im}\nolimits} \left[ {{\cal G}_\omega  \left( {\left. {\bf{r}} \right|{\bf{r}}'} \right)} \right],
\end{equation}
becomes
\begin{equation} \label{[E,E+]}
\left[ {{\bf{\hat E}}_\omega  \left( {\bf{r}} \right),{\bf{\hat E}}_{\omega '}^\dag  \left( {{\bf{r}}'} \right)} \right] = \delta \left( {\omega  - \omega '} \right)\frac{{\hbar k_\omega ^2 }}{{\pi \varepsilon _0 }}{\mathop{\rm Im}\nolimits} \left[ {{\cal G}_\omega  \left( {\left. {\bf{r}} \right|{\bf{r}}'} \right)} \right].
\end{equation}
The crucial point is that the commutation relations in Eqs.(\ref{[E,f+]}) and (\ref{[E,E+]}) coincide with the corresponding ones pertaining to the LNF and CQME where Eqs.(\ref{CQME_Com_Rel}) hold true so that, since the spectra of the canonical field operators in Eqs.(\ref{CQME_Spectra}) depend solely on ${{\bf{\hat E}}_\omega  }$ and ${{\bf{\hat f}}_\omega  }$, the canonical commutation relations are self-consistently recovered by assuming the validity of Eqs.(\ref{MLNF_Com_Rel}), as detailed in Ref.\cite{Ciatt1}.

In the last step of the indirect justification of MLNF from CQME, the six canonical field operators in the Heisenberg picture, deduced from their spectra of Eqs.(\ref{CQME_Spectra}), are substituted in the CQME Hamiltonian of Eq.(\ref{CQME_H}) thus getting, as a result, the MLNF Hamiltonian in the first of Eqs.(\ref{MLNF_HE}), as detailed in Ref.\cite{Ciatt1}.

\section{Polariton operators}
As emphasized in Sec.IIC, the polariton operators ${\bf{\hat g}}_{\omega s}$, ${\bf{\hat f}}_{\omega e}$ and ${\bf{\hat f}}_{\omega m}$ were introduced in Ref.\cite{Ciatt1} as mere integration constants without establishing their formal definitions within the CQME framework. In this section we fill this gap by using the aforementioned formalism to derive the analytical expressions of the polariton operators in terms of the six canonical field operators. To achieve this goal, we start by evaluating the commutators between the six canonical field operators, obtained from their spectra of Eqs.(\ref{CQME_Spectra}) through Eq.(\ref{SpOpDe}), and the polariton creation operators. After noting that the first of Eqs.(\ref{MLNF_Com_Rel}) and Eq.(\ref{Eom}) yield the relation $[ {{\bf{\hat E}}_\omega  \left( {\bf{r}} \right),{\bf{\hat g}}_{\omega 's}^\dag  \left( {\bf{n}} \right)} ] = \delta \left( {\omega  - \omega '} \right){\cal F}_{\omega s} \left( {\left. {\bf{r}} \right|{\bf{n}}} \right)$, it is straightforward to prove the commutation relations 
\begin{align} \label{Com_g+}
\left[ {{\bf{\hat A}}\left( {\bf{r}} \right),{\bf{\hat g}}_{\omega s}^\dag  \left( {\bf{n}} \right)} \right] =& \frac{1}{{i\omega }}\int {d^3 {\bf{r}}'} \,{\cal D}^ \bot  \left( {{\bf{r}} - {\bf{r}}'} \right) \cdot {\cal F}_{\omega s} \left( {\left. {{\bf{r}}'} \right|{\bf{n}}} \right), &
\left[ {{\bf{\hat \Pi }}_A \left( {\bf{r}} \right),{\bf{\hat g}}_{\omega s}^\dag  \left( {\bf{n}} \right)} \right] =&  - \varepsilon _0 \varepsilon _\omega  \left( {\bf{r}} \right){\cal F}_{\omega s} \left( {\left. {\bf{r}} \right|{\bf{n}}} \right), \nonumber  \\
\left[ {{\bf{\hat X}}^\Omega  \left( {\bf{r}} \right),{\bf{\hat g}}_{\omega s}^\dag  \left( {\bf{n}} \right)} \right] =& \frac{{\alpha ^\Omega  \left( {\bf{r}} \right)}}{{\Omega ^2  - \left( {\omega  + i\eta } \right)^2 }}{\cal F}_{\omega s} \left( {\left. {\bf{r}} \right|{\bf{n}}} \right), & 
\left[ {{\bf{\hat \Pi }}_X^\Omega  \left( {\bf{r}} \right),{\bf{\hat g}}_{\omega s}^\dag  \left( {\bf{n}} \right)} \right] =&  - i\omega \frac{{\alpha ^\Omega  \left( {\bf{r}} \right)}}{{\Omega ^2  - \left( {\omega  + i\eta } \right)^2 }}{\cal F}_{\omega s} \left( {\left. {\bf{r}} \right|{\bf{n}}} \right), \nonumber \\
\left[ {{\bf{\hat Y}}^\Omega  \left( {\bf{r}} \right),{\bf{\hat g}}_{\omega s}^\dag  \left( {\bf{n}} \right)} \right] =& \frac{1}{{i\omega }}\frac{{\beta ^\Omega  \left( {\bf{r}} \right)}}{{\Omega ^2  - \left( {\omega  + i\eta } \right)^2 }}\nabla _{\bf{r}}  \times {\cal F}_{\omega s} \left( {\left. {\bf{r}} \right|{\bf{n}}} \right), & 
\left[ {{\bf{\hat \Pi }}_Y^\Omega  \left( {\bf{r}} \right),{\bf{\hat g}}_{\omega s}^\dag  \left( {\bf{n}} \right)} \right] =& - \frac{{\beta ^\Omega  \left( {\bf{r}} \right)}}{{\Omega ^2  - \left( {\omega  + i\eta } \right)^2 }}\nabla _{\bf{r}}  \times {\cal F}_{\omega s} \left( {\left. {\bf{r}} \right|{\bf{n}}} \right),
\end{align}
involving the $s$ polariton creation operator. Analogously, by using Eq.(\ref{[E,f+]}) it is also straightforward to prove the commutation relations 
\begin{align} \label{Com_f+}
\left[ {{\bf{\hat A}}\left( {\bf{r}} \right),{\bf{\hat f}}_{\omega \nu }^\dag  \left( {{\bf{r}}'} \right)} \right] =& \frac{1}{{i\omega }}\int {d^3 {\bf{s}}} \,{\cal D}^ \bot  \left( {{\bf{r}} - {\bf{s}}} \right) \cdot {\cal G}_{\omega \nu } \left( {\left. {\bf{s}} \right|{\bf{r}}'} \right), & 
\left[ {{\bf{\hat \Pi }}_A \left( {\bf{r}} \right),{\bf{\hat f}}_{\omega \nu }^\dag  \left( {{\bf{r}}'} \right)} \right] =&  - \varepsilon _0 \varepsilon _\omega  \left( {\bf{r}} \right){\cal G}_{\omega \nu } \left( {\left. {\bf{r}} \right|{\bf{r}}'} \right) \nonumber  \\
& &
&- i\sqrt {\frac{\hbar }{{2\omega }}} \alpha ^\omega  \left( {\bf{r}} \right)\delta \left( {{\bf{r}} - {\bf{r}}'} \right)\delta _{\nu e} {\cal I}, \nonumber  \\
\left[ {{\bf{\hat X}}^\Omega  \left( {\bf{r}} \right),{\bf{\hat f}}_{\omega \nu }^\dag  \left( {{\bf{r}}'} \right)} \right] =& \frac{{\alpha ^\Omega  \left( {\bf{r}} \right)}}{{\Omega ^2  - \left( {\omega  + i\eta } \right)^2 }}{\cal G}_{\omega \nu } \left( {\left. {\bf{r}} \right|{\bf{r}}'} \right) & 
\left[ {{\bf{\hat \Pi }}_X^\Omega  \left( {\bf{r}} \right),{\bf{\hat f}}_{\omega \nu }^\dag  \left( {{\bf{r}}'} \right)} \right] =&  - i\omega \frac{{\alpha ^\Omega  \left( {\bf{r}} \right)}}{{\Omega ^2  - \left( {\omega  + i\eta } \right)^2 }}{\cal G}_{\omega \nu } \left( {\left. {\bf{r}} \right|{\bf{r}}'} \right) \nonumber \\
&+ i\sqrt {\frac{\hbar }{{2\omega }}} \delta \left( {\omega  - \Omega } \right)\delta \left( {{\bf{r}} - {\bf{r}}'} \right)\delta _{\nu e} {\cal I}, &
&+ \sqrt {\frac{{\hbar \omega }}{2}} \delta \left( {\omega  - \Omega } \right)\delta \left( {{\bf{r}} - {\bf{r}}'} \right)\delta _{\nu e} {\cal I}, \nonumber \\
\left[ {{\bf{\hat Y}}^\Omega  \left( {\bf{r}} \right),{\bf{\hat f}}_{\omega \nu }^\dag  \left( {{\bf{r}}'} \right)} \right] =& \frac{1}{{i\omega }}\frac{{\beta ^\Omega  \left( {\bf{r}} \right)}}{{\Omega ^2  - \left( {\omega  + i\eta } \right)^2 }}\nabla _{\bf{r}}  \times {\cal G}_{\omega \nu } \left( {\left. {\bf{r}} \right|{\bf{r}}'} \right) &  \left[ {{\bf{\hat \Pi }}_Y^\Omega  \left( {\bf{r}} \right),{\bf{\hat f}}_{\omega \nu }^\dag  \left( {{\bf{r}}'} \right)} \right] =&  - \frac{{\beta ^\Omega  \left( {\bf{r}} \right)}}{{\Omega ^2  - \left( {\omega  + i\eta } \right)^2 }}\nabla _{\bf{r}}  \times {\cal G}_{\omega \nu } \left( {\left. {\bf{r}} \right|{\bf{r}}'} \right) \nonumber  \\
& + \sqrt {\frac{\hbar }{{2\omega }}} \delta \left( {\omega  - \Omega } \right)\delta \left( {{\bf{r}} - {\bf{r}}'} \right)\delta _{\nu m} {\cal I}, & 
&  - i\sqrt {\frac{{\hbar \omega }}{2}} \delta \left( {\omega  - \Omega } \right)\delta \left( {{\bf{r}} - {\bf{r}}'} \right)\delta _{\nu m} {\cal I},
\end{align}
involving the $e$ and $m$ polariton creation operators. A crucial point is that Eqs.(\ref{Com_g+}) and (\ref{Com_f+}) can alternatively be regarded as equations for the unknown polariton operators ${\bf{\hat g}}_{\omega s}$ and ${\bf{\hat f}}_{\omega \nu}$ and they are satisfied by
\begin{eqnarray} \label{gf1}
&& {\bf{\hat g}}_{\omega s} \left( {\bf{n}} \right) = \frac{i}{\hbar }\int {d^3 {\bf{r}}} \; {\cal F}_{\omega s}^{T*} \left( {\left. {\bf{r}} \right|{\bf{n}}} \right) \cdot \left\{ {\varepsilon _0 \varepsilon _\omega ^* \left( {\bf{r}} \right){\bf{\hat A}}\left( {\bf{r}} \right) - \frac{1}{{i\omega }}{\bf{\hat \Pi }}_A \left( {\bf{r}} \right) + \int {d\Omega } \;\frac{{\alpha ^\Omega  \left( {\bf{r}} \right)\left[ { - i\omega {\bf{\hat X}}^\Omega  \left( {\bf{r}} \right) + {\bf{\hat \Pi }}_X^\Omega  \left( {\bf{r}} \right)} \right]}}{{\Omega ^2  - \left( {\omega  - i\eta } \right)^2 }}} \right\} \nonumber \\ 
&& \quad \quad + \frac{1}{{\hbar \omega }}\int {d^3 {\bf{r}}} \left[ {{\cal F}_{\omega s}^{T*} \left( {\left. {\bf{r}} \right|{\bf{n}}} \right) \times \mathord{\buildrel{\lower3pt\hbox{$\scriptscriptstyle\leftarrow$}} 
\over \nabla } _{\bf{r}} } \right] \cdot \int {d\Omega } \;\frac{{\beta ^\Omega  \left( {\bf{r}} \right)\left[ { - i\omega {\bf{\hat Y}}^\Omega  \left( {\bf{r}} \right) + {\bf{\hat \Pi }}_Y^\Omega  \left( {\bf{r}} \right)} \right]}}{{\Omega ^2  - \left( {\omega  - i\eta } \right)^2 }}, \nonumber \\
&& {\bf{\hat f}}_{\omega \nu } \left( {\bf{r}} \right) = \frac{i}{\hbar }\int {d^3 {\bf{r}}'} {\cal G}_{\omega \nu }^{T*} \left( {\left. {{\bf{r}}'} \right|{\bf{r}}} \right) \cdot \left\{ {\varepsilon _0 \varepsilon _\omega ^* \left( {{\bf{r}}'} \right){\bf{\hat A}}\left( {{\bf{r}}'} \right) - \frac{1}{{i\omega }}{\bf{\hat \Pi }}_A \left( {{\bf{r}}'} \right) + \int  {d\Omega } \frac{{\alpha ^\Omega  \left( {{\bf{r}}'} \right)\left[ { - i\omega {\bf{\hat X}}^\Omega  \left( {{\bf{r}}'} \right) + {\bf{\hat \Pi }}_X^\Omega  \left( {{\bf{r}}'} \right)} \right]}}{{\Omega ^2  - \left( {\omega  - i\eta } \right)^2 }}} \right\} \nonumber \\ 
&& \quad \quad  + \frac{1}{{\hbar \omega }}\int {d^3 {\bf{r}}'} \left[ {{\cal G}_{\omega \nu }^{T*} \left( {\left. {{\bf{r}}'} \right|{\bf{r}}} \right) \times \mathord{\buildrel{\lower3pt\hbox{$\scriptscriptstyle\leftarrow$}} 
\over \nabla } _{{\bf{r}}'} } \right] \cdot \int {d\Omega } \frac{{\beta ^\Omega  \left( {{\bf{r}}'} \right)\left[ { - i\omega {\bf{\hat Y}}^\Omega  \left( {{\bf{r}}'} \right) + {\bf{\hat \Pi }}_Y^\Omega  \left( {{\bf{r}}'} \right)} \right]}}{{\Omega ^2  - \left( {\omega  - i\eta } \right)^2 }} \nonumber \\ 
&& \quad \quad  + \frac{{\delta _{\nu e} }}{{\sqrt {2\hbar \omega } }}\left[ {\alpha ^\omega  \left( {\bf{r}} \right){\bf{\hat A}}\left( {\bf{r}} \right) - i\omega {\bf{\hat X}}^\omega  \left( {\bf{r}} \right) + {\bf{\hat \Pi }}_X^\omega  \left( {\bf{r}} \right)} \right] + \frac{{i\delta _{\nu m} }}{{\sqrt {2\hbar \omega } }}\left[ { - i\omega {\bf{\hat Y}}^\omega  \left( {\bf{r}} \right) + {\bf{\hat \Pi }}_Y^\omega  \left( {\bf{r}} \right)} \right],
\end{eqnarray}
as can be easily verified by inserting Eqs.(\ref{gf1}) into Eqs.(\ref{Com_g+}) and (\ref{Com_f+}) and using  the canonical commutation relations of Eqs.(\ref{CQME_Com_Rel}). Equations (\ref{gf1}) can be given a more concise and transparent form after integrating by parts their contributions containing the curls $ \times \mathord{\buildrel{\lower3pt\hbox{$\scriptscriptstyle\leftarrow$}} 
\over \nabla } _{\bf{r}}$ and ${ \times \mathord{\buildrel{\lower3pt\hbox{$\scriptscriptstyle\leftarrow$}} 
\over \nabla } _{{\bf{r}}'} }$ by using the second of Eqs.(\ref{IntPart}) and the asymptotic magnetic condition $\beta ^\Omega  \left( \infty  \right) = 0$, yielding
\begin{eqnarray} \label{gf2}
 {\bf{\hat g}}_{\omega s} \left( {\bf{n}} \right) &=& \int {d^3 {\bf{r}}} \,{\cal F}_{\omega s}^{T*} \left( {\left. {\bf{r}} \right|{\bf{n}}} \right) \cdot {\bf{\hat F}}_\omega  \left( {\bf{r}} \right), \nonumber \\
{\bf{\hat f}}_{\omega \nu } \left( {\bf{r}} \right) &=& \int {d^3 {\bf{r}}'} \;{\cal G}_{\omega \nu }^{T*} \left( {\left. {{\bf{r}}'} \right|{\bf{r}}} \right) \cdot {\bf{\hat F}}_\omega  \left( {{\bf{r}}'} \right) + {\bf{\hat h}}_{\omega \nu } \left( {\bf{r}} \right)
\end{eqnarray}
where we have set
\begin{eqnarray} \label{Fom}
{\bf{\hat F}}_\omega   &=& \frac{i}{\hbar }\left( {\varepsilon _0 \varepsilon _\omega ^* {\bf{\hat A}} - \frac{1}{{i\omega }}{\bf{\hat \Pi }}_A } \right) + \frac{1}{{\hbar \omega }}\int {d\Omega } \frac{{i\omega \alpha ^\Omega  \left( { - i\omega {\bf{\hat X}}^\Omega   + {\bf{\hat \Pi }}_X^\Omega  } \right) - \nabla  \times \left[ {\beta ^\Omega  \left( { - i\omega {\bf{\hat Y}}^\Omega   + {\bf{\hat \Pi }}_Y^\Omega  } \right)} \right]}}{{\Omega ^2  - \left( {\omega  - i\eta } \right)^2 }}, \nonumber \\
{\bf{\hat h}}_{\omega \nu }  &=& \frac{1}{{\sqrt {2\hbar \omega } }}\left[ {\delta _{\nu e} \left( {\alpha ^\omega  {\bf{\hat A}} - i\omega {\bf{\hat X}}^\omega   + {\bf{\hat \Pi }}_X^\omega  } \right) + i \delta _{\nu m} \left( { - i\omega {\bf{\hat Y}}^\omega   + {\bf{\hat \Pi }}_Y^\omega  } \right)} \right].
\end{eqnarray}
Equations (\ref{gf2}) combined with Eqs.(\ref{Fom}) provide the precise analytical mapping between the canonical CQME field operators in the Schr\"odinger picture and the polariton operators. This mapping is structured around two fundamental operators. The first is the macroscopic excitation density ${\bf{\hat F}}_\omega$, which encapsulates the non-local collective optical response of the dressed electromagnetic field. In the vacuum limit, where the material reservoir contributions vanish, ${\bf{\hat F}}_\omega$ reduces to the standard linear combination of the bare vector potential and its conjugate momentum used in ordinary quantum electrodynamics to extract the free-photon annihilation operators. In the presence of a dispersive object, ${\bf{\hat F}}_\omega$ naturally generalizes this concept, coupling the purely electromagnetic degrees of freedom with the continuum of reservoir fields. The second is the bare material excitation ${\bf{\hat h}}_{\omega \nu }$, which accounts for the local state of the internal reservoirs at the specific frequency $\omega$ and represents the bare mechanical oscillators decoupled from the macroscopic electromagnetic field. In the electric component ${\bf{\hat h}}_{\omega e}$, the local vector potential ${\bf{\hat A}}$ implements the minimal-coupling shift required to convert the bare canonical momentum ${\bf{\hat \Pi}}_X^\omega$ into the physical momentum of the electric dipoles. Correspondingly, the magnetic component ${\bf{\hat h}}_{\omega m}$ exhibits no such shift, reflecting the structure of the macroscopic Lagrangian where magnetic reservoirs couple directly to the curl of the vector potential.

The structural distinction between the non-local macroscopic excitation density ${\bf{\hat F}}_\omega$ and the strictly local bare material excitation ${\bf{\hat h}}_{\omega \nu }$ clarifies the asymmetry between the scattering polaritons and the electric and magnetic polaritons in Eqs.(\ref{gf2}). The scattering polariton operator ${\bf{\hat g}}_{\omega s}$ is determined exclusively by the spatial projection of ${\bf{\hat F}}_\omega$ onto the scattering kernel ${\cal F}_{\omega s}$, without any direct contribution from ${\bf{\hat h}}_{\omega \nu }$. Because scattering modes describe asymptotic free-space radiation interacting with the object, they lack a local material source; therefore, the projection of the macroscopic dressed field fully defines their state. Conversely, the electric and magnetic polariton operators ${\bf{\hat f}}_{\omega \nu }$ describe active material sources localized within the lossy medium. Their definition requires both the bare material excitation ${\bf{\hat h}}_{\omega \nu }$ and the non-local collective contribution propagated by the kernels ${\cal G}_{\omega \nu }$, which accounts for the macroscopic radiation reaction and the electromagnetic dressing. Furthermore, the local operator ${\bf{\hat h}}_{\omega \nu }$ transfers the spatial independence of the bare reservoir fields ${\bf{\hat X}}^\Omega$ and ${\bf{\hat Y}}^\Omega$ to the internal polaritons, ensuring that their fundamental commutation relations correctly yield the required spatial Dirac delta function.

To verify the consistency of this inversion, we note that, we observe that the first of Eqs.(\ref{gf2}) inherently guarantees the transversality condition ${{\bf{n}} \cdot {\bf{\hat g}}_{\omega s} \left( {\bf{n}} \right) = 0}$ due to the structural property of the modal dyadic ${\cal F}_\omega  \left( {\left. {\bf{r}} \right|{\bf{n}}} \right) \cdot {\bf{n}} = 0$. Similarly, the second of Eqs.(\ref{gf2}) rigorously yields ${\bf{\hat f}}_{\omega \nu } \left( {\bf{r}} \right) = 0$ for coordinates outside the object's volume, correctly reflecting the spatial confinement of the material degrees of freedom, since the dyadic Green's function ${\cal G}_{\omega } \left( {\left. {{\bf{r}}'} \right|{\bf{r}}} \right)$ and the local material coefficients identically vanish for ${\bf{r}} \notin V$.

\section{Bosonic algebra of the polariton operators}
The expressions for the polariton operators in Eqs.(\ref{gf2}) were derived in Sec.III within the MLNF framework, relying on the bosonic commutation relations of Eqs.(\ref{MLNF_Com_Rel}) to obtain Eqs.(\ref{Com_g+}) and (\ref{Com_f+}). Irrespective of their derivation, Eqs.(\ref{gf2}) serve as the starting point for a direct derivation of the MLNF from the CQME, which is the main focus of the present paper. The key insight is that Eqs.(\ref{gf2}) depend exclusively on the six canonical field operators of the CQME and the scattering, electric, and magnetic dyadic kernels of macroscopic classical electrodynamics; consequently, they can be used to define polariton operators directly within the CQME framework. In this context, a rigorous derivation of the MLNF from the CQME requires demonstrating that the polariton operators in Eqs.(\ref{gf2}) satisfy the bosonic commutation relations of Eqs.(\ref{MLNF_Com_Rel}) as a direct consequence of the canonical commutation relations in Eqs.(\ref{CQME_Com_Rel}). Furthermore, it must be shown that the CQME Hamiltonian in Eq.(\ref{CQME_H}) coincides with the MLNF Hamiltonian in the first of Eqs.(\ref{MLNF_HE}).

This logical sequence conceptually mirrors the standard algebraic treatment of the elementary quantum harmonic oscillator. There, for an oscillator of mass $m$ and frequency $\omega$, one introduces the annihilation and creation operators explicitly as the linear combinations $\hat{a}=\frac{1}{\sqrt{2\hbar m\omega}}(m\omega\hat{q}+i\hat{p})$ and $\hat{a}^\dag=\frac{1}{\sqrt{2\hbar m\omega}}(m\omega\hat{q}-i\hat{p})$ of the canonical coordinate $\hat{q}$ and momentum $\hat{p}$. The theoretical framework is then established by demonstrating that the canonical relation $[\hat{q},\hat{p}]=i\hbar$ directly yields the bosonic commutator $[\hat{a},\hat{a}^\dag]=1$, and that these operators diagonalize the Hamiltonian into the form $\frac{1}{2}\hbar\omega(\hat{a}^\dag\hat{a} +\hat{a}\hat{a}^\dag)$.

By strict analogy, the present section derives the macroscopic bosonic commutation relations for the polaritons from the CQME canonical commutators, while Sec.V addresses the exact diagonalization of the CQME Hamiltonian. Although the subsequent algebraic derivation is highly involved, this mathematical procedure is indispensable to dispel any conceptual doubts and definitively validate the theoretical consistency of the MLNF. For notational convenience, we hereafter denote the electric and magnetic kernels in Eqs.(\ref{Kernels}) (and their conjugate transposes) by
\begin{align} \label{G-T}
{\cal G}_{\omega \nu } \left( {\left. {\bf{r}} \right|{\bf{r}}'} \right) =& {\cal G}_\omega  \left( {\left. {\bf{r}} \right|{\bf{r}}'} \right) \mathord{\buildrel{\lower3pt\hbox{$\scriptscriptstyle\leftarrow$}} 
\over T} _{\omega \nu } \left( {{\bf{r}}'} \right), & \mathord{\buildrel{\lower3pt\hbox{$\scriptscriptstyle\leftarrow$}} 
\over T} _{\omega \nu } \left( {{\bf{r}}'} \right) =& i\sqrt {\frac{{\hbar k_\omega ^4 }}{{\pi \varepsilon _0 }}} \left\{ {\sqrt {{\mathop{\rm Im}\nolimits} \left[ {\varepsilon _\omega  \left( {{\bf{r}}'} \right)} \right]} \delta _{\nu e}  - \frac{{ \times \mathord{\buildrel{\lower3pt\hbox{$\scriptscriptstyle\leftarrow$}} 
\over \nabla } _{\bf{r}'} }}{{k_\omega  }}\sqrt {{\mathop{\rm Im}\nolimits} \left[ {\frac{{ - 1}}{{\mu _\omega  \left( {{\bf{r}}'} \right)}}} \right]} \delta _{\nu m} } \right\}, \nonumber \\
{\cal G}_{\omega \nu }^{T*} \left( {\left. {\bf{r}} \right|{\bf{r}}'} \right) =& \vec T_{\omega \nu } \left( {{\bf{r}}'} \right){\cal G}_\omega ^{T*} \left( {\left. {\bf{r}} \right|{\bf{r}}'} \right), &
\vec T_{\omega \nu } \left( {{\bf{r}}'} \right) =& i\sqrt {\frac{{\hbar k_\omega ^4 }}{{\pi \varepsilon _0 }}} \left\{ { - \delta _{\nu e} \sqrt {{\mathop{\rm Im}\nolimits} \left[ {\varepsilon _\omega  \left( {{\bf{r}}'} \right)} \right]}  - \delta _{\nu m} \sqrt {{\mathop{\rm Im}\nolimits} \left[ {\frac{{ - 1}}{{\mu _\omega  \left( {{\bf{r}}'} \right)}}} \right]} \frac{{\nabla _{{\bf{r}}'}  \times }}{{k_\omega  }}} \right\}
\end{align}
where $\mathord{\buildrel{\lower3pt\hbox{$\scriptscriptstyle\leftarrow$}} 
\over T} _{\omega \nu }$ and $\vec T_{\omega \nu }$ are vector operators acting from the right and from the left, respectively.

\subsection{Annihilation-creation commutator for scattering polaritons $\left[ {{\bf{\hat g}}_{\omega s} \left( {\bf{n}} \right),{\bf{\hat g}}_{\omega 's}^\dag  \left( {{\bf{n}}'} \right)} \right] = \delta \left( {\omega  - \omega '} \right)\delta \left( {o_{\bf{n}}  - o_{{\bf{n}}'} } \right){\cal I}_{\bf{n}}$}
From the first of Eqs.(\ref{gf2}) and the first of Eqs.(\ref{Kernels}) we get 
\begin{equation} \label{COMM_A_0}
\left[ {{\bf{\hat g}}_{\omega s} \left( {\bf{n}} \right),{\bf{\hat g}}_{\omega 's}^\dag  \left( {{\bf{n}}'} \right)} \right] = \frac{{\sqrt {k_\omega  k_{\omega '} } }}{{16\pi ^3 c}}\left( {\frac{{\hbar c}}{{\varepsilon _0 }}k_\omega  k_{\omega '} } \right)\int {d^3 {\bf{r}}} \int {d^3 {\bf{r}}'} {\cal F}_\omega ^{T*} \left( {\left. {\bf{r}} \right|{\bf{n}}} \right) \cdot \left[ {{\bf{\hat F}}_\omega  \left( {\bf{r}} \right),{\bf{\hat F}}_{\omega '}^\dag  \left( {{\bf{r}}'} \right)} \right] \cdot {\cal F}_{\omega'}  \left( {\left. {{\bf{r}}'} \right|{\bf{n}}} \right)
\end{equation}
which, resorting to Eq.(\ref{F_Fc}) with ${\cal Q}_\omega  \left( {\bf{r}} \right) = {\cal F}_{\omega } \left( {\left. {\bf{r}} \right|{\bf{n}}} \right)$ and ${\cal P}_\omega  \left( {\bf{r}} \right) = {\cal F}_{\omega } \left( {\left. {\bf{r}} \right|{\bf{n}}} \right)$, becomes
\begin{eqnarray} \label{COMM_A_1}
&& \left[ {{\bf{\hat g}}_{\omega s} \left( {\bf{n}} \right),{\bf{\hat g}}_{\omega 's}^\dag  \left( {{\bf{n}}'} \right)} \right] =   \frac{{\sqrt {k_\omega  k_{\omega '} } }}{{16\pi ^3 c}}\left\{ {k_\omega  \int {d^3 {\bf{r}}} \left\{ {\int {d^3 {\bf{r}}'\;{\cal D}^ \bot  \left( {{\bf{r}} - {\bf{r}}'} \right) \cdot } \left[ {\varepsilon _\omega  \left( {{\bf{r}}'} \right){\cal F}_{\omega s} \left( {\left. {{\bf{r}}'} \right|{\bf{n}}} \right)} \right]} \right\}^{T*}  \cdot {\cal F}_{\omega '} \left( {\left. {\bf{r}} \right|{\bf{n}}} \right)} \right. \nonumber \\ 
&& \quad \quad  + k_{\omega '} \int {d^3 {\bf{r}}\;} {\cal F}_\omega ^{T*} \left( {\left. {\bf{r}} \right|{\bf{n}}} \right) \cdot \left\{ {\int {d^3 {\bf{r}}'} \;{\cal D}^ \bot  \left( {{\bf{r}} - {\bf{r}}'} \right) \cdot \left[ {\varepsilon _{\omega '} \left( {{\bf{r}}'} \right){\cal F}_{\omega '} \left( {\left. {\bf{r}} \right|{\bf{n}}} \right)} \right]} \right\} \nonumber \\ 
&& \quad \quad   - \frac{1}{{k_\omega   - k_{\omega '}  - i\eta }}\left\{ {\left( {k_\omega   - k_{\omega '} } \right)\int {d^3 {\bf{r}}} \left[ {k_\omega  \varepsilon _\omega ^* \left( {\bf{r}} \right) + k_{\omega '} \varepsilon _{\omega '} \left( {\bf{r}} \right)} \right]{\cal F}_\omega ^{T*} \left( {\left. {\bf{r}} \right|{\bf{n}}} \right) \cdot {\cal F}_{\omega '} \left( {\left. {\bf{r}} \right|{\bf{n}}} \right)} \right\} \nonumber \\ 
&& \quad \quad   - \frac{1}{{k_\omega   - k_{\omega '}  - i\eta }}\int {d^3 {\bf{r}}} \;\left\{ {\left[ {\left( {\nabla _{\bf{r}}  \times \frac{1}{{\mu _\omega  \left( {\bf{r}} \right)}}\nabla _{\bf{r}}  \times } \right) - k_\omega ^2 \varepsilon _\omega  \left( {\bf{r}} \right)} \right]{\cal F}_{\omega s} \left( {\left. {\bf{r}} \right|{\bf{n}}} \right)} \right\}^{T*}  \cdot {\cal F}_{\omega '} \left( {\left. {\bf{r}} \right|{\bf{n}}} \right) \nonumber \\ 
&& \quad \quad   + \frac{1}{{k_\omega   - k_{\omega '}  - i\eta }}\int {d^3 {\bf{r}}} \;{\cal F}_\omega ^{T*} \left( {\left. {\bf{r}} \right|{\bf{n}}} \right) \cdot \left\{ {\left[ {\left( {\nabla _{\bf{r}}  \times \frac{1}{{\mu _{\omega '} \left( {\bf{r}} \right)}}\nabla _{\bf{r}}  \times } \right) - k_{\omega '}^2 \varepsilon _{\omega '} \left( {\bf{r}} \right)} \right]{\cal F}_{\omega '} \left( {\left. {\bf{r}} \right|{\bf{n}}} \right)} \right\} \nonumber \\ 
&& \quad \quad  \left. { + \int\limits_{S_\infty  } {dS_r } \;\frac{{\left[ {{\bf{u}}_r  \times {\cal F}_\omega  \left( {\left. {\bf{r}} \right|{\bf{n}}} \right)} \right]^{T*}  \cdot \left[ {\nabla _{\bf{r}}  \times {\cal F}_{\omega '} \left( {\left. {\bf{r}} \right|{\bf{n}}'} \right)} \right] - \left[ {\nabla _{\bf{r}}  \times {\cal F}_\omega  \left( {\left. {\bf{r}} \right|{\bf{n}}} \right)} \right]^{T*}  \cdot \left[ {{\bf{u}}_r  \times {\cal F}_{\omega '} \left( {\left. {\bf{r}} \right|{\bf{n}}'} \right)} \right]}}{{k_\omega   - k_{\omega '}  - i\eta }}} \right\}.
\end{eqnarray}
This complex expression can be significantly simplified by observing that the integrations over ${\bf r}'$ in the first and second terms are governed by the transverse property of the modal dyadics in Eq.(\ref{Modal_Tra}). Furthermore, the fourth and fifth contributions identically vanish as a direct consequence of the homogeneous Helmholtz equations satisfied by the modal dyadics in Eq.(\ref{GS}), while the asymptotic surface integral in the last term vanishes according to the detailed derivation in Appendix E. Consequently, we obtain
\begin{eqnarray} \label{COMM_A_2}
&& \left[ {{\bf{\hat g}}_{\omega s} \left( {\bf{n}} \right),{\bf{\hat g}}_{\omega 's}^\dag  \left( {{\bf{n}}'} \right)} \right] = \frac{{\sqrt {k_\omega  k_{\omega '} } }}{{16\pi ^3 c}}\left\{ {\int {d^3 {\bf{r}}} \left[ {k_\omega  \varepsilon _\omega ^* \left( {\bf{r}} \right) + k_{\omega '} \varepsilon _{\omega '} \left( {\bf{r}} \right)} \right]{\cal F}_\omega ^{T*} \left( {\left. {\bf{r}} \right|{\bf{n}}} \right) \cdot {\cal F}_{\omega '} \left( {\left. {\bf{r}} \right|{\bf{n}}} \right)} \right. \nonumber \\ 
&& \quad \quad  \left. { - \frac{1}{{k_\omega   - k_{\omega '}  - i\eta }}\left\{ {\left( {k_\omega   - k_{\omega '} } \right)\int {d^3 {\bf{r}}} \left[ {k_\omega  \varepsilon _\omega ^* \left( {\bf{r}} \right) + k_{\omega '} \varepsilon _{\omega '} \left( {\bf{r}} \right)} \right]{\cal F}_\omega ^{T*} \left( {\left. {\bf{r}} \right|{\bf{n}}} \right) \cdot {\cal F}_{\omega '} \left( {\left. {\bf{r}} \right|{\bf{n}}} \right)} \right\}} \right\}.
\end{eqnarray}
At this point, the spatial integral warrants careful analysis. Since $\varepsilon _\omega  \left( \infty  \right) = 1$ and the modal dyadics $\mathcal{F}_\omega$ contain an asymptotic plane-wave contribution, the integral of the terms proportional to $\varepsilon_\omega$ do not converge in the ordinary sense and must be interpreted as a generalized function within the framework of distribution theory. Similarly, the Plemelj factor
\begin{equation} \label{Plemelj}
\frac{1}{{k_\omega   - k_{\omega '}  - i\eta }} = P\frac{1}{{k_\omega   - k_{\omega '} }} + i\pi \delta \left( {k_\omega   - k_{\omega '} } \right)
\end{equation}
in Eq.(\ref{COMM_A_2}) is a generalized function. This requires strict adherence to the sequence of operations dictated by the inner curly brackets, exactly as it naturally emerged from the derivation, which strictly forbids the naive algebraic cancellation $(k_\omega - k_{\omega'})/(k_\omega - k_{\omega'} - i\eta) = 1$. Such a premature simplification would incorrectly lead to a vanishing result. To properly handle these singular terms, we observe that the integral 
\begin{equation}
\int {d^3 {\bf{r}}} \left\{ {k_\omega  \left[ {1 - \varepsilon _\omega ^* \left( {\bf{r}} \right)} \right] + k_{\omega '} \left[ {1 - \varepsilon _{\omega '} \left( {\bf{r}} \right)} \right]} \right\}{\cal F}_\omega ^{T*} \left( {\left. {\bf{r}} \right|{\bf{n}}} \right) \cdot {\cal F}_{\omega '} \left( {\left. {\bf{r}} \right|{\bf{n}}} \right)
\end{equation}
is convergent, as its integrand vanishes at infinity. By adding and subtracting this term within the main brackets of Eq.(\ref{COMM_A_2}), we obtain
\begin{eqnarray} \label{COMM_A_3}
&& \left[ {{\bf{\hat g}}_{\omega s} \left( {\bf{n}} \right),{\bf{\hat g}}_{\omega 's}^\dag  \left( {{\bf{n}}'} \right)} \right] = \frac{{\sqrt {k_\omega  k_{\omega '} } }}{{16\pi ^3 c}}\left( {k_\omega   + k_{\omega '} } \right)\left\{ {\int {d^3 {\bf{r}}} {\cal F}_\omega ^{T*} \left( {\left. {\bf{r}} \right|{\bf{n}}} \right) \cdot {\cal F}_{\omega '} \left( {\left. {\bf{r}} \right|{\bf{n}}} \right)} \right. \nonumber \\ 
&& \quad \quad  \left. { - \frac{1}{{k_\omega   - k_{\omega '}  - i\eta }}\left[ {\left( {k_\omega   - k_{\omega '} } \right)\int {d^3 {\bf{r}}} {\cal F}_\omega ^{T*} \left( {\left. {\bf{r}} \right|{\bf{n}}} \right) \cdot {\cal F}_{\omega '} \left( {\left. {\bf{r}} \right|{\bf{n}}} \right)} \right]} \right\}.
\end{eqnarray}
To complete the proof, we employ the integral relation from Eq.(\ref{F-F_int}):
\begin{equation} \label{F-F-int_main}
\int {d^3 {\bf{r}}} {\cal F}_\omega ^{T*} \left( {\left. {\bf{r}} \right|{\bf{n}}} \right) \cdot {\cal F}_{\omega '} \left( {\left. {\bf{r}} \right|{\bf{n}}} \right) = \left( {2\pi } \right)^3 \frac{{\delta \left( {k_\omega   - k_{\omega '} } \right)}}{{k_\omega ^2 }}\delta \left( {o_{\bf{n}}  - o_{{\bf{n}}'} } \right){\cal I}_{\bf{n}}  + \frac{{{\cal C}_{\omega ,\omega '} \left( {\left. {\bf{n}} \right|{\bf{n}}'} \right)}}{{k_\omega ^2  - k_{\omega '}^2  - i\eta }}.
\end{equation}
Crucially, since the dyadic ${\cal C}_{\omega ,\omega '}$ is a regular function that does not contain a delta contribution $\delta(k_\omega - k_{\omega'})$, and using the identity $\left( {k_\omega ^2  - k_{\omega '}^2  - i\eta } \right)^{-1} = (k_\omega + k_{\omega'})^{-1} [P(k_\omega - k_{\omega'})^{-1} + i\pi \delta(k_\omega - k_{\omega'})]$, it follows that:
\begin{equation} \label{F-F-dis_main}
\frac{1}{{k_\omega   - k_{\omega '}  - i\eta }}\left[ {\left( {k_\omega   - k_{\omega '} } \right)\int {d^3 {\bf{r}}} {\cal F}_\omega ^{T*} \left( {\left. {\bf{r}} \right|{\bf{n}}} \right) \cdot {\cal F}_{\omega '} \left( {\left. {\bf{r}} \right|{\bf{n}}} \right)} \right] = \frac{{{\cal C}_{\omega ,\omega '} \left( {\left. {\bf{n}} \right|{\bf{n}}'} \right)}}{{k_\omega ^2  - k_{\omega '}^2  - i\eta }}.
\end{equation}
By substituting Eqs.(\ref{F-F-int_main}) and (\ref{F-F-dis_main}) into Eq.(\ref{COMM_A_3}), the regular parts involving ${\cal C}_{\omega ,\omega '}$ cancel each other perfectly. The remaining delta-function term yields the final bosonic commutation relation $\left[ {{\bf{\hat g}}_{\omega s} \left( {\bf{n}} \right),{\bf{\hat g}}_{\omega 's}^\dag  \left( {{\bf{n}}'} \right)} \right] = \delta \left( {\omega  - \omega '} \right)\delta \left( {o_{\bf{n}}  - o_{{\bf{n}}'} } \right){\cal I}_{\bf{n}}$. Physically, this result originates from the infinite extent of the vacuum relative to the finite-size object. The plane-wave component of the modal dyadic ${\cal F}_{\omega s}$, which represents the term remaining in the vacuum limit where the object is absent, dominates the spatial integration. Although the dispersive response of the object introduces localized contributions to the macroscopic excitation density $\hat{\bf F}_\omega$, these terms are regular and are exactly compensated within the integral relations.

\subsection{Annihilation-annihilation commutator for scattering polaritons $\left[ {{\bf{\hat g}}_{\omega s} \left( {\bf{n}} \right),{\bf{\hat g}}_{\omega 's} \left( {{\bf{n}}'} \right)} \right] = 0$}
Again from the first of Eqs.(\ref{gf2}) and the first of Eqs.(\ref{Kernels}) we get 
\begin{equation} \label{COMM_B_0}
\left[ {{\bf{\hat g}}_{\omega s} \left( {\bf{n}} \right),{\bf{\hat g}}_{\omega 's} \left( {{\bf{n}}'} \right)} \right] = \frac{{\sqrt {k_\omega  k_{\omega '} } }}{{16\pi ^3 c}}\left( {\frac{{\hbar c}}{{\varepsilon _0 }}k_\omega  k_{\omega '} } \right)\int {d^3 {\bf{r}}} \int {d^3 {\bf{r}}'} {\cal F}_\omega ^{T*} \left( {\left. {\bf{r}} \right|{\bf{n}}} \right) \cdot \left[ {{\bf{\hat F}}_\omega  \left( {\bf{r}} \right),{\bf{\hat F}}_{\omega '} \left( {{\bf{r}}'} \right)} \right] \cdot {\cal F}_{\omega'}^* \left( {\left. {{\bf{r}}'} \right|{\bf{n}}'} \right)
\end{equation}
which, resorting to  Eq.(\ref{F_F}) with ${\cal Q}_\omega  \left( {\bf{r}} \right) = {\cal F}_{\omega } \left( {\left. {\bf{r}} \right|{\bf{n}}} \right)$ and ${\cal P}_\omega  \left( {\bf{r}} \right) = {\cal F}_{\omega } \left( {\left. {\bf{r}} \right|{\bf{n}}'} \right)$, becomes 
\begin{eqnarray} \label{COMM_B_1}
&& \left[ {{\bf{\hat g}}_{\omega s} \left( {\bf{n}} \right),{\bf{\hat g}}_{\omega 's} \left( {{\bf{n}}'} \right)} \right] = \frac{{\sqrt {k_\omega  k_{\omega '} } }}{{16\pi ^3 c}}\left\{ { - k_\omega  \int {d^3 {\bf{r}}} \left\{ {\int {d^3 {\bf{r}}'} {\cal D}^\parallel  \left( {{\bf{r}} - {\bf{r}}'} \right) \cdot \left[ {\varepsilon _\omega  \left( {{\bf{r}}'} \right){\cal F}_\omega  \left( {\left. {{\bf{r}}'} \right|{\bf{n}}} \right)} \right]} \right\}^{T*}  \cdot {\cal F}_{\omega '}^* \left( {\left. {\bf{r}} \right|{\bf{n}}'} \right)} \right. \nonumber \\ 
&& \quad \quad   + k_{\omega '} \int {d^3 {\bf{r}}} \;{\cal F}_\omega ^{T*} \left( {\left. {\bf{r}} \right|{\bf{n}}} \right) \cdot \left\{ {\int {d^3 {\bf{r}}'} {\cal D}^\parallel  \left( {{\bf{r}} - {\bf{r}}'} \right) \cdot \left[ {\varepsilon _{\omega '} \left( {{\bf{r}}'} \right){\cal F}_{\omega '} \left( {\left. {{\bf{r}}'} \right|{\bf{n}}'} \right)} \right]} \right\}^*  \nonumber \\ 
&& \quad \quad   - \frac{1}{{\left( {k_\omega   + k_{\omega '} } \right)}}\int {d^3 {\bf{r}}\;} \left\{ {\left[ {\left( {\nabla  \times \frac{1}{{\mu _\omega  \left( {\bf{r}} \right)}}\nabla _{\bf{r}}  \times } \right) - k_\omega ^2 \varepsilon _\omega  \left( {\bf{r}} \right)} \right]{\cal F}_\omega  \left( {\left. {\bf{r}} \right|{\bf{n}}} \right)} \right\}^{T*}  \cdot {\cal F}_{\omega '}^* \left( {\left. {\bf{r}} \right|{\bf{n}}'} \right) \nonumber \\ 
&& \quad \quad   + \frac{1}{{\left( {k_\omega   + k_{\omega '} } \right)}}\int {d^3 {\bf{r}}} \;{\cal F}_\omega ^{T*} \left( {\left. {\bf{r}} \right|{\bf{n}}} \right) \cdot \left\{ {\left[ {\left( {\nabla _{\bf{r}}  \times \frac{1}{{\mu _{\omega '} \left( {\bf{r}} \right)}}\nabla _{\bf{r}}  \times } \right) - k_{\omega '}^2 \varepsilon _{\omega '} \left( {\bf{r}} \right)} \right]{\cal F}_{\omega '} \left( {\left. {\bf{r}} \right|{\bf{n}}'} \right)} \right\}^*  \nonumber \\ 
&& \quad \quad  \left. { + \int\limits_{S_\infty  } {dS_r } \frac{{\left\{ {\left[ {{\bf{u}}_r  \times {\cal F}_\omega  \left( {\left. {\bf{r}} \right|{\bf{n}}} \right)} \right]^T  \cdot \left[ {\nabla _{\bf{r}}  \times {\cal F}_{\omega '} \left( {\left. {\bf{r}} \right|{\bf{n}}'} \right)} \right] - \left[ {\nabla _{\bf{r}}  \times {\cal F}_\omega  \left( {\left. {\bf{r}} \right|{\bf{n}}} \right)} \right]^T  \cdot \left[ {{\bf{u}}_r  \times {\cal F}_{\omega '} \left( {\left. {\bf{r}} \right|{\bf{n}}'} \right)} \right]} \right\}^* }}{{\left( {k_\omega   + k_{\omega '} } \right)}}} \right\}. 
\end{eqnarray}
Similarly to Eq.(\ref{COMM_A_1}), Eq.(\ref{COMM_B_1}) can be significantly simplified since the first and second contributions vanish due to the longitudinal property of the modal dyadic in the second of Eqs.(\ref{Modal_Tra}), the third and fourth terms vanish due to the Helmholtz equation satisfied by the modal dyadic in the first of Eqs.(\ref{GS}) and the asymptotic surface integral in the fifth contribution vanish as detailed in Appendix E (see the second of Eqs.(\ref{Asy_Int})). Therefore we conclude that $\left[ {{\bf{\hat g}}_{\omega s} \left( {\bf{n}} \right),{\bf{\hat g}}_{\omega 's} \left( {{\bf{n}}'} \right)} \right] = 0$. Physically, the vanishing of this commutator is due to the transversality of the macroscopic radiation. The commutator $[\hat{\bf F}_\omega, \hat{\bf F}_{\omega'} ]$ yields a longitudinal contribution, characterized by the dyadic ${\cal D}^\parallel$. Because the material response associated with the scattering modes, $\varepsilon_\omega {\cal F}_{\omega s}$, is transverse, its spatial projection onto this longitudinal component is zero. This ensures that the Bose-Einstein statistics of the scattering polaritons $\hat{\bf g}_{\omega s}$ are maintained.

\subsection{Annihilation-creation commutator for material polaritons $\left[ {{\bf{\hat f}}_{\omega \nu } \left( {\bf{r}} \right),{\bf{\hat f}}_{\omega '\nu '}^\dag  \left( {{\bf{r}}'} \right)} \right] = \delta _{\nu \nu '} \delta \left( {\omega  - \omega '} \right)\delta \left( {{\bf{r}} - {\bf{r}}'} \right){\cal I}$}
From Eqs.(\ref{gf2}) and the second of Eqs.(\ref{G-T}) we get 
\begin{equation} \label{COMM_C_1}
\left[ {{\bf{\hat f}}_{\omega \nu } \left( {\bf{r}} \right),{\bf{\hat f}}_{\omega '\nu '}^\dag  \left( {{\bf{r}}'} \right)} \right] = \sum\limits_{j = 1}^4 {{\cal L}_{\omega \nu ,\omega '\nu '}^{\left( j \right)} \left( {\left. {\bf{r}} \right|{\bf{r}}'} \right)} 
\end{equation}
where
\begin{eqnarray} \label{L_dyad}
 {\cal L}_{\omega \nu ,\omega '\nu '}^{\left( 1 \right)} \left( {\left. {\bf{r}} \right|{\bf{r}}'} \right) &=& \vec T_{\omega \nu } \left( {\bf{r}} \right)\left\{ {\int {d^3 {\bf{s}}} \int {d^3 {\bf{s}}'} \;{\cal G}_\omega ^{T*} \left( {\left. {\bf{s}} \right|{\bf{r}}} \right) \cdot \left[ {{\bf{\hat F}}_\omega  \left( {\bf{s}} \right),{\bf{\hat F}}_{\omega '}^\dag  \left( {{\bf{s}}'} \right)} \right] \cdot {\cal G}_{\omega '} \left( {\left. {{\bf{s}}'} \right|{\bf{r}}'} \right)} \right\}\mathord{\buildrel{\lower3pt\hbox{$\scriptscriptstyle\leftarrow$}} 
\over T} _{\omega '\nu '} \left( {{\bf{r}}'} \right), \nonumber \\ 
 {\cal L}_{\omega \nu ,\omega '\nu '}^{\left( 2 \right)} \left( {\left. {\bf{r}} \right|{\bf{r}}'} \right) &=& \vec T_{\omega \nu } \left( {\bf{r}} \right)\left\{ {\int {d^3 {\bf{s}}} \;{\cal G}_\omega ^{T*} \left( {\left. {\bf{s}} \right|{\bf{r}}} \right) \cdot \left[ {{\bf{\hat F}}_\omega  \left( {\bf{s}} \right),{\bf{\hat h}}_{\omega '\nu '}^\dag  \left( {{\bf{r}}'} \right)} \right]} \right\}, \nonumber \\ 
 {\cal L}_{\omega \nu ,\omega '\nu '}^{\left( 3 \right)} \left( {\left. {\bf{r}} \right|{\bf{r}}'} \right) &=& \left\{ {\int {d^3 {\bf{s}}} \;\left[ {{\bf{\hat h}}_{\omega \nu } \left( {\bf{r}} \right),{\bf{\hat F}}_{\omega '}^\dag  \left( {\bf{s}} \right)} \right] \cdot {\cal G}_{\omega '} \left( {\left. {\bf{s}} \right|{\bf{r}}'} \right)} \right\}\mathord{\buildrel{\lower3pt\hbox{$\scriptscriptstyle\leftarrow$}} 
\over T} _{\omega '\nu '} \left( {{\bf{r}}'} \right), \nonumber \\ 
 {\cal L}_{\omega \nu ,\omega '\nu '}^{\left( 4 \right)} \left( {\left. {\bf{r}} \right|{\bf{r}}'} \right) &=& \left[ {{\bf{\hat h}}_{\omega \nu } \left( {\bf{r}} \right),{\bf{\hat h}}_{\omega '\nu '}^\dag  \left( {{\bf{r}}'} \right)} \right]. 
\end{eqnarray}
To evaluate the term ${\cal L}_{\omega \nu ,\omega '\nu '}^{\left( 1 \right)} \left( {\left. {\bf{r}} \right|{\bf{r}}'} \right)$ we resort to the same strategy adopted in Secs.IVA and IVB, i.e. we use Eq.(\ref{F_Fc}) with the variables relabelling ${\bf{r}} \to {\bf{s}},{\bf{r}}' \to {\bf{s}}'$ followed by the substitutions ${\cal Q}_\omega  \left( {\bf{s}} \right) = {\cal G}_\omega  \left( {\left. {\bf{s}} \right|{\bf{r}}} \right)$ and ${\cal P}_\omega  \left( {\bf{s}} \right) = {\cal G}_\omega  \left( {\left. {\bf{s}} \right|{\bf{r}}'} \right)$, thus getting
\begin{eqnarray} \label{L1_1}
&& {\cal L}_{\omega \nu ,\omega '\nu '}^{\left( 1 \right)} \left( {\left. {\bf{r}} \right|{\bf{r}}'} \right) = \vec T_{\omega \nu } \left( {\bf{r}} \right)\left\{ {\frac{{\varepsilon _0 }}{{\hbar c}}\frac{1}{{k_\omega  k_{\omega '} \left( {k_\omega   - k_{\omega '}  - i\eta } \right)}}\int {d^3 {\bf{s}}} \left\{ {{\cal G}_\omega ^{T*} \left( {\left. {\bf{s}} \right|{\bf{r}}} \right) \cdot \left[ {{\cal D}^ \bot  \left( {{\bf{s}} - {\bf{r}}'} \right) + \frac{{k_\omega  }}{{k_{\omega '} }}{\cal D}^\parallel  \left( {{\bf{s}} - {\bf{r}}'} \right)} \right]} \right.} \right.\nonumber \\ 
&& \quad \quad  \left. {\left. { + \left[ { - {\cal D}^ \bot  \left( {{\bf{s}} - {\bf{r}}} \right) - \frac{{k_{\omega '} }}{{k_\omega  }}{\cal D}^\parallel  \left( {{\bf{s}} - {\bf{r}}} \right)} \right] \cdot {\cal G}_{\omega '} \left( {\left. {\bf{s}} \right|{\bf{r}}'} \right)} \right\}} \right\}\mathord{\buildrel{\lower3pt\hbox{$\scriptscriptstyle\leftarrow$}} 
\over T} _{\omega '\nu '} \left( {{\bf{r}}'} \right)
\end{eqnarray}
where we have used the transverse property in the third of Eqs.(\ref{Modal_Tra}) and the inhomogeneous Helmholtz equations in the third of Eqs.(\ref{GS}) for the dyadic Green's function and we have dropped the vanishing contribution containing the asymptotic surface integral in the third of Eqs.(\ref{Asy_Int}). It is also essential to note that, in contrast to the modal dyadic, the dyadic Green's function does not exhibit a plane wave contribution in its asymptotic behavior (see the fourth of Eq.(\ref{GS})); consequently, no divergent integral has appeared in the derivation of Eq.(\ref{L1_1}), and the generalized function $1/\left( {k_\omega - k_{\omega '} - i\eta } \right)$ has been handled in a purely algebraic manner. Now using the first of Eqs.(\ref{[Fh]_[Fhc]}) and Eqs.(\ref{[hh]_[hhc]}) we directly get
\begin{eqnarray}
 {\cal L}_{\omega \nu ,\omega '\nu '}^{\left( 2 \right)} \left( {\left. {\bf{r}} \right|{\bf{r}}'} \right) &=& \vec T_{\omega \nu } \left( {\bf{r}} \right)\left\{ {\frac{{\varepsilon _0 }}{{\hbar c}}\frac{1}{{k_\omega  k_{\omega '} \left( {k_\omega   - k_{\omega '}  - i\eta } \right)}}\int  {d^3 {\bf{s}}} {\cal G}_{\omega \nu }^{T*} \left( {\left. {\bf{s}} \right|{\bf{r}}} \right) \cdot \left[ { - {\cal D}^ \bot  \left( {{\bf{s}} - {\bf{r}}'} \right) - \frac{{k_\omega  }}{{k_{\omega '} }}{\cal D}^\parallel  \left( {{\bf{s}} - {\bf{r}}'} \right)} \right]} \right\}\mathord{\buildrel{\lower3pt\hbox{$\scriptscriptstyle\leftarrow$}} 
\over T} _{\omega '\nu '} \left( {{\bf{r}}'} \right), \nonumber \\ 
 {\cal L}_{\omega \nu ,\omega '\nu '}^{\left( 3 \right)} \left( {\left. {\bf{r}} \right|{\bf{r}}'} \right) &=& \vec T_{\omega \nu } \left( {\bf{r}} \right)\left\{ {\frac{{\varepsilon _0 }}{{\hbar c}}\frac{1}{{k_\omega  k_{\omega '} \left( {k_\omega   - k_{\omega '}  - i\eta } \right)}}\int {d^3 {\bf{s}}} \left[ {{\cal D}^ \bot  \left( {{\bf{r}} - {\bf{s}}} \right) + \frac{{k_{\omega '} }}{{k_\omega  }}{\cal D}^\parallel  \left( {{\bf{r}} - {\bf{s}}} \right)} \right] \cdot {\cal G}_{\omega '} \left( {\left. {\bf{s}} \right|{\bf{r}}'} \right)} \right\}\mathord{\buildrel{\lower3pt\hbox{$\scriptscriptstyle\leftarrow$}} 
\over T} _{\omega '\nu '} \left( {{\bf{r}}'} \right), \nonumber \\
{\cal L}_{\omega \nu ,\omega '\nu '}^{\left( 4 \right)} \left( {\left. {\bf{r}} \right|{\bf{r}}'} \right) &=& \delta _{\nu \nu '} \delta \left( {\omega  - \omega '} \right)\delta \left( {{\bf{r}} - {\bf{r}}'} \right){\cal I},
\end{eqnarray}
which, inserted into Eq.(\ref{COMM_C_1}) together with Eq.(\ref{L1_1}), yield $\left[ {{\bf{\hat f}}_{\omega \nu } \left( {\bf{r}} \right),{\bf{\hat f}}_{\omega '\nu '}^\dag  \left( {{\bf{r}}'} \right)} \right] = \delta _{\nu \nu '} \delta \left( {\omega  - \omega '} \right)\delta \left( {{\bf{r}} - {\bf{r}}'} \right){\cal I}$. From a physical perspective, the recovery of the spatial delta function rigorously proves the locality of the internal excitations. Physically, the recovery of the spatial delta function indicates the locality of the internal excitations $\hat{\bf f}_{\omega \nu}$. The derivation shows that the non-local terms associated with the macroscopic excitation density $\hat{\bf F}_\omega$ and propagated by the dyadic Green’s function are exactly cancelled by the commutation properties of the bare material excitation $\hat{\bf h}_{\omega \nu}$. This ensures that the dressed polaritons $\hat{\bf f}_{\omega \nu}$ maintain the spatial independence of the bare mechanical oscillators, preserving their local bosonic identity from the macroscopic radiation reaction.

\subsection{Annihilation-annihilation commutator for material polaritons $\left[ {{\bf{\hat f}}_{\omega \nu } \left( {\bf{r}} \right),{\bf{\hat f}}_{\omega '\nu '} \left( {{\bf{r}}'} \right)} \right] = 0$}
From Eqs.(\ref{gf2}) and the second of Eqs.(\ref{G-T}) we get 
\begin{equation} \label{COMM_D_1}
\left[ {{\bf{\hat f}}_{\omega \nu } \left( {\bf{r}} \right),{\bf{\hat f}}_{\omega '\nu '} \left( {{\bf{r}}'} \right)} \right] = \sum\limits_{j = 1}^4 {{\cal M}_{\omega \nu ,\omega '\nu '}^{\left( j \right)} \left( {\left. {\bf{r}} \right|{\bf{r}}'} \right)} 
\end{equation}
where
\begin{eqnarray} \label{M_dyad}
{\cal M}_{\omega \nu ,\omega '\nu '}^{\left( 1 \right)} \left( {\left. {\bf{r}} \right|{\bf{r}}'} \right) &=&   \vec T_{\omega \nu } \left( {\bf{r}} \right)\left\{ - {\int {d^3 {\bf{s}}} \int {d^3 {\bf{s}}'} {\cal G}_\omega ^{T*} \left( {\left. {\bf{s}} \right|{\bf{r}}} \right) \cdot \left[ {{\bf{\hat F}}_\omega  \left( {\bf{s}} \right),{\bf{\hat F}}_{\omega '} \left( {{\bf{s}}'} \right)} \right] \cdot {\cal G}_{\omega '}^* \left( {\left. {{\bf{s}}'} \right|{\bf{r}}'} \right)} \right\}\mathord{\buildrel{\lower3pt\hbox{$\scriptscriptstyle\leftarrow$}} 
\over T} _{\omega '\nu '} \left( {{\bf{r}}'} \right), \nonumber \\ 
 {\cal M}_{\omega \nu ,\omega '\nu '}^{\left( 2 \right)} \left( {\left. {\bf{r}} \right|{\bf{r}}'} \right) &=& \vec T_{\omega \nu } \left( {\bf{r}} \right)\left\{ {\int {d^3 {\bf{s}}} {\cal G}_\omega ^{T*} \left( {\left. {\bf{s}} \right|{\bf{r}}} \right) \cdot \left[ {{\bf{\hat F}}_\omega  \left( {\bf{s}} \right),{\bf{\hat h}}_{\omega '\nu '} \left( {{\bf{r}}'} \right)} \right]} \right\}, \nonumber \\ 
 {\cal M}_{\omega \nu ,\omega '\nu '}^{\left( 3 \right)} \left( {\left. {\bf{r}} \right|{\bf{r}}'} \right) &=&   \left\{  - {\int {d^3 {\bf{s}}} \left[ {{\bf{\hat h}}_{\omega \nu } \left( {\bf{r}} \right),{\bf{\hat F}}_{\omega '} \left( {\bf{s}} \right)} \right] \cdot {\cal G}_{\omega '}^* \left( {\left. {\bf{s}} \right|{\bf{r}}'} \right)} \right\}\mathord{\buildrel{\lower3pt\hbox{$\scriptscriptstyle\leftarrow$}} 
\over T} _{\omega '\nu '} \left( {{\bf{r}}'} \right), \nonumber \\ 
 {\cal M}_{\omega \nu ,\omega '\nu '}^{\left( 4 \right)} \left( {\left. {\bf{r}} \right|{\bf{r}}'} \right) &=& \left[ {{\bf{\hat h}}_{\omega \nu } \left( {\bf{r}} \right),{\bf{\hat h}}_{\omega '\nu '} \left( {{\bf{r}}'} \right)} \right].
\end{eqnarray}
Now using Eq.(\ref{F_F}) with the variables relabelling ${\bf{r}} \to {\bf{s}},{\bf{r}}' \to {\bf{s}}'$ followed by the substitutions ${\cal Q}_\omega  \left( {\bf{s}} \right) = {\cal G}_\omega  \left( {\left. {\bf{s}} \right|{\bf{r}}} \right)$ and ${\cal P}_\omega  \left( {\bf{s}} \right) = {\cal G}_\omega  \left( {\left. {\bf{s}} \right|{\bf{r}}'} \right)$, the first of Eqs.(\ref{M_dyad}) becomes
\begin{eqnarray} \label{M1_1}
&& {\cal M}_{\omega \nu ,\omega '\nu '}^{\left( 1 \right)} \left( {\left. {\bf{r}} \right|{\bf{r}}'} \right) = \vec T_{\omega \nu } \left( {\bf{r}} \right)\left\{ {\frac{{\varepsilon _0 }}{{\hbar c}}\frac{1}{{k_\omega  k_{\omega '} \left( {k_\omega   + k_{\omega '} } \right)}}\int {d^3 {\bf{s}}} \;{\cal G}_\omega ^{T*} \left( {\left. {\bf{s}} \right|{\bf{r}}} \right) \cdot \left[ { - {\cal D}^ \bot  \left( {{\bf{s}} - {\bf{r}}'} \right) + \frac{{k_\omega  }}{{k_{\omega '} }}{\cal D}^\parallel  \left( {{\bf{s}} - {\bf{r}}'} \right)} \right]} \right. \nonumber \\ 
&& \quad \quad  \left. { + \left[ {{\cal D}^ \bot  \left( {{\bf{s}} - {\bf{r}}} \right) - \frac{{k_{\omega '} }}{{k_\omega  }}{\cal D}^\parallel  \left( {{\bf{s}} - {\bf{r}}} \right)} \right] \cdot {\cal G}_{\omega '}^* \left( {\left. {\bf{s}} \right|{\bf{r}}'} \right)} \right\}\mathord{\buildrel{\lower3pt\hbox{$\scriptscriptstyle\leftarrow$}} 
\over T} _{\omega '\nu '} \left( {{\bf{r}}'} \right) 
\end{eqnarray}
where we have used the longitudinal property in the fourth of Eqs.(\ref{Modal_Tra}) and the inhomogeneous Helmholtz equations in the third of Eqs.(\ref{GS}) for the dyadic Green's function and we have dropped the vanishing contribution containing the asymptotic surface integral in the fourth of Eqs.(\ref{Asy_Int}). Now using the second of Eqs.(\ref{[Fh]_[Fhc]}) and Eqs.(\ref{[hh]_[hhc]}) we directly get
\begin{eqnarray}
 {\cal M}_{\omega \nu ,\omega '\nu '}^{\left( 2 \right)} \left( {\left. {\bf{r}} \right|{\bf{r}}'} \right) &=& \vec T_{\omega \nu } \left( {\bf{r}} \right)\left\{ {\frac{{\varepsilon _0 }}{{\hbar c}}\frac{1}{{k_\omega  k_{\omega '} \left( {k_\omega   + k_{\omega '} } \right)}}\int {d^3 {\bf{s}}} {\cal G}_\omega ^{T*} \left( {\left. {\bf{s}} \right|{\bf{r}}} \right) \cdot \left[ {{\cal D}^ \bot  \left( {{\bf{s}} - {\bf{r}}'} \right) - \frac{{k_\omega  }}{{k_{\omega '} }}{\cal D}^\parallel  \left( {{\bf{s}} - {\bf{r}}'} \right)} \right]} \right\}\mathord{\buildrel{\lower3pt\hbox{$\scriptscriptstyle\leftarrow$}} 
\over T} _{\omega '\nu '} \left( {{\bf{r}}'} \right) , \nonumber \\ 
 {\cal M}_{\omega \nu ,\omega '\nu '}^{\left( 3 \right)} \left( {\left. {\bf{r}} \right|{\bf{r}}'} \right) &=& \vec T_{\omega \nu } \left( {\bf{r}} \right)\left\{ {\frac{{\varepsilon _0 }}{{\hbar c}}\frac{1}{{k_\omega  k_{\omega '} \left( {k_\omega   + k_{\omega '} } \right)}}\int {d^3 {\bf{s}}} \left[ { - {\cal D}^ \bot  \left( {{\bf{s}} - {\bf{r}}} \right) + \frac{{k_{\omega '} }}{{k_\omega  }}{\cal D}^\parallel  \left( {{\bf{s}} - {\bf{r}}} \right)} \right] \cdot {\cal G}_{\omega '}^* \left( {\left. {\bf{s}} \right|{\bf{r}}'} \right)} \right\}\mathord{\buildrel{\lower3pt\hbox{$\scriptscriptstyle\leftarrow$}} 
\over T} _{\omega '\nu '} \left( {{\bf{r}}'} \right), \nonumber \\ 
 {\cal M}_{\omega \nu ,\omega '\nu '}^{\left( 4 \right)} \left( {\left. {\bf{r}} \right|{\bf{r}}'} \right) &=& 0 ,
\end{eqnarray}
which, inserted into Eq.(\ref{COMM_D_1}) together with Eq.(\ref{M1_1}), directly yield $\left[ {{\bf{\hat f}}_{\omega \nu } \left( {\bf{r}} \right),{\bf{\hat f}}_{\omega '\nu '} \left( {{\bf{r}}'} \right)} \right] = 0$. Physically, the vanishing of this commutator complements the locality result in Sec.IVD and ensures the consistency of the bosonic statistics. The derivation indicates that all non-local terms, both transverse and longitudinal (represented by ${\cal D}^\perp$ and ${\cal D}^\parallel$), arising from the macroscopic excitation density $\hat{\bf F}_\omega$ are exactly compensated by the bare material excitation $\hat{\bf h}_{\omega \nu}$. This compensation ensures that the dressed polaritons $\hat{\bf f}_{\omega \nu}$ remain independent bosonic degrees of freedom, preserving the Bose-Einstein statistics of the material reservoir.

\subsection{Annihilation-creation cross-commutator between scattering and material polaritons $\left[ {{\bf{\hat g}}_{\omega s} \left( {\bf{n}} \right),{\bf{\hat f}}_{\omega '\nu }^\dag  \left( {\bf{r}} \right)} \right] = 0$}
From Eqs.(\ref{gf2}), the first of Eqs.(\ref{Kernels}) and the second of Eqs.(\ref{G-T}) we get 
\begin{equation} \label{COMM_E_1}
\left[ {{\bf{\hat g}}_{\omega s} \left( {\bf{n}} \right),{\bf{\hat f}}_{\omega '\nu }^\dag  \left( {\bf{r}} \right)} \right] = \sqrt {\frac{{\hbar k_\omega ^3 }}{{16\pi ^3 \varepsilon _0 }}} \sum\limits_{j = 1}^2 {{\cal N}_{\omega ,\omega '\nu }^{\left( j \right)} \left( {\left. {\bf{n}} \right|{\bf{r}}} \right)} 
\end{equation}
where
\begin{eqnarray} \label{N_dyad}
 {\cal N}_{\omega ,\omega '\nu }^{\left( 1 \right)} \left( {\left. {\bf{n}} \right|{\bf{r}}} \right) &=& \left\{ {\int {d^3 {\bf{s}}} \int {d^3 {\bf{s}}'} \,{\cal F}_\omega ^{T*} \left( {\left. {\bf{s}} \right|{\bf{n}}} \right) \cdot \left[ {{\bf{\hat F}}_\omega  \left( {\bf{s}} \right),{\bf{\hat F}}_{\omega '}^\dag  \left( {{\bf{s}}'} \right)} \right] \cdot {\cal G}_\omega  \left( {\left. {{\bf{s}}'} \right|{\bf{r}}} \right)} \right\}\mathord{\buildrel{\lower3pt\hbox{$\scriptscriptstyle\leftarrow$}} 
\over T} _{\omega \nu } \left( {\bf{r}} \right), \nonumber \\ 
 {\cal N}_{\omega ,\omega '\nu }^{\left( 2 \right)} \left( {\left. {\bf{n}} \right|{\bf{r}}} \right) &=& \int {d^3 {\bf{s}}} \,{\cal F}_\omega ^{T*} \left( {\left. {\bf{s}} \right|{\bf{n}}} \right) \cdot \left[ {{\bf{\hat F}}_\omega  \left( {\bf{s}} \right),{\bf{\hat h}}_{\omega '\nu }^\dag  \left( {\bf{r}} \right)} \right].
\end{eqnarray}
Using Eq.(\ref{F_Fc}) with the variables relabelling ${\bf{r}} \to {\bf{s}},{\bf{r}}' \to {\bf{s}}'$ followed by the substitutions ${\cal Q}_\omega  \left( {\bf{s}} \right) = {\cal F}_\omega  \left( {\left. {\bf{s}} \right|{\bf{n}}} \right)$ and ${\cal P}_\omega \left( {\bf{s}} \right) = {\cal G}_\omega  \left( {\left. {\bf{s}} \right|{\bf{r}}} \right)$, the first of Eqs.(\ref{M_dyad}) becomes
\begin{equation} \label{N1_1}
{\cal N}_{\omega ,\omega '\nu }^{\left( 1 \right)} \left( {\left. {\bf{n}} \right|{\bf{r}}} \right) = \left\{ {\frac{{\varepsilon _0 }}{{\hbar c}}\frac{1}{{k_\omega  k_{\omega '} \left( {k_\omega   - k_{\omega '}  - i\eta } \right)}}\int {d^3 {\bf{s}}} \,{\cal F}_\omega ^{T*} \left( {\left. {\bf{s}} \right|{\bf{n}}} \right) \cdot \left[ {{\cal D}^ \bot  \left( {{\bf{s}} - {\bf{r}}} \right) + \frac{{k_\omega  }}{{k_{\omega '} }}{\cal D}^\parallel  \left( {{\bf{s}} - {\bf{r}}} \right)} \right]} \right\}\mathord{\buildrel{\lower3pt\hbox{$\scriptscriptstyle\leftarrow$}} 
\over T} _{\omega '\nu } \left( {\bf{r}} \right)
\end{equation}
where we have used the transverse properties in the first and third of Eqs.(\ref{Modal_Tra}) and the homogeneous and inhomogeneous Helmholtz equations in the first and third of Eqs.(\ref{GS}) for the modal dyadic and the dyadic Green's function and we have dropped the vanishing contribution containing the asymptotic surface integral in the fifth of Eqs.(\ref{Asy_Int}). Now using the first of Eqs.(\ref{[Fh]_[Fhc]}) we directly get
\begin{equation}
{\cal N}_{\omega ,\omega '\nu }^{\left( 2 \right)} \left( {\left. {\bf{n}} \right|{\bf{r}}} \right) = \left\{ {\frac{{\varepsilon _0 }}{{\hbar c}}\frac{1}{{k_\omega  k_{\omega '} \left( {k_\omega   - k_{\omega '}  - i\eta } \right)}}\int {d^3 {\bf{s}}} \,{\cal F}_\omega ^{T*} \left( {\left. {\bf{s}} \right|{\bf{n}}} \right) \cdot \left[ { - {\cal D}^ \bot  \left( {{\bf{s}} - {\bf{r}}} \right) - \frac{{k_\omega  }}{{k_{\omega '} }}{\cal D}^\parallel  \left( {{\bf{s}} - {\bf{r}}} \right)} \right]} \right\}\mathord{\buildrel{\lower3pt\hbox{$\scriptscriptstyle\leftarrow$}} 
\over T} _{\omega '\nu } \left( {\bf{r}} \right)
\end{equation}
which, inserted into Eq.(\ref{COMM_E_1}) together with Eq.(\ref{N1_1}), directly yield $\left[ {{\bf{\hat g}}_{\omega s} \left( {\bf{n}} \right),{\bf{\hat f}}_{\omega '\nu }^\dag  \left( {\bf{r}} \right)} \right] = 0$. Physically, the vanishing of this cross-commutator indicates the quantum independence between the asymptotic radiation and the internal material excitations. The derivation shows that, although both operators are constructed from the macroscopic excitation density $\hat{\bf F}_\omega$, they act on orthogonal sectors of the Hilbert space. This ensures that scattering processes are decoupled from the internal material dynamics.

\subsection{Annihilation-annihilation cross-commutator between scattering and material polaritons $\left[ {{\bf{\hat g}}_{\omega s} \left( {\bf{n}} \right),{\bf{\hat f}}_{\omega '\nu } \left( {\bf{r}} \right)} \right] = 0$}
From Eqs.(\ref{gf2}), the first of Eqs.(\ref{Kernels}) and the second of Eqs.(\ref{G-T}) we get 
\begin{equation} \label{COMM_F_1}
{\left[ {{\bf{\hat g}}_{\omega s} \left( {\bf{n}} \right),{\bf{\hat f}}_{\omega '\nu } \left( {\bf{r}} \right)} \right] = \sqrt {\frac{{\hbar k_\omega ^3 }}{{16\pi ^3 \varepsilon _0 }}} \sum\limits_{j = 1}^2 {{\cal O}_{\omega ,\omega '\nu }^{\left( j \right)} \left( {\left. {\bf{n}} \right|{\bf{r}}} \right)} }
\end{equation}
where
\begin{eqnarray} \label{O_dyad}
 {\cal O}_{\omega ,\omega '\nu }^{\left( 1 \right)} \left( {\left. {\bf{n}} \right|{\bf{r}}} \right) &= & \left\{- {\int {d^3 {\bf{s}}} \int {d^3 {\bf{s}}'\,} {\cal F}_\omega ^{T*} \left( {\left. {\bf{s}} \right|{\bf{n}}} \right) \cdot \left[ {{\bf{\hat F}}_\omega  \left( {\bf{s}} \right),{\bf{\hat F}}_{\omega '} \left( {{\bf{s}}'} \right)} \right] \cdot {\cal G}_{\omega '}^* \left( {\left. {{\bf{s}}'} \right|{\bf{r}}} \right)} \right\}\mathord{\buildrel{\lower3pt\hbox{$\scriptscriptstyle\leftarrow$}} 
\over T} _{\omega '\nu } \left( {\bf{r}} \right), \nonumber \\ 
 {\cal O}_{\omega ,\omega '\nu }^{\left( 2 \right)} \left( {\left. {\bf{n}} \right|{\bf{r}}} \right) &=& \int {d^3 {\bf{s}}} \,{\cal F}_\omega ^{T*} \left( {\left. {\bf{s}} \right|{\bf{n}}} \right) \cdot \left[ {{\bf{\hat F}}_\omega  \left( {\bf{s}} \right),{\bf{\hat h}}_{\omega '\nu } \left( {\bf{r}} \right)} \right].
\end{eqnarray}
Now using Eq.(\ref{F_F}) with the variables relabelling ${\bf{r}} \to {\bf{s}},{\bf{r}}' \to {\bf{s}}'$ followed by the substitutions ${\cal Q}_\omega  \left( {\bf{s}} \right) = {\cal F}_\omega  \left( {\left. {\bf{s}} \right|{\bf{n}}} \right)$ and ${\cal P}_\omega  \left( {\bf{s}} \right) = {\cal G}_\omega  \left( {\left. {\bf{s}} \right|{\bf{r}}} \right)$, the first of Eqs.(\ref{O_dyad}) becomes
\begin{equation} \label{O1_1}
{\cal O}_{\omega ,\omega '\nu }^{\left( 1 \right)} \left( {\left. {\bf{n}} \right|{\bf{r}}} \right) = \left\{ {\frac{{\varepsilon _0 }}{{\hbar c}}\frac{1}{{k_\omega  k_{\omega '} \left( {k_\omega   + k_{\omega '} } \right)}}\int {d^3 {\bf{s}}} {\cal F}_\omega ^{T*} \left( {\left. {\bf{s}} \right|{\bf{n}}} \right) \cdot \left[ { - {\cal D}^ \bot  \left( {{\bf{s}} - {\bf{r}}} \right) + \frac{{k_\omega  }}{{k_{\omega '} }}{\cal D}^\parallel  \left( {{\bf{s}} - {\bf{r}}} \right)} \right]} \right\}\mathord{\buildrel{\lower3pt\hbox{$\scriptscriptstyle\leftarrow$}} 
\over T} _{\omega '\nu } \left( {\bf{r}} \right)
\end{equation}
where we have used the longitudinal properties in the second and fourth of Eqs.(\ref{Modal_Tra}) and the homogeneous and inhomogeneous Helmholtz equations in the first and third of Eqs.(\ref{GS}) for the modal dyadic and the dyadic Green's function and we have dropped the vanishing contribution containing the asymptotic surface integral in the sixth of Eqs.(\ref{Asy_Int}). Now using the second of Eqs.(\ref{[Fh]_[Fhc]}) we directly get
\begin{equation}
{\cal O}_{\omega ,\omega '\nu }^{\left( 2 \right)} \left( {\left. {\bf{n}} \right|{\bf{r}}} \right) = \left\{ {\frac{{\varepsilon _0 }}{{\hbar c}}\frac{1}{{k_\omega  k_{\omega '} \left( {k_\omega   + k_{\omega '} } \right)}}\int {d^3 {\bf{s}}} \,{\cal F}_\omega ^{T*} \left( {\left. {\bf{s}} \right|{\bf{n}}} \right) \cdot \left[ {{\cal D}^ \bot  \left( {{\bf{s}} - {\bf{r}}} \right) - \frac{{k_\omega  }}{{k_{\omega '} }}{\cal D}^\parallel  \left( {{\bf{s}} - {\bf{r}}} \right)} \right]} \right\}\mathord{\buildrel{\lower3pt\hbox{$\scriptscriptstyle\leftarrow$}} 
\over T} _{\omega '\nu } \left( {\bf{r}} \right)
\end{equation}
which, inserted into Eq.(\ref{COMM_F_1}) together with Eq.(\ref{O1_1}), directly yield $\left[ {{\bf{\hat g}}_{\omega s} \left( {\bf{n}} \right),{\bf{\hat f}}_{\omega '\nu } \left( {\bf{r}} \right)} \right] = 0$. Consistent with the results in Sec.IVE, the vanishing of this commutator indicates the decoupling between the scattering channels $\hat{\bf g}_{\omega s}$ and the material sources $\hat{\bf f}_{\omega \nu}$. This result confirms that these operators represent independent quantum degrees of freedom, despite being constructed from the same macroscopic excitation density $\hat{\bf F}_\omega$.

\section{Diagonalization of the macroscopic Hamiltonian}
Having established that the polariton operators in Eqs.(\ref{gf2}) satisfy the bosonic commutation relations of Eqs.(\ref{MLNF_Com_Rel}) by virtue of the canonical commutation relations in Eqs.(\ref{CQME_Com_Rel}), we now complete the direct derivation of the MLNF from the CQME. Specifically, we demonstrate that the CQME Hamiltonian in Eq.(\ref{CQME_H}) is diagonalized by these polariton operators, thereby coinciding with the MLNF Hamiltonian defined in Eq.(\ref{MLNF_HE}). 

We begin by noting that, since both Eqs.(\ref{MLNF_Com_Rel}) and (\ref{CQME_Com_Rel}) hold, the procedure used in Sec.III to obtain the polariton operators is inherently invertible, i.e. the spectra of the canonical field operators in Eqs.(\ref{CQME_Spectra}), where now $\mathbf{\hat{E}}_\omega$ is defined by Eq.(\ref{Eom}), can be deduced from the polariton operators in Eqs.(\ref{gf2}). In the context of the present direct derivation, the canonical field operators obtained from these spectra through Eq.(\ref{SpOpDe}), namely
\begin{align} \label{Direct_Can_Fie_Ope}
 {\bf{\hat A}} =&  \int {d\omega } \frac{{{\bf{\hat E}}_\omega ^ \bot  }}{{i\omega }}  + {\rm H.c.}, &
 {\bf{\hat \Pi }}_{A }  =&  \int {d\omega } \left( { - \varepsilon _0 \varepsilon _\omega  {\bf{\hat E}}_\omega   - i\alpha ^\omega  \sqrt {\frac{\hbar }{{2\omega }}} {\bf{\hat f}}_{\omega e} } \right)   + {\rm H.c.}, \nonumber \\
 {\bf{\hat X}}^\Omega   =&  \left[ \int {d\omega } \frac{{\alpha ^\Omega{\bf{\hat E}}_\omega  }}{{\Omega ^2  - \left( {\omega  + i\eta } \right)^2 }} + i\sqrt {\frac{\hbar }{{2\Omega }}} {\bf{\hat f}}_{\Omega e}  \right] + {\rm H.c.}, & 
 {\bf{\hat \Pi }}_X^\Omega   =&  \left[ \int {d\omega } \frac{{ - i\omega \alpha ^\Omega  {\bf{\hat E}}_\omega  }}{{\Omega ^2  - \left( {\omega  + i\eta } \right)^2 }} + \sqrt {\frac{{\hbar \Omega }}{2}} {\bf{\hat f}}_{\Omega e} \right]  + {\rm H.c.}, \nonumber \\
{\bf{\hat Y}}^\Omega   =&  \left[  \int {d\omega } \frac{{\frac{1}{{i\omega }} \beta ^\Omega \nabla  \times {\bf{\hat E}}_\omega  }}{{\Omega ^2  - \left( {\omega  + i\eta } \right)^2 }} + \sqrt {\frac{\hbar }{{2\Omega }}} {\bf{\hat f}}_{\Omega m} \right] + {\rm H.c.}, &
 {\bf{\hat \Pi }}_Y^\Omega   =&  \left[ \int {d\omega } \frac{{ - \beta ^\Omega  \nabla  \times {\bf{\hat E}}_\omega  }}{{\Omega ^2  - \left( {\omega  + i\eta } \right)^2 }} - i\sqrt {\frac{{\hbar \Omega }}{2}} {\bf{\hat f}}_{\Omega m} \right] + {\rm H.c.} ,
\end{align}
where
\begin{equation} \label{Eom2}
{\bf{\hat E}}_\omega  \left( {\bf{r}} \right) = \int {do_{\bf{n}} } {\cal F}_{\omega s} \left( {\left. {\bf{r}} \right|{\bf{n}}} \right) \cdot {\bf{\hat g}}_{\omega s} \left( {\bf{n}} \right) + \sum\limits_\nu  {\int {d^3 {\bf{r}}'\,} {\cal G}_{\omega \nu } \left( {\left. {\bf{r}} \right|{\bf{r}}'} \right) \cdot {\bf{\hat f}}_{\omega \nu } \left( {{\bf{r}}'} \right)},
\end{equation}
constitute the formal inverse of the relations in Eqs.(\ref{gf2}). As a consistency check of this inversion, we verify that the operator $\mathbf{\hat{E}}_\omega$  in Eq.(\ref{Eom2}) correctly represents  the spectrum of the CQME electric field operator $\mathbf{\hat{E}}$ in the first of Eqs.(\ref{CQME_EB}), i.e. ${\bf{\hat E}} = \int {d\omega } {\bf{\hat E}}_\omega   + {\rm H.c.}$; this is demonstrated by first inserting the expressions for the conjugate momentum ${\bf{\hat \Pi }}_A $ and the reservoir field ${\bf{\hat X}}^\Omega$ of Eqs.(\ref{Direct_Can_Fie_Ope}) into the first of Eqs.(\ref{CQME_EB}), i.e. ${\bf{\hat E}} = \int {d\omega } \left[ {\varepsilon _\omega   - \frac{1}{{\varepsilon _0 }}\int{d\Omega } \;\frac{{\left( {\alpha ^\Omega  } \right)^2 }}{{\Omega ^2  - \left( {\omega  + i\eta } \right)^2 }}} \right]{\bf{\hat E}}_\omega   + {\rm H.c.}$, and then using the first of Eqs.(\ref{Int_1}) to replace the integral over $\Omega$ with $\varepsilon _0 \left( {\varepsilon _\omega   - 1} \right)$.

We are now in a position to substitute the canonical field operators from Eqs.(\ref{Direct_Can_Fie_Ope}) into the CQME Hamiltonian of Eq.(\ref{CQME_H}). Because of the extensive algebra involved we have outlined the evaluation of the different terms of the Hamiltonian density in Appendix F and we here report and comment on the results. The total electric Hamiltonian density (first contribution in Eq.(\ref{CQME_H})) is
\begin{equation} \label {HHH_1}
\frac{1}{{2\varepsilon _0 }}\left( {{\bf{\hat \Pi }}_A  + \int {d\Omega } \;\alpha ^\Omega  {\bf{\hat X}}^\Omega  } \right)^2  = \frac{1}{2}\varepsilon _0 {\bf{\hat E}}^2 
\end{equation}
as expected, since the interaction with the continuum of electric reservoir fields (accounting for the medium local response) fully recovers the familiar quadratic dependence on the total macroscopic electric field. The total magnetic Hamiltonian density (second contribution in Eq.(\ref{CQME_H})) is
\begin{eqnarray} \label{HHH_2}
&& \left( {\nabla  \times {\bf{\hat A}}} \right) \cdot \left( {\frac{1}{{2\mu _0 }}\nabla  \times {\bf{\hat A}} - \int {d\Omega } \;\beta ^\Omega  {\bf{\hat Y}}^\Omega  } \right) = - \frac{1}{{2\mu _0 }}\left( {\nabla  \times {\bf{\hat A}}} \right)^2  \nonumber \\
&& \quad \quad  - \frac{1}{2}\int {d\omega } \int {d\omega '} \frac{1}{{\omega '}}\left[ {\left( {\nabla  \times {\bf{\hat E}}_{\omega '}  - \nabla  \times {\bf{\hat E}}_{\omega '}^\dag  } \right) \cdot {\bf{\hat Q}}_{\omega m}  + {\bf{\hat Q}}_{\omega m}  \cdot \left( {\nabla  \times {\bf{\hat E}}_{\omega '}  - \nabla  \times {\bf{\hat E}}_{\omega '}^\dag  } \right) + {\rm H.c.}} \right], 
\end{eqnarray}
where we have defined the auxiliary operator
\begin{equation} \label{Qw_def}
{\bf{\hat Q}}_{\omega m}  = \frac{{\nabla  \times {\bf{\hat E}}_\omega  }}{{\omega \mu _0 \mu _\omega  }} - i\sqrt {\frac{\hbar }{{2\omega }}} \beta ^\omega  {\bf{\hat f}}_{\omega m},
\end{equation}
which effectively acts as a spectral macroscopic magnetizing field  combining the permeability-dressed response with the magnetic polariton operator. This combination yields a strong dressing effect that dynamically renormalizes the vacuum contribution, directly explaining the negative sign of the bare magnetic energy emerging in Eq.(\ref{HHH_2}). The electric reservoir Hamiltonian density (third contribution in Eq.(\ref{CQME_H})) yields
\begin{eqnarray} \label{HHH_3}
&& \frac{1}{2}\int {d\Omega } \left( {{\bf{\hat \Pi }}_X^{\Omega 2}  + \Omega ^2 {\bf{\hat X}}^{\Omega 2} } \right) = \frac{1}{2}\int {d\omega } \hbar \omega \left( {{\bf{\hat f}}_{\omega e}^\dag   \cdot {\bf{\hat f}}_{\omega e}  + {\bf{\hat f}}_{\omega e}  \cdot {\bf{\hat f}}_{\omega e}^\dag  } \right) - \frac{1}{2}\varepsilon _0 {\bf{\hat E}}^2  \nonumber \\
&& \quad \quad   + \frac{1}{2}\int {d\omega } \int {d\omega '} \left[ {\left( {\frac{{{\bf{\hat Q}}_{\omega e}  \cdot {\bf{\hat E}}_{\omega '}  + {\bf{\hat E}}_{\omega '}  \cdot {\bf{\hat Q}}_{\omega e} }}{{\omega  + \omega '}} + \frac{{{\bf{\hat Q}}_{\omega e}  \cdot {\bf{\hat E}}_{\omega '}^\dag   + {\bf{\hat E}}_{\omega '}^\dag   \cdot {\bf{\hat Q}}_{\omega e} }}{{\omega  - \omega ' + i\eta }}} \right) + {\rm H.c.}} \right]  
\end{eqnarray}
where we have introduced the auxiliary operator
\begin{equation} \label{Qwe_def}
{\bf{\hat Q}}_{\omega e}  = \omega \varepsilon _0 \varepsilon _\omega  {\bf{\hat E}}_\omega   + i\sqrt {\frac{{\hbar \omega }}{2}} \alpha ^\omega  {\bf{\hat f}}_{\omega e},
\end{equation}
which operates as a spectral macroscopic displacement field incorporating both the dispersive dielectric response and the electric polariton operator. The structure of this operator allows the integrated reservoir energy in Eq.(\ref{HHH_3}) to explicitly isolate the bare free-evolution energy of the electric polaritons (first term) while concurrently generating a negative macroscopic field term (second term) that precisely cancels the total electric energy of Eq.(\ref{HHH_1}). Remarkably, the auxiliary operators are connected through the relation
\begin{equation} \label{Maxwell_Ampere}
 \nabla  \times {\bf{\hat Q}}_{\omega m} ={\bf{\hat Q}}_{\omega e} ,
\end{equation}
representing the quantum operator analogue of the macroscopic Amp\`{e}re-Maxwell equation in the frequency domain. Finally, the evaluation of the magnetic reservoir Hamiltonian density (fourth contribution in Eq.(\ref{CQME_H})) yields
\begin{eqnarray} \label{HHH_4}
&& \frac{1}{2}\int {d\Omega } \left( {{\bf{\hat \Pi }}_Y^{\Omega 2}  + \Omega ^2 {\bf{\hat Y}}^{\Omega 2} } \right) = \frac{1}{2}\int {d\omega } \hbar \omega \left( {{\bf{\hat f}}_{\omega m}^\dag   \cdot {\bf{\hat f}}_{\omega m}  + {\bf{\hat f}}_{\omega m}  \cdot {\bf{\hat f}}_{\omega m}^\dag  } \right) + \frac{1}{{2\mu _0 }}\left( {\nabla  \times {\bf{\hat A}}} \right)^2  \nonumber \\ 
&& \quad \quad   + \frac{1}{2}\int {d\omega } \int {d\omega '} \left[ {\frac{\omega }{{\omega '}}\left( {\frac{{{\bf{\hat Q}}_{\omega m}  \cdot \nabla  \times {\bf{\hat E}}_{\omega '}  + \nabla  \times {\bf{\hat E}}_{\omega '}  \cdot {\bf{\hat Q}}_{\omega m} }}{{\omega  + \omega '}} - \frac{{{\bf{\hat Q}}_{\omega m}  \cdot \nabla  \times {\bf{\hat E}}_{\omega '}^\dag   + \nabla  \times {\bf{\hat E}}_{\omega '}^\dag   \cdot {\bf{\hat Q}}_{\omega m} }}{{\omega  - \omega ' + i\eta }}} \right) + {\rm H.c.}} \right]. 
\end{eqnarray}
Frequency integration here  primarily isolates the free-evolution energy of the magnetic polaritons (first term). Concurrently, it extracts a positive bare magnetic energy term that perfectly compensates the negative vacuum contribution in Eq.(\ref{HHH_2}). Furthermore, it gives rise to a set of cross-terms of the same nature as those encountered in the other Hamiltonian density contributions.  Summing these contributions, the total Hamiltonian is expressed as
\begin{eqnarray} \label{H_diag_1}
&& \hat H = \frac{1}{2}\int {d\omega } \hbar \omega \sum\limits_\nu  {\int {d^3 {\bf{r}}} \left( {{\bf{\hat f}}_{\omega \nu }^\dag   \cdot {\bf{\hat f}}_{\omega \nu }  + {\bf{\hat f}}_{\omega \nu }  \cdot {\bf{\hat f}}_{\omega \nu }^\dag  } \right)}  \nonumber \\ 
&& \quad \quad  + \frac{1}{2}\int {d\omega } \int {d\omega '} \int {d^3 {\bf{r}}}  \left[ {\frac{{{\bf{\hat Q}}_{\omega e}  \cdot {\bf{\hat E}}_{\omega '}  - {\bf{\hat Q}}_{\omega m}  \cdot \nabla  \times {\bf{\hat E}}_{\omega '}  + {\bf{\hat E}}_{\omega '}  \cdot {\bf{\hat Q}}_{\omega e}  - \nabla  \times {\bf{\hat E}}_{\omega '}  \cdot {\bf{\hat Q}}_{\omega m} }}{{\omega  + \omega '}}} \right. \nonumber \\ 
&& \quad \quad  \left. { + \frac{{{\bf{\hat Q}}_{\omega e}  \cdot {\bf{\hat E}}_{\omega '}^\dag   - {\bf{\hat Q}}_{\omega m}  \cdot \nabla  \times {\bf{\hat E}}_{\omega '}^\dag   + {\bf{\hat E}}_{\omega '}^\dag   \cdot {\bf{\hat Q}}_{\omega e}  - \nabla  \times {\bf{\hat E}}_{\omega '}^\dag   \cdot {\bf{\hat Q}}_{\omega m} }}{{\omega  - \omega ' + i\eta }} + {\rm H.c.}} \right]  .
\end{eqnarray}
By applying the first of Eqs.(\ref{IntPart}) to perform spatial integration by parts, the curl terms on the RHS of Eq.(\ref{H_diag_1}) generate two distinct contributions: a volume integral, where the curl operator is transferred from the electric field spectrum to the auxiliary field $\hat{\bf{Q}}_{\omega m}$, and a surface integral over the asymptotic sphere $S_\infty$. By further utilizing the Amp\`ere-Maxwell operator relation in Eq.(\ref{Maxwell_Ampere}), the resulting volume integrals exactly cancel the pre-existing terms containing the auxiliary field $\hat{\bf{Q}}_{\omega e}$, leaving only the asymptotic surface integrals. Within these surface terms, one can employ the relation $\hat{\bf{Q}}_{\omega m} = \frac{1}{\omega \mu_0} \nabla \times \hat{\bf{E}}_\omega$, which holds valid just outside the finite object. For instance, for the first integral contribution in Eq.(\ref{H_diag_1}), we obtain:
\begin{eqnarray}
&& \int {d^3 {\bf{r}}} \left( {{\bf{\hat Q}}_{\omega e}  \cdot {\bf{\hat E}}_{\omega '}  - {\bf{\hat Q}}_{\omega m}  \cdot \nabla  \times {\bf{\hat E}}_{\omega '} } \right) = \int {d^3 {\bf{r}}} \left( {{\bf{\hat Q}}_{\omega e}  - \nabla  \times {\bf{\hat Q}}_{\omega m} } \right) \cdot {\bf{\hat E}}_{\omega '}  + \int\limits_{S_\infty  } {dS_r } \;{\bf{u}}_r  \cdot \left( {{\bf{\hat Q}}_{\omega m}  \times {\bf{\hat E}}_{\omega '} } \right) \nonumber \\ 
&& \quad \quad  = \frac{1}{{\omega \mu _0 }}\int\limits_{S_\infty  } {dS_r } \;{\bf{u}}_r  \cdot \left[ {\left( {\nabla  \times {\bf{\hat E}}_\omega  } \right) \times {\bf{\hat E}}_{\omega '} } \right] 
\end{eqnarray}
Consequently, the Hamiltonian is expressed as
\begin{eqnarray} \label{H_diag_2}
&& \hat H = \frac{1}{2}\int {d\omega } \hbar \omega \sum\limits_\nu  {\int {d^3 {\bf{r}}} \left( {{\bf{\hat f}}_{\omega \nu }^\dag   \cdot {\bf{\hat f}}_{\omega \nu }  + {\bf{\hat f}}_{\omega \nu }  \cdot {\bf{\hat f}}_{\omega \nu }^\dag  } \right)}  \nonumber  \\ 
&& \quad \quad   + \frac{1}{{2\mu _0 c}}\int {d\omega } \frac{1}{\omega }\int {d\omega '} \int {do_{\bf{r}} } {\bf{u}}_r  \cdot \left[ {\mathop {\lim }\limits_{r \to  + \infty } r^2 \frac{{\;\left( {\nabla  \times {\bf{\hat E}}_\omega  } \right) \times {\bf{\hat E}}_{\omega '}  - {\bf{\hat E}}_{\omega '}  \times \left( {\nabla  \times {\bf{\hat E}}_\omega  } \right)}}{{k_\omega   + k_{\omega '} }}} \right.  \nonumber  \\ 
&& \quad \quad   + \mathop {\lim }\limits_{r \to  + \infty } \left. {r^2 \frac{{\left( {\nabla  \times {\bf{\hat E}}_\omega  } \right) \times {\bf{\hat E}}_{\omega '}^\dag   - {\bf{\hat E}}_{\omega '}^\dag   \times \left( {\nabla  \times {\bf{\hat E}}_\omega  } \right)}}{{k_\omega   - k_{\omega '}  + i\eta }} + {\rm H.c.}} \right] 
\end{eqnarray}
where the prescription in Eq.(\ref{asym_sur_int}) has been utilized to explicitly reveal the residual asymptotic surface integral. To evaluate the final integral, we note that the electric field spectrum in Eq.(\ref{Eom2}), as demonstrated in Ref\cite{Ciatt2}, exhibits a far-field behavior characterized by the superposition of ingoing and outgoing spherical waves, i.e.
\begin{equation} \label{Ing-Out}
\hat{\bf{E}}_\omega ({\bf{r}}) \approx_{r \to +\infty} \frac{e^{-ik_\omega r}}{r} \hat{\bf{U}}_\omega^- ({\bf{u}}_r) + \frac{e^{ik_\omega r}}{r} \hat{\bf{U}}_\omega^+ ({\bf{u}}_r)
\end{equation}
where ${\bf{u}}_r \cdot \hat{\bf{U}}_\omega^\pm ({\bf{u}}_r) = 0$. Consequently, the $r^2$ factors in Eq.(\ref{H_diag_2}) are cancelled by the $1/r$ dependence of the spherical waves, while the four exponential terms $e^{i(k_\omega + k_{\omega'})r}$, $e^{i(k_\omega - k_{\omega'})r}$, $e^{-i(k_\omega + k_{\omega'})r}$, and $e^{-i(k_\omega - k_{\omega'})r}$ appear within the $r \to +\infty$ limits. These observations, combined with the first of Eqs.(\ref{Dist_FUND}), imply that the contribution to Eq.(\ref{H_diag_2}) arising from the $1/(k_\omega + k_{\omega'})$ factor in the integrand vanishes. Regarding the contribution from the $(k_\omega - k_{\omega'} + i\eta)^{-1}$ factor, an analysis similar to that in Appendix E reveals that the combination of both relations in Eqs.(\ref{Dist_FUND}) implies that the only non-vanishing term is the one associated with the $e^{-i(k_\omega - k_{\omega'})r}$ factor. Specifically, we have:
\begin{eqnarray}
&& \lim_{r \to +\infty} r^2 \frac{(\nabla \times \hat{\bf{E}}_\omega) \times \hat{\bf{E}}_{\omega'}^\dagger - \hat{\bf{E}}_{\omega'}^\dagger \times (\nabla \times \hat{\bf{E}}_\omega)}{k_\omega - k_{\omega'} + i\eta} \nonumber \\ 
&& \quad = ik_\omega \lim_{r \to +\infty} \frac{e^{-i(k_\omega - k_{\omega'})r}}{k_\omega - k_{\omega'} + i\eta} \left\{ [\hat{\bf{U}}_{\omega'}^- ({\bf{u}}_r)]^\dagger \times [{\bf{u}}_r \times \hat{\bf{U}}_\omega^- ({\bf{u}}_r)] - [{\bf{u}}_r \times \hat{\bf{U}}_\omega^- ({\bf{u}}_r)] \times [\hat{\bf{U}}_{\omega'}^- ({\bf{u}}_r)]^\dagger \right\} \nonumber \\ 
&& \quad = 2\pi k_\omega \delta(k_\omega - k_{\omega'}) {\bf{u}}_r \left\{ [\hat{\bf{U}}_\omega^- ({\bf{u}}_r)]^\dagger \cdot [\hat{\bf{U}}_\omega^- ({\bf{u}}_r)] + [\hat{\bf{U}}_\omega^- ({\bf{u}}_r)] \cdot [\hat{\bf{U}}_\omega^- ({\bf{u}}_r)]^\dagger \right\}
\end{eqnarray}
where the second step follows from the second of Eqs.(\ref{Dist_FUND}) and the expansion of the triple vector products. Accordingly, the Hamiltonian in Eq.(\ref{H_diag_2}) becomes
\begin{equation} \label{H_diag_3}
\hat H = \frac{1}{2}\int {d\omega } \hbar \omega \sum\limits_\nu  {\int {d^3 {\bf{r}}} \left( {{\bf{\hat f}}_{\omega \nu }^\dag   \cdot {\bf{\hat f}}_{\omega \nu }  + {\bf{\hat f}}_{\omega \nu }  \cdot {\bf{\hat f}}_{\omega \nu }^\dag  } \right)}  + \frac{{2\pi }}{{\mu _0 c}}\int {d\omega } \int {do_{\bf{n}} } \left\{ {\left[ {{\bf{\hat U}}_\omega ^ -  \left( {\bf{n}} \right)} \right]^\dag   \cdot \left[ {{\bf{\hat U}}_\omega ^ -  \left( {\bf{n}} \right)} \right] + \left[ {{\bf{\hat U}}_\omega ^ -  \left( {\bf{n}} \right)} \right] \cdot \left[ {{\bf{\hat U}}_\omega ^ -  \left( {\bf{n}} \right)} \right]^\dag  } \right\}
\end{equation}
where we have set ${\bf{n}} = {\bf{u}}_r$ for convenience. The residual term thus depends solely on the operator amplitude ${\bf{\hat U}}_\omega ^ -  \left( {\bf{n}} \right)$ of the ingoing spherical wave appearing in the asymptotic far-field spectrum of Eq.(\ref{Ing-Out}). From the general form of the spectrum in Eq.(\ref{Eom2}), it is evident that this contribution arises exclusively from the scattering term associated with the kernel ${\cal F}_{\omega s}$; indeed, the term involving the kernel ${\cal G}_{\omega \nu}$ contains only an outgoing spherical wave contribution due to the radiation condition of the dyadic Green's function (see the fourth of Eqs.(\ref{GS})). Conversely, the asymptotic behavior of the modal dyadic contains a plane-wave term (see the first of Eqs.(\ref{GS})) which involves an ingoing spherical wave contribution, as it is evident from the distributional form of Jones' lemma in Eq.(\ref{JonLemDist}). Using such lemma, after some algebra we get
\begin{equation}
\hat{\bf{U}}_\omega^- ({\bf{n}}) = i \sqrt{\frac{\hbar k_\omega}{4\pi \varepsilon_0}} \hat{\bf{g}}_{\omega s} (-{\bf{n}}).
\end{equation}
Substituting this into Eq.(\ref{H_diag_3}) yields the final expression for the Hamiltonian
\begin{equation} \label{H_diag_4}
\hat H = \frac{1}{2}\int {d\omega } \hbar \omega \left[ {\sum\limits_\nu  {\int {d^3 {\bf{r}}} \left( {{\bf{\hat f}}_{\omega \nu }^\dag   \cdot {\bf{\hat f}}_{\omega \nu }  + {\bf{\hat f}}_{\omega \nu }  \cdot {\bf{\hat f}}_{\omega \nu }^\dag  } \right)}  + \int {do_{\bf{n}} } \left( {{\bf{\hat g}}_{\omega s}^\dag   \cdot {\bf{\hat g}}_{\omega s}  + {\bf{\hat g}}_{\omega s}  \cdot {\bf{\hat g}}_{\omega s}^\dag  } \right)} \right]
\end{equation}
which concludes our direct derivation of the MLNF from the CQME since it coincides with the MLNF Hamiltonian in the first of Eqs.(\ref{MLNF_HE}) upon omitting the zero-point energy of the oscillators, as is standard practice in quantum electrodynamics.

\section{Resolution of the double second-quantization paradox}
In the Introduction, we identified the double second-quantization paradox: the CQME Hamiltonian was previously shown to yield the standard LNF, whereas the present derivation demonstrates that it rigorously leads to the MLNF. The resolution of this apparent contradiction originates directly from the restricted mathematical ansatz initially chosen to perform the standard Fano diagonalization. 

Specifically, the standard LNF is derived in Ref.\cite{Philbi} by assuming that the canonical field operators can be expanded entirely in terms of a continuous set of internal material polariton operators (${\bf{\hat f}}_{\omega e}$ and ${\bf{\hat f}}_{\omega m}$). Although not explicitly stated as a spatial constraint in the original formulation, this premise inherently restricts the physical validity of the procedure to an unbounded dissipative medium, thereby artificially limiting the structure of the resulting Fock space. Indeed, for an unbounded lossy medium, this approach is mathematically exact: any electromagnetic radiation is ultimately absorbed, the dyadic Green's function ${\cal G}_{\omega}$ exhibits an exponentially decaying asymptotic behavior, and the modal dyadic ${\cal F}_{\omega}$ identically vanishes everywhere, as no plane waves can propagate from or to spatial infinity. Consequently, all asymptotic surface integrals strictly vanish, justifying the complete omission of the scattering channels.

However, for a finite-size lossy or transparent object surrounded by vacuum, there are asymptotic surface integrals evaluated over $S_\infty$ that do not vanish. A critical manifestation of this fundamental difference is found in our exact diagonalization of the Hamiltonian (Sec.V). As shown in Eq.(\ref{H_diag_2}), the residual asymptotic surface integral does not evaluate to zero; instead, it explicitly generates the free-evolution energy term of the scattering polaritons ${\bf{\hat g}}_{\omega s}$. Ignoring these asymptotic surface contributions artificially truncates the full Fock space of the field-matter system by omitting the independent degrees of freedom associated with free-space radiation.

To understand why the standard ansatz fails to capture these radiative degrees of freedom, it is crucial to recall that applying the Fano diagonalization method ``essentially amounts to solving the Hamilton equations for the quantum canonical operators'' (i.e., the macroscopic quantum Maxwell equations), as correctly stated by Philbin in Ref.\cite{Philbi}. For an unbounded medium, the solution to these equations consists solely of the particular solution, namely the radiative contribution governed entirely by the dyadic Green's function. This is because any nontrivial solution to the homogeneous equations is physically inadmissible, as it would exhibit an exponential divergence in certain spatial directions (leaky modes). Conversely, for a bounded medium, the complete solution must inherently encompass the general solution of the homogeneous equations, which is generated by the modal dyadic. Consequently, if one were to apply the Fano diagonalization procedure to a finite-size object using an expanded ansatz that includes the scattering polariton operators alongside the electric and magnetic ones, the method would rigorously and exactly yield the MLNF. This is strongly corroborated by the fact that the expressions for the material polariton operators ${{\bf{\hat f}}_{\omega \nu } }$ in terms of the canonical fields, as derived in Ref.\cite{Philbi}, identically match those obtained in our Eqs.(\ref{gf2}). The fundamental distinction, therefore, lies entirely in the absence of the scattering polaritons ${\bf{\hat g}}_{\omega s}$ in Philbin's original diagonalization procedure.

This definitively resolves the double second-quantization paradox. While the standard LNF provides the correct theoretical framework for unbounded dissipative media, the MLNF constitutes the general second-quantized formulation of the CQME valid for arbitrary bounded or unbounded, lossy or transparent, magneto-dielectric system. In the spatial limit of an unbounded medium, the scattering polariton contributions identically vanish, and the MLNF formally reduces to the LNF.

\section{Conclusion}
In this work, we have presented a rigorous derivation of the modified Langevin noise formalism (MLNF) directly from the canonical quantization of macroscopic electromagnetism (CQME). By introducing the analytical expressions for the scattering and material polaritons as postulated constitutive operators, we have formally demonstrated that they rigorously satisfy the macroscopic bosonic algebra and exactly diagonalize the fundamental field-matter Hamiltonian. This fully direct and exact procedure definitively dispels any potential conceptual doubts left open by the previous indirect derivation of Ref.\cite{Ciatt1}, unequivocally establishing the rigorous theoretical validity of the MLNF.

From a physical standpoint, the exact analytical mapping between the canonical fields and the polariton operators reveals the fundamental physical distinction between the localized material excitations and the free-space radiative channels. The polaritons are constructed from two fundamental physical entities: the macroscopic excitation density ${\bf{\hat F}}_\omega$, which encapsulates the non-local collective response of the dressed electromagnetic field, and the bare material excitation ${\bf{\hat h}}_{\omega \nu}$, representing the strictly local decoupled mechanical oscillators. The analytical expressions for the material polaritons ${\bf{\hat f}}_{\omega \nu }$ are constructed as the spatial projection of ${\bf{\hat F}}_\omega$ onto the dyadic Green's function augmented by the bare material excitation ${\bf{\hat h}}_{\omega \nu}$, identically recovering those derived in the original CQME diagonalization of Ref.\cite{Philbi}. In contrast, the scattering polaritons ${\bf{\hat g}}_{\omega s}$ are determined exclusively by the spatial projection of ${\bf{\hat F}}_\omega$ onto the scattering modal dyadic. Their explicit analytical definition and inclusion constitute the crucial mathematical extension missing in the standard diagonalization procedure.

The extensive algebraic derivations required to prove the bosonic algebra and diagonalize the Hamiltonian mathematically reflect a fundamental physical interplay: the dynamic coexistence of free-space radiation, governed by the modal dyadic ${\cal F}_{\omega}$, and the electromagnetic field generated by internal quantum sources, propagated by the dyadic Green's function ${\cal G}_{\omega}$. This interplay is regulated by the distinct asymptotic behaviors of these fundamental classical electrodynamics dyadics, which give rise to mixed asymptotic surface integrals each fulfilling a distinct role throughout the theoretical derivation. In the derivation of the bosonic commutation relations, such asymptotic surface integrals consistently vanish, physically ensuring that the internal material dynamics and the asymptotic scattering channels remain independent and orthogonal bosonic degrees of freedom. Conversely, in diagonalizing the CQME Hamiltonian, a residual asymptotic surface integral explicitly survives to yield the scattering polaritons' free-evolution energy. This recovers a contribution structurally unattainable through the dyadic Green's function alone, as its strictly outgoing radiation condition intrinsically fails to capture incoming free-space modes. 

Ultimately, the observation that the MLNF encompasses all independent electromagnetic degrees of freedom provides the definitive physical resolution to the double second-quantization paradox. Clarifying that the Fano diagonalization procedure in Ref.\cite{Philbi} implicitly relies on a restricted ansatz that artificially truncates the Fock space by omitting free-space radiation (an assumption perfectly valid only for unbounded dissipative media), we establish the MLNF as the fully general second-quantized formulation of the CQME, valid for any arbitrary bounded or unbounded, lossy or transparent, magneto-dielectric system.

\appendix

\section{Dyadic and asymptotic relations}
We denote with ${\bf{AB}}$ the dyad formed by the vectors ${\bf{A}}$ and ${\bf{B}}$. The identity dyadic is ${\cal I} = {\bf{u}}_x {\bf{u}}_x  + {\bf{u}}_y {\bf{u}}_y  + {\bf{u}}_z {\bf{u}}_z$, where ${\bf{u}}_x$, ${\bf{u}}_y$ and ${\bf{u}}_z$ are Cartesian unit vectors. Any dyadic ${\cal Q}$ admits the cartesian expansion ${\cal Q} = Q_{ij} {\bf{u}}_i {\bf{u}}_j $, where $Q_{ij}  = {\bf{u}}_i  \cdot {\cal Q} \cdot {\bf{u}}_j$, and its transpose is the dyadic ${\cal Q}^T  = Q_{ji} {\bf{u}}_i {\bf{u}}_j$ which, for any vector ${\bf V} = V_i {\bf u}_i$, satisfies the relation ${\cal Q}^T  \cdot {\bf{V}} = {\bf{V}} \cdot {\cal Q}$. 

If  ${\bf{n}}$ is a unit vector, we denote with ${\cal I}_{\bf{n}}  = {\cal I} - {\bf{nn}}$ the dyadic projector on the plane orthogonal to ${\bf{n}}$ which is also given by the relation ${\cal I}_{\bf{n}}  = \sum\nolimits_\lambda  {{\bf{e}}_{{\bf{n}}1} {\bf{e}}_{{\bf{n}}2} }$ where ${\bf{e}}_{{\bf{n}}1}$ and ${\bf{e}}_{{\bf{n}}2}$ are arbitrary orthonormal vectors spanning the plane orthogonal to ${\bf{n}}$.

Given a vector field ${\bf{V}}$ and the dyadic field  ${\cal Q}$,  left ${\nabla  \times }$ and right ${ \times \mathord{\buildrel{\lower3pt\hbox{$\scriptscriptstyle\leftarrow$}} \over \nabla } }$ curls can be defined according to
\begin{align}
\nabla  \times {\bf{V}} &= \varepsilon _{ijk} \partial _i V_j {\bf{u}}_k , & 
\nabla  \times {\cal Q} &= \varepsilon _{kip} \partial _k Q_{ij} {\bf{u}}_p {\bf{u}}_j, &
{\bf{V}} \times \mathord{\buildrel{\lower3pt\hbox{$\scriptscriptstyle\leftarrow$}} 
\over \nabla }  &= \varepsilon _{jik} \partial _i V_j {\bf{u}}_k , &
{\cal Q} \times \mathord{\buildrel{\lower3pt\hbox{$\scriptscriptstyle\leftarrow$}} 
\over \nabla }  &= \varepsilon _{jkp} \partial _k Q_{ij} {\bf{u}}_i {\bf{u}}_p , 
\end{align}
whose connection is defined by the relations 
\begin{align}
{\bf{V}} \times \mathord{\buildrel{\lower3pt\hbox{$\scriptscriptstyle\leftarrow$}} 
\over \nabla }  =&  - \nabla  \times {\bf{V}}, &
{\cal Q} \times \mathord{\buildrel{\lower3pt\hbox{$\scriptscriptstyle\leftarrow$}} 
\over \nabla }  =&  - \left( {\nabla  \times {\cal Q}^T } \right)^T.
\end{align} 

If the vector fields ${\bf{V}}$ and ${\bf{U}}$ and the dyadic fields ${\cal Q}$ and ${\cal P}$ are piecewise continuous, the divergence theorem intended in the sense of distributions \cite{Kanwal} readily provides the relations
\begin{eqnarray} \label{IntPart}
 \int {d^3 {\bf{r}}} \;{\bf{V}} \cdot \left( {\nabla  \times {\bf{U}}} \right) &=& \int {d^3 {\bf{r}}} \left( {\nabla  \times {\bf{V}}} \right) \cdot {\bf{U}} + \int\limits_{S_\infty  } {dS}_r \;{\bf{V}} \cdot \left( {{\bf{u}}_r  \times {\bf{U}}} \right), \nonumber \\ 
 \int {d^3 {\bf{r}}} \;{\cal Q}^T  \cdot \left( {\nabla  \times {\bf{U}}} \right) &=& \int {d^3 {\bf{r}}} \left( {\nabla  \times {\cal Q}} \right)^T  \cdot {\bf{U}} + \int\limits_{S_\infty  } {dS}_r \;{\cal Q}^T  \cdot \left( {{\bf{u}}_r  \times {\bf{U}}} \right), \nonumber \\ 
 \int {d^3 {\bf{r}}} \;{\cal Q}^T  \cdot \left( {\nabla  \times {\cal P}} \right) &=& \int {d^3 {\bf{r}}} \left( {\nabla  \times {\cal Q}} \right)^T  \cdot {\cal P} + \int\limits_{S_\infty  } {dS}_r \;{\cal Q}^T  \cdot \left( {{\bf{u}}_r  \times {\cal P}} \right)
\end{eqnarray}
where ${\bf{u}}_r = {\bf r}/r$ is the radial unit vector and the surface integrals are performed over the sphere $S_\infty$ of infinite radius by means of the limiting prescription 
\begin{equation} \label{asym_sur_int}
\int_{S_\infty  } {dS}_r \;f\left( {\bf{r}} \right) = \mathop {\lim }\limits_{r \to \infty } r^2 \int {do_r f\left( {r{\bf{u}}_r } \right)}.
\end{equation}
where ${do_r }$ is the differential of the solid angle around the radial unit vector ${{\bf{u}}_r }$.

The dyadic transverse $\bot$ and longitudinal $\parallel$ delta functions, together with their divergence and curl, are 
\begin{align} \label{trs-long-delt}
{\cal D}^ \bot  \left( {\bf{r}} \right) =& \frac{1}{{\left( {2\pi } \right)^3 }}\int {d^3 {\bf{k}}} \;e^{i{\bf{k}} \cdot {\bf{r}}} \left( {{\cal I} - \frac{{{\bf{kk}}}}{{k^2 }}} \right), & 
{\cal D}^\parallel  \left( {\bf{r}} \right) =& \frac{1}{{\left( {2\pi } \right)^3 }}\int {d^3 {\bf{k}}} \;e^{i{\bf{k}} \cdot {\bf{r}}} \left( {\frac{{{\bf{kk}}}}{{k^2 }}} \right), \nonumber \\
\nabla _{\bf{r}}  \cdot {\cal D}^ \bot  \left( {\bf{r}} \right) =& 0, &
\nabla _{\bf{r}}  \cdot {\cal D}^\parallel  \left( {\bf{r}} \right) =& \nabla _{\bf{r}}  \cdot \left[ {\delta \left( {\bf{r}} \right){\cal I}} \right], \nonumber \\
\nabla _{\bf{r}}  \times {\cal D}^ \bot  \left( {\bf{r}} \right) =& \nabla_{\bf{r}}  \times \left[ {\delta \left( {\bf{r}} \right){\cal I}} \right], &
\nabla _{\bf{r}}  \times {\cal D}^\parallel  \left( {\bf{r}} \right) =& 0
\end{align}
and they are the dyadic kernels of projection operators owing to the relations
\begin{align}
{\cal D}^ \bot  \left( {\bf{r}} \right) + {\cal D}^\parallel  \left( {\bf{r}} \right) =& \delta \left( {\bf{r}} \right){\cal I}, & 
\int {d^3 {\bf{s}}} \;{\cal D}^\alpha  \left( {{\bf{r}} - {\bf{s}}} \right) \cdot {\cal D}^\beta  \left( {{\bf{s}} - {\bf{r}}'} \right) =& \delta _{\alpha \beta } {\cal D}^\alpha  \left( {{\bf{r}} - {\bf{r}}'} \right).
\end{align}
Besides, for any vector field $\bf V$, the relations
\begin{align} \label{Tra_vec}
\int {d^3 {\bf{r}}'} \;{\cal D}^ \bot  \left( {{\bf{r}} - {\bf{r}}'} \right) \cdot \left[ {\nabla _{{\bf{r}}'}  \times {\bf{V}}\left( {{\bf{r}}'} \right)} \right] =& \nabla _{\bf{r}}  \times {\bf{V}}\left( {\bf{r}} \right), &
\int {d^3 {\bf{r}}'} \;{\cal D}^\parallel  \left( {{\bf{r}} - {\bf{r}}'} \right) \cdot \left[ {\nabla _{{\bf{r}}'}  \times {\bf{V}}\left( {{\bf{r}}'} \right)} \right] =& 0
\end{align}
hold.

The commutator of two vector operators ${\bf{\hat V}}= \hat V_i {\bf{u}}_i $ and ${\bf{\hat U}} = \hat U_i {\bf{u}}_i$ is the dyadic operator $[ {{\bf{\hat V}},{\bf{\hat U}}} ] = [ {\hat V_i ,\hat U_j } ]{\bf{u}}_i {\bf{u}}_j$ and it satisfies the relations
\begin{align} \label{Com_Prop}
\left[ {{\bf{\hat V}},{\bf{\hat U}}} \right]^T  =&  - \left[ {{\bf{\hat U}},{\bf{\hat V}}} \right], & 
\left[ {{\bf{\hat V}},{\bf{\hat U}}} \right]^\dag   =&  - \left[ {{\bf{\hat V}}^\dag  ,{\bf{\hat U}}^\dag  } \right].
\end{align}
If ${\cal Q}$ and ${\cal P}$ are two dyadics, it is easy to prove the relation
\begin{equation}
[ {{\cal Q} \cdot {\bf{\hat V}},{\cal P} \cdot {\bf{\hat U}}} ] = {\cal Q} \cdot [ {{\bf{\hat V}},{\bf{\hat U}}} ] \cdot {\cal P}^T.
\end{equation}

If ${\bf n}$ and ${\bf m}$ are unit vectors and $f({\bf m})$ is a function defined on the unit sphere, the classical Jones' lemma (see Appendix XII of Ref.\cite{Bornn1}) states that
\begin{equation} \label{JonLem}
\int {do_{\bf{m}} } e^{i\left( {k_\omega  {\bf{n}}} \right) \cdot \left( {r{\bf{m}}} \right)} f\left( {\bf{m}} \right)\mathop  \approx \limits_{r \to  + \infty } \frac{{2\pi i}}{{k_\omega  }}\left[ {\frac{{e^{ - ik_\omega  r} }}{r}f\left( { - {\bf{n}}} \right) - \frac{{e^{ik_\omega  r} }}{r}f\left( {\bf{n}} \right)} \right],
\end{equation}
which can be written in distributional form as
\begin{equation} \label{JonLemDist}
e^{i\left( {k_\omega  {\bf{n}}} \right) \cdot \left( {r{\bf{m}}} \right)} \mathop  \approx \limits_{r \to  + \infty } \frac{{2\pi i}}{{k_\omega  }}\left[ {\frac{{e^{ - ik_\omega  r} }}{r}\delta \left( {o_{\bf{m}}  - o_{ - {\bf{n}}} } \right) - \frac{{e^{ik_\omega  r} }}{r}\delta \left( {o_{\bf{m}}  - o_{\bf{n}} } \right)} \right].
\end{equation}
Two generalized functions that play an essential role in the present study are defined by the limits 
\begin{align} \label{Dist_FUND}
 \mathop {\lim }\limits_{r \to  + \infty } e^{iKr}  =& 0, &
\mathop {\lim }\limits_{r \to  + \infty } \frac{{e^{ - iKr} }}{{K + i\sigma \eta }} =&  - 2\pi i\delta _{\sigma ,1} \delta \left( K \right),
\end{align}
where $\sigma = \pm 1$. The first of Eqs.(\ref{Dist_FUND}) is a direct consequence of the Riemann-Lebesgue lemma which, for any test function $f(K)$, ensures that 
\begin{equation}
{\mathop {\lim }\limits_{r \to  + \infty } \int\limits_{ - \infty }^{ + \infty } {dK} e^{iKr} f\left( K \right) = 0}.
\end{equation}
The second of Eqs.(\ref{Dist_FUND}) follows from the Plemelj identity, $\frac{1}{{K + i\sigma \eta }} = P\frac{1}{K} - i\pi \sigma \delta \left( K \right)$, which for any test function $f(K)$ yields
\begin{equation}
\mathop {\lim }\limits_{r \to  + \infty } \int\limits_{ - \infty }^{ + \infty } {dK} \;\frac{{e^{ - iKr} }}{{K + i\sigma \eta }}f\left( K \right) = \mathop {\lim }\limits_{r \to  + \infty } \left[ {P\int\limits_{ - \infty }^{ + \infty } {dK} \;e^{ - iKr} \frac{{f\left( K \right)}}{K}} \right] - i\pi \sigma f\left( 0 \right).
\end{equation}
By employing the well-known principal value integral $P\int\limits_{ - \infty }^{ + \infty } {dK} \frac{{e^{ - iKr} }}{K} =  - i\pi \theta \left( r \right)$ (for $r > 0$), the expression can be rewritten as
\begin{equation}
\mathop {\lim }\limits_{r \to  + \infty } \int\limits_{ - \infty }^{ + \infty } {dK} \;\frac{{e^{ - iKr} }}{{K + i\sigma \eta }}f\left( K \right) = \mathop {\lim }\limits_{r \to  + \infty } \left[ {P\int\limits_{ - \infty }^{ + \infty } {dK} \;e^{ - iKr} \frac{{f\left( K \right) - f\left( 0 \right)}}{K}} \right] - i\pi \left( {1 + \sigma } \right)f\left( 0 \right),
\end{equation}
which leads directly to the second of Eqs.(\ref{Dist_FUND}), as the function $\left[ {f\left( K \right) - f\left( 0 \right)} \right]/K$ is non-singular at $K=0$ and the Riemann-Lebesgue lemma ensures the vanishing of the first contribution.

\section{Properties of the modal dyadic and the dyadic Green's function}
In this Appendix we focus on the modal dyadic ${\cal F}_{\omega } \left( {\left. {\bf{r}} \right|{\bf{n}}} \right)$ and the dyadic Green's function ${\cal G}_\omega  \left( {\left. {\bf{r}} \right|{\bf{r}}'} \right)$ defined in Eqs.(\ref{GS}) and we review some of their properties which are essential in the present paper. The scattering dyadic ${\cal S}_\omega  \left( {{\bf{m}}\left| {\bf{n}} \right.} \right)$ satisfies the orthogonality relations ${\bf{m}} \cdot {\cal S}_\omega  \left( {{\bf{m}}\left| {\bf{n}} \right.} \right) = 0$ and ${\cal S}_\omega  \left( {{\bf{m}}\left| {\bf{n}} \right.} \right) \cdot {\bf{n}} = 0$. As a consequence, the modal dyadic satisfies the orthogonality relation ${\cal F}_\omega  \left( {\left. {\bf r} \right|{\bf{n}}} \right) \cdot {\bf{n}} = 0$ and accordingly it is convenient to introduce two mutually orthogonal unit vectors ${\bf{e}}_{{\bf{n}}1}$, ${\bf{e}}_{{\bf{n}}2}$ orthogonal to ${\bf n}$ so that, since  ${\cal I}_{\bf{n}}  = \sum\nolimits_\lambda  {{\bf{e}}_{{\bf{n}} \lambda} {\bf{e}}_{{\bf{n}} \lambda} }$, the field ${\bf{F}}_{\omega {\bf{n}}\lambda } \left( {\bf{r}} \right) = {\cal F}_{\omega} \left( {\left. {\bf{r}} \right|{\bf{n}}} \right) \cdot {\bf{e}}_{{\bf{n}}\lambda }$ satisfies the boundary value problem 
\begin{align}
\left[ {\left( {\nabla _{\bf{r}}  \times \frac{1}{{\mu _\omega  \left( {\bf{r}} \right)}}\nabla _{\bf{r}}  \times } \right) - k_\omega ^2 \varepsilon _\omega  \left( {\bf{r}} \right)} \right]{\bf{F}}_{\omega {\bf{n}}\lambda } \left( {\bf{r}} \right) &= 0, &
{\bf{F}}_{\omega {\bf{n}}\lambda } \left( {r{\bf{m}}} \right) & \mathop  \approx \limits_{r \to  + \infty } e^{i\left( {k_\omega  {\bf{n}}} \right) \cdot \left( {r{\bf{m}}} \right)} {\bf{e}}_{{\bf{n}}\lambda }  + \frac{{e^{ik_\omega  r} }}{r}{\cal S}_\omega  \left( {{\bf{m}}\left| {\bf{n}} \right.} \right) \cdot {\bf{e}}_{{\bf{n}}\lambda } 
\end{align}
or, in other words, ${\bf{F}}_{\omega {\bf{n}}\lambda } \left( {\bf{r}} \right)$ is the overall field scattered by the object (scattering mode) when illuminated by the modal plane wave of frequency $\omega$, direction $\bf n$ and polarization ${\bf{e}}_{{\bf{n}}\lambda }$. It is also evident that the modal dyadic admits the decomposition 
${\cal F}_\omega  \left( {\left. {\bf{r}} \right|{\bf{n}}} \right) = \sum\nolimits_\lambda  {{\bf{F}}_{\omega {\bf{n}}\lambda } \left( {\bf{r}} \right){\bf{e}}_{{\bf{n}}\lambda } }$ which further elucidates its physical meaning. The dyadic ${\cal W}_\omega  \left( {{\bf{m}}\left| {\bf{r}} \right.} \right)$ satisfies the orthogonality relation ${\bf{m}} \cdot {\cal W}_\omega  \left( {{\bf{m}}\left| {\bf{r}} \right.} \right) = 0$. Two essential properties of the dyadic Green's function are the reciprocity relation ${\cal G}_\omega ^T \left( {\left. {\bf{r}} \right|{\bf{r}}'} \right) = {\cal G}_\omega  \left( {\left. {{\bf{r}}'} \right|{\bf{r}}} \right)$ and the fundamental integral relation
\begin{equation} \label{Gom_fun}
\frac{{\hbar k_\omega ^3 }}{{\pi \varepsilon _0 }}\int {do_{\bf{m}} } {\cal W}_\omega ^T \left( {\left. {\bf{m}} \right|{\bf{r}}} \right) \cdot {\cal W}_\omega ^* \left( {\left. {\bf{m}} \right|{\bf{r}}'} \right) + \sum\limits_\nu  {\int {d^3 {\bf{s}}} \,{\cal G}_{\omega \nu } \left( {\left. {\bf{r}} \right|{\bf{s}}} \right) \cdot {\cal G}_{\omega \nu }^{T*} \left( {\left. {{\bf{r}}'} \right|{\bf{s}}} \right)}  = \frac{{\hbar k_\omega ^2 }}{{\pi \varepsilon _0 }}{\mathop{\rm Im}\nolimits} \left[ {{\cal G}_\omega  \left( {\left. {\bf{r}} \right|{\bf{r}}'} \right)} \right],
\end{equation}
where ${\cal G}_{\omega e} \left( {\left. {\bf{r}} \right|{\bf{r}}'} \right)$ and ${\cal G}_{\omega m} \left( {\left. {\bf{r}} \right|{\bf{r}}'} \right)$ are the dyadic kernels defined in Eqs.(\ref{Kernels}). Such essential properties have been rederived in Ref.\cite{Ciatt1} where, in addition, it  has been shown that the asymptotic amplitude of the dyadic Green's function is related to the modal dyadic by the relation 
\begin{equation}
{\cal W}_\omega  \left( {{\bf{m}}\left| {\bf{r}}\right.} \right) = \frac{1}{{4\pi }}{\cal F}_\omega ^T \left( {\left. {\bf{r}} \right| - {\bf{m}}} \right)
\end{equation}
 which, together with Eq.(\ref{Gom_fun}), yields the integral relation 
\begin{equation}
\int {do_{\bf{n}} } {\cal F}_{\omega s} \left( {\left. {\bf{r}} \right|{\bf{n}}} \right) \cdot {\cal F}_{\omega s}^{T*} \left( {\left. {{\bf{r}}'} \right|{\bf{n}}} \right) + \sum\limits_\nu  {\int {d^3 {\bf{s}}} \,{\cal G}_{\omega \nu } \left( {\left. {\bf{r}} \right|{\bf{s}}} \right) \cdot {\cal G}_{\omega \nu }^{T*} \left( {\left. {{\bf{r}}'} \right|{\bf{s}}} \right)}  = \frac{{\hbar k_\omega ^2 }}{{\pi \varepsilon _0 }}{\mathop{\rm Im}\nolimits} \left[ {{\cal G}_\omega  \left( {\left. {\bf{r}} \right|{\bf{r}}'} \right)} \right]
\end{equation}
connecting the modal dyadic and the dyadic Green's function, where ${\cal F}_{\omega s} \left( {\left. {\bf{r}} \right|{\bf{n}}} \right)$ is the dyadic kernel defined in the first Eqs.(\ref{Kernels}). The modal dyadic and the dyadic Green's function provide the description of any classical scattering and radiation process in the presence of the object since, if ${\bf{E}}_\omega ^{\left( {in} \right)} \left( {\bf{r}} \right) = \int {do_{\bf{n}} } e^{i\left( {k_\omega  {\bf{n}}} \right) \cdot {\bf{r}}} {\bf{\tilde E}}_\omega ^{\left( {in} \right)} \left( {\bf{n}} \right)$ (with ${\bf{n}} \cdot {\bf{\tilde E}}_\omega ^{\left( {in} \right)} \left( {\bf{n}} \right)= 0$), is any vacuum free field impinging onto the object and ${\bf{J}}_\omega  \left( {\bf{r}} \right)$ is the current density of any radiation source, the inhomogeneous Helmholtz equation 
\begin{equation}
\left[ {\left( {\nabla  \times \frac{1}{{\mu _\omega  }}\nabla  \times } \right) - k_\omega ^2 \varepsilon _\omega  } \right]{\bf{E}}_\omega   = i\omega \mu _0 {\bf{J}}_\omega  
\end{equation}
for the field throughout the whole space has the unique solution
\begin{equation} \label{Clas_Sol}
{\bf{E}}_\omega  \left( {\bf{r}} \right) = \int {do_{\bf{n}} } {\cal F}_\omega  \left( {\left. {\bf{r}} \right|{\bf{n}}} \right) \cdot {\bf{\tilde E}}_\omega ^{\left( {in} \right)} \left( {\bf{n}} \right) + i\omega \mu _0 \int {d^3 {\bf{r}}'} \;{\cal G}_\omega  \left( {\left. {\bf{r}} \right|{\bf{r}}'} \right) \cdot {\bf{J}}_\omega  \left( {{\bf{r}}'} \right),
\end{equation}
which is the superposition of the scattered (first contribution) and radiated (second contribution) fields. From the right column of Eqs.(\ref{GS}), the asymptotic behavior of such field is  ${\bf{E}}_\omega  \left( {r{\bf{m}}} \right)\mathop  \approx \limits_{r \to  + \infty } {\bf{E}}_\omega ^{\left( {in} \right)} \left( {r{\bf{m}}} \right) + \frac{{e^{ik_\omega  r} }}{r}\left[ {{\bf{W}}_\omega ^{\left( {sca} \right)} \left( {\bf{m}} \right) + {\bf{W}}_\omega ^{\left( {rad} \right)} \left( {\bf{m}} \right)} \right]$ where ${\bf{W}}_\omega ^{\left( {sca} \right)} \left( {\bf{m}} \right) = \int {do_{\bf{n}} } {\cal S}_\omega  \left( {\left. {\bf{m}} \right|{\bf{n}}} \right) \cdot {\bf{\tilde E}}_\omega ^{\left( {in} \right)} \left( {\bf{n}} \right)$ and ${\bf{W}}_\omega ^{\left( {rad} \right)} \left( {\bf{m}} \right) = i\omega \mu _0 \int {d^3 {\bf{r}}} \;{\cal W}_\omega  \left( {\left. {\bf{m}} \right|{\bf{r}}} \right) \cdot {\bf{J}}_\omega  \left( {\bf{r}} \right)$ are its far field scattering and radiation amplitudes.

The transverse and longitudinal properties of the modal dyadic and of the dyadic Green's function are also important in our analysis. Such properties are easily obtained by combining Eqs.(\ref{GS}) with the dyadic counterpart of Eqs.(\ref{Tra_vec}) thus getting after some simple algebra
\begin{eqnarray} \label{Modal_Tra}
 \int {d^3 {\bf{s}}} \,{\cal D}^ \bot  \left( {{\bf{r}} - {\bf{s}}} \right) \cdot \left[ {\varepsilon _\omega  \left( {\bf{s}} \right){\cal F}_\omega  \left( {\left. {\bf{s}} \right|{\bf{n}}} \right)} \right] &=& \varepsilon _\omega  \left( {\bf{r}} \right){\cal F}_\omega  \left( {\left. {\bf{r}} \right|{\bf{n}}} \right), \nonumber \\ 
 \int {d^3 {\bf{s}}} \,{\cal D}^\parallel  \left( {{\bf{r}} - {\bf{s}}} \right) \cdot \left[ {\varepsilon _\omega  \left( {\bf{s}} \right){\cal F}_\omega  \left( {\left. {\bf{s}} \right|{\bf{n}}} \right)} \right] &=& 0, \nonumber \\ 
 \int {d^3 {\bf{s}}\,} {\cal D}^ \bot  \left( {{\bf{r}} - {\bf{s}}} \right) \cdot \left[ {\varepsilon _\omega  \left( {\bf{s}} \right){\cal G}_\omega  \left( {\left. {\bf{s}} \right|{\bf{r}}'} \right)} \right] &=& \varepsilon _\omega  \left( {\bf{r}} \right){\cal G}_\omega  \left( {\left. {\bf{r}} \right|{\bf{r}}'} \right) + \frac{1}{{k_\omega ^2 }}{\cal D}^\parallel  \left( {{\bf{r}} - {\bf{r}}'} \right), \nonumber \\ 
 \int {d^3 {\bf{s}}} \,{\cal D}^\parallel  \left( {{\bf{r}} - {\bf{s}}} \right) \cdot \left[ {\varepsilon _\omega  \left( {\bf{s}} \right){\cal G}_\omega  \left( {\left. {\bf{s}} \right|{\bf{r}}'} \right)} \right] &=&  - \frac{1}{{k_\omega ^2 }}{\cal D}^\parallel  \left( {{\bf{r}} - {\bf{r}}'} \right).
\end{eqnarray}

We conclude this Appendix by analyzing the integral $\int {d^3 {\bf{r}}} {\cal F}_\omega ^{T*} \left( {\left. {\bf{r}} \right|{\bf{n}}} \right) \cdot {\cal F}_{\omega '} \left( {\left. {\bf{r}} \right|{\bf{n}}'} \right)$ which plays a crucial role in the present paper. We start from the well-known free-space dyadic Green function
\begin{equation} \label{G0}
{\cal G}_\omega ^{\left( 0 \right)} \left( {\left. {\bf{r}} \right|{\bf{r}}'} \right) = \frac{1}{{\left( {2\pi } \right)^3 }}\int {d^3 {\bf{k}}} \frac{{e^{i{\bf{k}} \cdot \left( {{\bf{r}} - {\bf{r}}'} \right)} }}{{k^2  - k_\omega ^2  - i\eta }}\left( {{\cal I} - \frac{{{\bf{kk}}}}{{k_\omega ^2 }}} \right)
\end{equation}
which satisfies the Helmholtz equation $\left( {\nabla _{\bf{r}}  \times \nabla _{\bf{r}}  \times  - k_\omega ^2 } \right){\cal G}_\omega ^{\left( 0 \right)} \left( {\left. {\bf{r}} \right|{\bf{r}}'} \right) = \delta \left( {\left. {\bf{r}} \right|{\bf{r}}'} \right)$ and has the correct outgoing spherical wave behavior in the far field $r \rightarrow +\infty$ as a consequence of the prescription $\eta \rightarrow 0^+$. The free-space dyadic Green's function here plays a major role since, using its properties, it is straightforward showing that the modal dyadic satisfies the integral equation
\begin{equation} \label{F_int_equ}
{\cal F}_\omega  \left( {\left. {\bf{r}} \right|{\bf{n}}} \right) = e^{i\left( {k_\omega  {\bf{n}}} \right) \cdot {\bf{r}}} {\cal I}_{\bf{n}}  + \int\limits_{V^ +  } {d{\bf{r}}'} \;{\cal G}_\omega ^{\left( 0 \right)} \left( {\left. {\bf{r}} \right|{\bf{r}}'} \right) \cdot \left\{ {\left\{ {\nabla _{{\bf{r}}'}  \times \left[ {1 - \frac{1}{{\mu _\omega  \left( {{\bf{r}}'} \right)}}} \right]\nabla _{{\bf{r}}'}  \times } \right\} - k_\omega ^2 \left[ {1 - \varepsilon _\omega  \left( {{\bf{r}}'} \right)} \right]} \right\}{\cal F}_\omega  \left( {\left. {{\bf{r}}'} \right|{\bf{n}}} \right)
\end{equation}
where the integration is performed over a spatial region $V^+$ slightly larger than the region $V$ filled by the object since the integrand vanishes just outside $V$ as a consequence of Eqs.(\ref{eps_mu}). Now, after noting that the free-space dyadic Green's function of Eq.(\ref{G0}) satisfies the relations
\begin{eqnarray}
\int {d^3 {\bf{s}}} \;e^{ - i\left( {k_\omega  {\bf{n}}} \right) \cdot {\bf{s}}} {\cal I}_{\bf{n}}  \cdot {\cal G}_{\omega '}^{\left( 0 \right)} \left( {\left. {\bf{s}} \right|{\bf{r}}'} \right) &=& \frac{{e^{ - i\left( {k_\omega  {\bf{n}}} \right) \cdot {\bf{r}}'} }}{{k_\omega ^2  - k_{\omega '}^2  - i\eta }}{\cal I}_{\bf{n}}, \nonumber \\
\int {d^3 {\bf{s}}} \;{\cal G}_\omega ^{\left( 0 \right)T*} \left( {\left. {\bf{s}} \right|{\bf{r}}} \right) \cdot {\cal G}_{\omega '}^{\left( 0 \right)} \left( {\left. {\bf{s}} \right|{\bf{r}}'} \right) &=& \frac{{{\cal G}_\omega ^{\left( 0 \right)T*} \left( {\left. {{\bf{r}}'} \right|{\bf{r}}} \right) - {\cal G}_{\omega '}^{\left( 0 \right)} \left( {\left. {\bf{r}} \right|{\bf{r}}'} \right)}}{{k_\omega ^2  - k_{\omega '}^2  - i\eta }}
\end{eqnarray}
from Eq.(\ref{F_int_equ}), after some tedious dyadic algebra, it is possible to prove that 
\begin{equation} \label{F-F_int}
\int {d^3 {\bf{r}}} {\cal F}_\omega ^{T*} \left( {\left. {\bf{r}} \right|{\bf{n}}} \right) \cdot {\cal F}_{\omega '} \left( {\left. {\bf{r}} \right|{\bf{n}}'} \right) = \left( {2\pi } \right)^3 \frac{{\delta \left( {k_\omega   - k_{\omega '} } \right)}}{{k_\omega ^2 }}\delta \left( {o_{\bf{n}}  - o_{{\bf{n}}'} } \right){\cal I}_{\bf{n}}  + \frac{{{\cal C}_{\omega ,\omega '} \left( {\left. {\bf{n}} \right|{\bf{n}}'} \right)}}{{k_\omega ^2  - k_{\omega '}^2  - i\eta }}
\end{equation}
where the first delta contribution arises from the asymptotic plane wave behavior of the modal dyadic whereas the dyadic
\begin{equation}
{\cal C}_{\omega ,\omega '} \left( {\left. {\bf{n}} \right|{\bf{n}}'} \right) = \int\limits_{V^ +  } {d{\bf{r}}} {\cal F}_\omega ^{T*} \left( {\left. {\bf{r}} \right|{\bf{n}}} \right) \cdot \left\{ {\left\{ {\nabla _{\bf{r}}  \times \left[ {\frac{1}{{\mu _\omega ^* \left( {\bf{r}} \right)}} - \frac{1}{{\mu _{\omega '} \left( {\bf{r}} \right)}}} \right]\nabla _{\bf{r}}  \times } \right\} + k_\omega ^2 \left[ {1 - \varepsilon _\omega ^* \left( {\bf{r}} \right)} \right] - k_{\omega '}^2 \left[ {1 - \varepsilon _{\omega '} \left( {\bf{r}} \right)} \right]} \right\}{\cal F}_{\omega '} \left( {\left. {\bf{r}} \right|{\bf{n}}'} \right)
\end{equation}
is a regular function, i.e. it does not yield a delta contribution $\delta \left( {k_\omega   - k_{\omega '} } \right)$ since the integration is performed over the bounded region $V^+$ and it can not diverge.

\section{Generalized dispersion integrals for the magneto-dielectric response}
In this Appendix, we derive several generalized dispersion relations by evaluating frequency integrals over the optical properties of the medium. Specifically, the integrands involve the permittivity $\varepsilon _\omega$ and the inverse 
permeability $\mu _\omega ^{ - 1}$ (see Eqs.(\ref{eps_mu})) which, for conciseness, we both denote with  $\varphi _\omega$. The dispersive optical response of the object is such that $
\varphi _\omega   = 1 + \chi _\omega$ where the susceptibility $\chi _\omega$ is such that $\chi _\omega ^*  = \chi _{ - \omega }$, it is an holomorphic function in the upper half-plane ${\mathop{\rm Im}\nolimits} \left( \omega  \right) > 0$ (due to causality) and $\chi _\omega   \to 0$ for $\omega  \to \infty$. The fundamental integrals, which serve to generate all others required for this work, are
\begin{eqnarray} \label{fund_int}
\frac{2}{\pi }\int\limits_0^{ + \infty } {d\Omega } \;\frac{{\Omega {\mathop{\rm Im}\nolimits} \left( {\varphi _\Omega  } \right)}}{{\Omega ^2  - \left( {\omega  + i\eta } \right)^2 }} &=& \varphi _\omega   - 1, \nonumber \\
\frac{2}{\pi }\int\limits_0^{ + \infty } {d\Omega } \frac{{\Omega ^{2n + 1} {\mathop{\rm Im}\nolimits} \left( {\varphi _\Omega  } \right)}}{{\left[ {\Omega ^2  - \left( {\omega  + i\sigma \eta } \right)^2 } \right]\left[ {\Omega ^2  - \left( {\omega ' + i\sigma '\eta } \right)^2 } \right]}} &=& \frac{{\omega ^{2n} \left( {\varphi _{\sigma \omega }  - 1} \right) - \omega '^{2n} \left( {\varphi _{\sigma '\omega '}  - 1} \right)}}{{\left( {\omega  + i\sigma \eta } \right)^2  - \left( {\omega ' + i\sigma '\eta } \right)^2 }}
\end{eqnarray}
where $\omega >0$, $\omega'>0$, $\sigma  =  \pm 1$, $\sigma'  =  \pm 1$ and $n =0,1$. The proof of Eqs.(\ref{fund_int}) is based on the fact that, for any 
odd function $F \left( \Omega  \right) =  - F\left( { - \Omega } \right)$ which decays at least as $\Omega ^ {-1}$, it is simple to show that
\begin{equation}
\int\limits_0^{ + \infty } {d\Omega } \;F\left( \Omega  \right){\mathop{\rm Im}\nolimits} \left( {\varphi _\Omega  } \right) = \frac{1}{{2i}}\int\limits_{ - \infty }^{ + \infty } {d\Omega } \;F\left( \Omega  \right)\chi _\Omega  
\end{equation}
so that, by the residue theorem, the integral is determined solely by the poles of $F(\Omega)$ in the upper half-plane. This is ensured by the analyticity of $\chi_\Omega$ and the vanishing contribution of the infinite semicircular contour, as the integrand $F(\Omega)\chi_\Omega$ decays faster than $\Omega^{-1}$ at infinity. Thus, the first integral in Eqs.(\ref{fund_int}) is evaluated via the single pole $\omega + i\eta$, while the second arises from the poles $\sigma\omega + i\eta$ and $\sigma'\omega' + i\eta$. Note that the first of Eqs.(\ref{fund_int}) is strictly equivalent to the well-known Kramers-Kronig relations whereas the second is a more complex generalized dispersion relation whose behavior as a generalized function depends significantly on the signs $\sigma$ and $\sigma'$. Using the first of Eqs.(\ref{fund_int}) and Eqs.(\ref{alf_bet}), it follows directly that
\begin{align} \label{Int_1}
 \int\limits_0^{ + \infty } {d\Omega } \;\frac{{\left( {\alpha ^\Omega  } \right)^2 }}{{\Omega ^2  - \left( {\omega  + i\eta } \right)^2 }} =& \varepsilon _0 \left( {\varepsilon _\omega   - 1} \right), &
 \int\limits_0^{ + \infty } {d\Omega } \;\frac{{\left( {\beta ^\Omega  } \right)^2 }}{{\Omega ^2  - \left( {\omega  + i\eta } \right)^2 }} =& \frac{1}{{\mu _0 }}\left( {1 - \frac{1}{{\mu _\omega  }}} \right).
\end{align}
From the second of Eqs.(\ref{fund_int}) with $n=0$, after some algebra, we obtain 
\begin{equation} \label{Int_2}
\frac{2}{\pi }\int\limits_0^{ + \infty } {d\Omega } \frac{{\left( {\omega  + \sigma '\omega '} \right)\Omega {\mathop{\rm Im}\nolimits} \left( {\varphi _\Omega  } \right)}}{{\left[ {\Omega ^2  - \left( {\omega  - i\eta } \right)^2 } \right]\left[ {\Omega ^2  - \left( {\omega ' + i\sigma '\eta } \right)^2 } \right]}} = \delta _{\sigma ', - 1} \left( {\frac{{\varphi _\omega ^*  - \varphi _{\omega '}^* }}{{\omega  + \omega '}}} \right) + \delta _{\sigma ',1} \left( {\frac{{\varphi _\omega ^*  - \varphi _{\omega '} }}{{\omega  - \omega ' - i\eta }}} \right).
\end{equation}
Eventually, resorting to the second of Eqs.(\ref{fund_int}) with both $n=0$ and $n=1$ and using 
Eqs.(\ref{alf_bet}), after some work, we get
\begin{eqnarray} \label{Int_3}
 \int\limits_0^{ + \infty } {d\Omega } \frac{{\left( {\Omega ^2  - \sigma '\omega \omega '} \right)\left( {\alpha ^\Omega  } \right)^2 }}{{\left[ {\Omega ^2  - \left( {\omega  + i\eta } \right)^2 } \right]\left[ {\Omega ^2  - \left( {\omega ' + i\sigma '\eta } \right)^2 } \right]}} &=& \varepsilon _0 \left[ {\delta _{\sigma ', - 1} \left( {\frac{{\omega \varepsilon _\omega   - \omega '\varepsilon _{\omega '}^* }}{{\omega  - \omega ' + i\eta }} - 1} \right) + \delta _{\sigma ',1} \left( {\frac{{\omega \varepsilon _\omega   + \omega '\varepsilon _{\omega '} }}{{\omega  + \omega '}} - 1} \right)} \right], \nonumber  \\
\int\limits_0^{ + \infty } {d\Omega } \frac{{\left( {\Omega ^2  - \sigma '\omega \omega '} \right)\left( {\beta ^\Omega  } \right)^2 }}{{\left[ {\Omega ^2  - \left( {\omega  + i\eta } \right)^2 } \right]\left[ {\Omega ^2  - \left( {\omega ' + i\sigma '\eta } \right)^2 } \right]}} &=& \frac{1}{{\mu _0 }}\left[ {\delta _{\sigma ', - 1} \left( {\frac{{ - \omega \mu _\omega ^{ - 1}  + \omega '\mu _{\omega '}^{ - 1*} }}{{\omega  - \omega ' + i\eta }} + 1} \right) + \delta _{\sigma ',1} \left( {\frac{{ - \omega \mu _\omega ^{ - 1}  - \omega '\mu _{\omega '}^{ - 1} }}{{\omega  + \omega '}} + 1} \right)} \right]. \nonumber \\
\end{eqnarray}

\section{Commutation relations involving the operators ${\bf{\hat F}}_\omega$ and ${\bf{\hat h}}_{\omega \nu }$}
The following commutators involving the operators ${\bf{\hat F}}_\omega$ and ${\bf{\hat h}}_{\omega \nu }$ defined in Eqs.(\ref{Fom}) are evaluated by using the canonical commutation relations of the CQME in Eqs.(\ref{CQME_Com_Rel}). The two commutators involving only ${\bf{\hat F}}_\omega$ operators are seen to be given by
\begin{eqnarray} \label{FF_FFc}
&& \frac{\hbar }{{\varepsilon _0 }}\left[ {{\bf{\hat F}}_\omega  \left( {\bf{r}} \right),{\bf{\hat F}}_{\omega '}^\dag  \left( {{\bf{r}}'} \right)} \right] = \left[ {\frac{{\varepsilon _\omega ^* \left( {\bf{r}} \right)}}{{\omega '}} + \frac{{\varepsilon _{\omega '} \left( {{\bf{r}}'} \right)}}{\omega }} \right]{\cal D}^ \bot  \left( {{\bf{r}} - {\bf{r}}'} \right) + \left\{ {\frac{2}{\pi }\int\limits_0^\infty  {d\Omega } \frac{{\left( {\omega  + \omega '} \right)\Omega {\mathop{\rm Im}\nolimits} \left[ {\varepsilon_	\Omega \left( {\bf{r}} \right)} \right]}}{{\left[ {\Omega ^2  - \left( {\omega  - i\eta } \right)^2 } \right]\left[ {\Omega ^2  - \left( {\omega ' + i\eta } \right)^2 } \right]}}} \right\}\delta \left( {{\bf{r}} - {\bf{r}}'} \right){\cal I} \nonumber \\ 
&& \quad \quad   + \frac{{c^2 }}{{\omega \omega '}}\nabla _{\bf{r}}  \times \left\{ {\left\{ {\frac{2}{\pi }\int\limits_0^\infty  {d\Omega } \frac{{\left( {\omega  + \omega '} \right)\Omega {\mathop{\rm Im}\nolimits} \left[ {\mu _\Omega ^{ - 1} \left( {\bf{r}} \right)} \right]}}{{\left[ {\Omega ^2  - \left( {\omega  - i\eta } \right)^2 } \right]\left[ {\Omega ^2  - \left( {\omega ' + i\eta } \right)^2 } \right]}}} \right\}\delta \left( {{\bf{r}} - {\bf{r}}'} \right){\cal I}} \right\} \times \mathord{\buildrel{\lower3pt\hbox{$\scriptscriptstyle\leftarrow$}} 
\over \nabla } _{{\bf{r}}'}, \nonumber \\ 
&& \frac{\hbar }{{\varepsilon _0 }}\left[ {{\bf{\hat F}}_\omega  \left( {\bf{r}} \right),{\bf{\hat F}}_{\omega '} \left( {{\bf{r}}'} \right)} \right] = \left[ {\frac{{\varepsilon _\omega ^* \left( {\bf{r}} \right)}}{{\omega '}} - \frac{{\varepsilon _{\omega '}^* \left( {{\bf{r}}'} \right)}}{\omega }} \right]{\cal D}^ \bot  \left( {{\bf{r}} - {\bf{r}}'} \right) - \left\{ {\frac{2}{\pi }\int\limits_0^\infty  {d\Omega } \frac{{\left( {\omega  - \omega '} \right)\Omega {\mathop{\rm Im}\nolimits} \left[ {\varepsilon _\Omega  \left( {\bf{r}} \right)} \right]}}{{\left[ {\Omega ^2  - \left( {\omega  - i\eta } \right)^2 } \right]\left[ {\Omega ^2  - \left( {\omega ' - i\eta } \right)^2 } \right]}}} \right\}\delta \left( {{\bf{r}} - {\bf{r}}'} \right){\cal I} \nonumber \\ 
&& \quad \quad   + \frac{{c^2 }}{{\omega \omega '}}\nabla _{\bf{r}}  \times \left\{ {\left\{ {\frac{2}{\pi }\int\limits_0^\infty  {d\Omega } \frac{{\left( {\omega  - \omega '} \right)\Omega {\mathop{\rm Im}\nolimits} \left[ {\mu _\Omega ^{ - 1} \left( {\bf{r}} \right)} \right]}}{{\left[ {\Omega ^2  - \left( {\omega  - i\eta } \right)^2 } \right]\left[ {\Omega ^2  - \left( {\omega ' - i\eta } \right)^2 } \right]}}} \right\}\delta \left( {{\bf{r}} - {\bf{r}}'} \right){\cal I}} \right\} \times \mathord{\buildrel{\lower3pt\hbox{$\scriptscriptstyle\leftarrow$}} 
\over \nabla } _{{\bf{r}}'}  
\end{eqnarray}
which, taking advantage of Eqs.(\ref{Int_2}) to perform the integrations over $\Omega$, becomes
\begin{eqnarray} \label{[FF]_[FFc]}
&& \frac{\hbar }{{\varepsilon _0 }}\left[ {{\bf{\hat F}}_\omega  \left( {\bf{r}} \right),{\bf{\hat F}}_{\omega '}^\dag  \left( {{\bf{r}}'} \right)} \right] = \left[ {\frac{{\varepsilon _\omega ^* \left( {\bf{r}} \right)}}{{\omega '}} + \frac{{\varepsilon _{\omega '} \left( {{\bf{r}}'} \right)}}{\omega }} \right]{\cal D}^ \bot  \left( {{\bf{r}} - {\bf{r}}'} \right) + \frac{{\left[ {\varepsilon _\omega ^* \left( {\bf{r}} \right) - \varepsilon _{\omega '} \left( {\bf{r}} \right)} \right]\delta \left( {{\bf{r}} - {\bf{r}}'} \right){\cal I}}}{{\omega  - \omega ' - i\eta }} \nonumber\\ 
&& \quad \quad   + \frac{{c^2 }}{{\omega \omega '}}\frac{{\nabla _{\bf{r}}  \times \left\{ {\left[ {\mu _\omega ^{ - 1*} \left( {\bf{r}} \right) - \mu _{\omega '}^{ - 1} \left( {\bf{r}} \right)} \right]\delta \left( {{\bf{r}} - {\bf{r}}'} \right){\cal I}} \right\} \times \mathord{\buildrel{\lower3pt\hbox{$\scriptscriptstyle\leftarrow$}} 
\over \nabla } _{{\bf{r}}'} }}{{\omega  - \omega ' - i\eta }}, \nonumber \\ 
&& \frac{\hbar }{{\varepsilon _0 }}\left[ {{\bf{\hat F}}_\omega  \left( {\bf{r}} \right),{\bf{\hat F}}_{\omega '} \left( {{\bf{r}}'} \right)} \right] = \left[ {\frac{{\varepsilon _\omega ^* \left( {\bf{r}} \right)}}{{\omega '}} - \frac{{\varepsilon _{\omega '}^* \left( {{\bf{r}}'} \right)}}{\omega }} \right]{\cal D}^ \bot  \left( {{\bf{r}} - {\bf{r}}'} \right) + \frac{{\left[ { - \varepsilon _\omega ^* \left( {\bf{r}} \right) + \varepsilon _{\omega '}^* \left( {\bf{r}} \right)} \right]\delta \left( {{\bf{r}} - {\bf{r}}'} \right){\cal I}}}{{\omega  + \omega '}} \nonumber \\ 
&& \quad \quad   + \frac{{c^2 }}{{\omega \omega '}}\frac{{\nabla _{\bf{r}}  \times \left\{ {\left[ {\mu _\omega ^{ - 1*} \left( {\bf{r}} \right) - \mu _{\omega '}^{ - 1*} \left( {\bf{r}} \right)} \right]\delta \left( {{\bf{r}} - {\bf{r}}'} \right){\cal I}} \right\} \times \mathord{\buildrel{\lower3pt\hbox{$\scriptscriptstyle\leftarrow$}} 
\over \nabla } _{{\bf{r}}'} }}{{\omega  + \omega '}} .
\end{eqnarray}
Now using such commutation relations combined with the third of Eqs.(\ref{IntPart}) to perform suitable integration by parts, it is tedious but straightforward to prove that, for any two dyadics ${\cal Q}_\omega  \left( {\bf{r}} \right)$ and ${\cal P}_\omega  \left( {\bf{r}} \right)$, the relations 
\begin{eqnarray} \label{F_Fc}
&& \left( {\frac{{\hbar c}}{{\varepsilon _0 }}k_\omega  k_{\omega '} } \right)\int {d^3 {\bf{r}}} \int {d^3 {\bf{r}}'\;} {\cal Q}_\omega ^{T*} \left( {\bf{r}} \right) \cdot \left[ {{\bf{\hat F}}_\omega  \left( {\bf{r}} \right),{\bf{\hat F}}_{\omega '}^\dag  \left( {{\bf{r}}'} \right)} \right] \cdot {\cal P}_{\omega '} \left( {{\bf{r}}'} \right) \nonumber \\ 
&& \quad \quad   = k_\omega  \int {d^3 {\bf{r}}} \left\{ {\int {d^3 {\bf{r}}'\;{\cal D}^ \bot  \left( {{\bf{r}} - {\bf{r}}'} \right) \cdot } \left[ {\varepsilon _\omega  \left( {{\bf{r}}'} \right){\cal Q}_\omega  \left( {{\bf{r}}'} \right)} \right]} \right\}^{T*}  \cdot {\cal P}_{\omega '} \left( {\bf{r}} \right) \nonumber \\ 
&& \quad \quad   + k_{\omega '} \int {d^3 {\bf{r}}\;} {\cal Q}_\omega ^{T*} \left( {\bf{r}} \right) \cdot \left\{ {\int {d^3 {\bf{r}}'} \;{\cal D}^ \bot  \left( {{\bf{r}} - {\bf{r}}'} \right) \cdot \left[ {\varepsilon _{\omega '} \left( {{\bf{r}}'} \right){\cal P}_{\omega '} \left( {{\bf{r}}'} \right)} \right]} \right\} \nonumber \\ 
&& \quad \quad   - \frac{1}{{k_\omega   - k_{\omega '}  - i\eta }}\left\{ {\left( {k_\omega   - k_{\omega '} } \right)\int {d^3 {\bf{r}}} \left[ {k_\omega  \varepsilon _\omega ^* \left( {\bf{r}} \right) + k_{\omega '} \varepsilon _{\omega '} \left( {\bf{r}} \right)} \right]{\cal Q}_\omega ^{T*} \left( {\bf{r}} \right) \cdot {\cal P}_{\omega '} \left( {\bf{r}} \right)} \right\} \nonumber \\ 
&& \quad \quad   - \frac{1}{{k_\omega   - k_{\omega '}  - i\eta }}\int {d^3 {\bf{r}}} \;\left\{ {\left[ {\left( {\nabla _{\bf{r}}  \times \frac{1}{{\mu _\omega  \left( {\bf{r}} \right)}}\nabla _{\bf{r}}  \times } \right) - k_\omega ^2 \varepsilon _\omega  \left( {\bf{r}} \right)} \right]{\cal Q}_\omega  \left( {\bf{r}} \right)} \right\}^{T*}  \cdot {\cal P}_{\omega '} \left( {\bf{r}} \right) \nonumber \\ 
&& \quad \quad   + \frac{1}{{k_\omega   - k_{\omega '}  - i\eta }}\int {d^3 {\bf{r}}} \;{\cal Q}_\omega ^{T*} \left( {\bf{r}} \right) \cdot \left\{ {\left[ {\left( {\nabla _{\bf{r}}  \times \frac{1}{{\mu _{\omega '} \left( {\bf{r}} \right)}}\nabla _{\bf{r}}  \times } \right) - k_{\omega '}^2 \varepsilon _{\omega '} \left( {\bf{r}} \right)} \right]{\cal P}_{\omega '} \left( {\bf{r}} \right)} \right\} \nonumber \\ 
&& \quad \quad   + \int\limits_{S_\infty  } {dS}_r \;\frac{{\left[ {{\bf{u}}_r  \times {\cal Q}_\omega  \left( {\bf{r}} \right)} \right]^{T*}  \cdot \left[ {\nabla _{\bf{r}}  \times {\cal P}_{\omega '} \left( {\bf{r}} \right)} \right] - \left[ {\nabla _{\bf{r}}  \times {\cal Q}_\omega  \left( {\bf{r}} \right)} \right]^{T*}  \cdot \left[ {{\bf{u}}_r  \times {\cal P}_{\omega '} \left( {\bf{r}} \right)} \right]}}{{k_\omega   - k_{\omega '}  - i\eta }}
\end{eqnarray}
and
\begin{eqnarray} \label{F_F}
&& \left( {\frac{{\hbar c}}{{\varepsilon _0 }}k_\omega  k_{\omega '} } \right)\int {d^3 {\bf{r}}} \int {d^3 {\bf{r}}'} {\cal Q}_\omega ^{T*} \left( {\bf{r}} \right) \cdot \left[ {{\bf{\hat F}}_\omega  \left( {\bf{r}} \right),{\bf{\hat F}}_{\omega '} \left( {{\bf{r}}'} \right)} \right] \cdot {\cal P}_{\omega '}^* \left( {{\bf{r}}'} \right) \nonumber \\ 
&& \quad \quad   =  - k_\omega  \int {d^3 {\bf{r}}} \left\{ {\int {d^3 {\bf{r}}'} {\cal D}^\parallel  \left( {{\bf{r}} - {\bf{r}}'} \right) \cdot \left[ {\varepsilon _\omega  \left( {{\bf{r}}'} \right){\cal Q}_\omega  \left( {{\bf{r}}'} \right)} \right]} \right\}^{T*}  \cdot {\cal P}_{\omega '}^* \left( {\bf{r}} \right) \nonumber \\ 
&& \quad \quad   + k_{\omega '} \int {d^3 {\bf{r}}} {\cal Q}_\omega ^{T*} \left( {\bf{r}} \right) \cdot \left\{ {\int {d^3 {\bf{r}}'} {\cal D}^\parallel  \left( {{\bf{r}} - {\bf{r}}'} \right) \cdot \left[ {\varepsilon _{\omega '} \left( {{\bf{r}}'} \right){\cal P}_{\omega '} \left( {{\bf{r}}'} \right)} \right]} \right\}^* \nonumber  \\ 
&& \quad \quad   - \frac{1}{{\left( {k_\omega   + k_{\omega '} } \right)}}\int {d^3 {\bf{r}}} \left\{ {\left[ {\left( {\nabla  \times \frac{1}{{\mu _\omega  \left( {\bf{r}} \right)}}\nabla _{\bf{r}}  \times } \right) - k_\omega ^2 \varepsilon _\omega  \left( {\bf{r}} \right)} \right]{\cal Q}\left( {\bf{r}} \right)} \right\}^{T*}  \cdot {\cal P}_{\omega '}^* \left( {\bf{r}} \right) \nonumber \\ 
&& \quad \quad   + \frac{1}{{\left( {k_\omega   + k_{\omega '} } \right)}}\int {d^3 {\bf{r}}} \;{\cal Q}_\omega ^{T*} \left( {\bf{r}} \right) \cdot \left\{ {\left[ {\left( {\nabla _{\bf{r}}  \times \frac{1}{{\mu _{\omega '} \left( {\bf{r}} \right)}}\nabla _{\bf{r}}  \times } \right) - k_{\omega '}^2 \varepsilon _{\omega '} \left( {\bf{r}} \right)} \right]{\cal P}_{\omega '} \left( {\bf{r}} \right)} \right\}^* \nonumber  \\ 
&& \quad \quad   + \frac{1}{{\left( {k_\omega   + k_{\omega '} } \right)}}\int\limits_{S_\infty  } {dS}_r \;\left\{ {\left[ {{\bf{u}}_r  \times {\cal Q}_\omega  \left( {\bf{r}} \right)} \right]^T  \cdot \left[ {\nabla _{\bf{r}}  \times {\cal P}_{\omega '} \left( {\bf{r}} \right)} \right] - \;\left[ {\nabla _{\bf{r}}  \times {\cal Q}\left( {\bf{r}} \right)} \right]^T  \cdot \left[ {{\bf{u}}_{r}  \times {\cal P}_{\omega '} \left( {\bf{r}} \right)} \right]} \right\}^*
\end{eqnarray}
hold. It is worth noting that, in the asymptotic surface integral contribution (last term) of Eq.(\ref{F_Fc}), the factor $\left( {k_\omega   - k_{\omega '}  - i\eta } \right)^{ - 1}$  was not moved outside the integral since the order of the two limiting procedures $r \rightarrow +\infty$ (see  
Eq.(\ref{asym_sur_int})) and $\eta \rightarrow 0^+$ can not be generally interchanged. A sufficient condition permitting this interchange is the convergence of the asymptotic surface integral in the absence of the factor $\left( {k_\omega   - k_{\omega '}  - i\eta } \right)^{ - 1}$ (see Appendix E).

The commutators involving both the operators ${\bf{\hat F}}_\omega$ and ${\bf{\hat h}}_{\omega \nu }$ are straightforwardly seen to be
\begin{eqnarray}
&& \left[ {{\bf{\hat F}}_\omega  \left( {\bf{r}} \right),{\bf{\hat h}}_{\omega '\nu '}^\dag  \left( {{\bf{r}}'} \right)} \right] = \frac{i}{{\sqrt {2\hbar \omega '} }}\left\{ {\delta _{\nu 'e} \left[ {\frac{1}{\omega }{\cal D}^ \bot  \left( {{\bf{r}} - {\bf{r}}'} \right) + \frac{{\omega ' + \omega }}{{\omega '^2  - \left( {\omega  - i\eta } \right)^2 }}\delta \left( {{\bf{r}} - {\bf{r}}'} \right){\cal I}} \right]\alpha ^{\omega '} \left( {{\bf{r}}'} \right)} \right. \nonumber \\ 
&& \quad \quad  \left. { + \delta _{\nu 'm} \nabla _{\bf{r}}  \times \left[ {\frac{{\omega ' + \omega }}{{\omega '^2  - \left( {\omega  - i\eta } \right)^2 }}\delta \left( {{\bf{r}} - {\bf{r}}'} \right){\cal I}} \right]\frac{{\beta ^{\omega '} \left( {{\bf{r}}'} \right)}}{\omega }} \right\}, \nonumber \\ 
&& \left[ {{\bf{\hat F}}_\omega  \left( {\bf{r}} \right),{\bf{\hat h}}_{\omega '\nu '} \left( {{\bf{r}}'} \right)} \right] = \frac{i}{{\sqrt {2\hbar \omega '} }}\left\{ {\delta _{\nu 'e} \left[ {\frac{1}{\omega }{\cal D}^ \bot  \left( {{\bf{r}} - {\bf{r}}'} \right) - \frac{{\omega ' - \omega }}{{\omega '^2  - \left( {\omega  - i\eta } \right)^2 }}\delta \left( {{\bf{r}} - {\bf{r}}'} \right){\cal I}} \right]\alpha ^{\omega '} \left( {{\bf{r}}'} \right)} \right. \nonumber \\ 
&&  + \left. {\delta _{\nu 'm} \nabla _{\bf{r}}  \times \left[ {\frac{{\omega ' - \omega }}{{\omega '^2  - \left( {\omega  - i\eta } \right)^2 }}\delta \left( {{\bf{r}} - {\bf{r}}'} \right){\cal I}} \right]\frac{{\beta ^{\omega '} \left( {{\bf{r}}'} \right)}}{\omega }} \right\}, 
\end{eqnarray}
which, by resorting to the properties of the 
 dyadic transverse  and longitudinal  delta functions of Eqs.(\ref{trs-long-delt}) and using the expressions of the coefficients in Eqs.(\ref{alf_bet}), after some algebra become
\begin{eqnarray} 
 \left[ {{\bf{\hat F}}_\omega  \left( {\bf{r}} \right),{\bf{\hat h}}_{\omega '\nu '}^\dag  \left( {{\bf{r}}'} \right)} \right] &=& \sqrt {\frac{{\varepsilon _0 }}{{\pi \hbar }}} \frac{i}{{\omega  - \omega ' - i\eta }}\left[ { - \frac{{\omega '}}{\omega }{\cal D}^ \bot  \left( {{\bf{r}} - {\bf{r}}'} \right) - {\cal D}^\parallel  \left( {{\bf{r}} - {\bf{r}}'} \right)} \right]\left\{ {\sqrt {{\mathop{\rm Im}\nolimits} \left[ {\varepsilon _{\omega '} \left( {{\bf{r}}'} \right)} \right]} \delta _{\nu 'e}  - \frac{{ \times \mathord{\buildrel{\lower3pt\hbox{$\scriptscriptstyle\leftarrow$}} 
\over \nabla } _{{\bf{r}}'} }}{{k_{\omega '} }}\sqrt {{\mathop{\rm Im}\nolimits} \left[ {\frac{{ - 1}}{{\mu _{\omega '} \left( {{\bf{r}}'} \right)}}} \right]} \delta _{\nu 'm} } \right\}, \nonumber \\ 
 \left[ {{\bf{\hat F}}_\omega  \left( {\bf{r}} \right),{\bf{\hat h}}_{\omega '\nu '} \left( {{\bf{r}}'} \right)} \right] &=& \sqrt {\frac{{\varepsilon _0 }}{{\pi \hbar }}} \frac{i}{{\omega  + \omega '}}\left[ {\frac{{\omega '}}{\omega }{\cal D}^ \bot  \left( {{\bf{r}} - {\bf{r}}'} \right) - {\cal D}^\parallel  \left( {{\bf{r}} - {\bf{r}}'} \right)} \right]\left\{ {\sqrt {{\mathop{\rm Im}\nolimits} \left[ {\varepsilon _{\omega '} \left( {{\bf{r}}'} \right)} \right]} \delta _{\nu 'e}  - \frac{{ \times \mathord{\buildrel{\lower3pt\hbox{$\scriptscriptstyle\leftarrow$}} 
\over \nabla } _{{\bf{r}}'} }}{{k_{\omega '} }}\sqrt {{\mathop{\rm Im}\nolimits} \left[ {\frac{{ - 1}}{{\mu _{\omega '} \left( {{\bf{r}}'} \right)}}} \right]} \delta _{\nu 'm} } \right\}, 
\end{eqnarray}
or, using the definition of the operator $\mathord{\buildrel{\lower3pt\hbox{$\scriptscriptstyle\leftarrow$}} 
\over T} _{\omega \nu } \left( {\bf{r}} \right)$ in Eqs.(\ref{G-T}),
\begin{eqnarray} \label{[Fh]_[Fhc]}
 \left[ {{\bf{\hat F}}_\omega  \left( {\bf{r}} \right),{\bf{\hat h}}_{\omega '\nu '}^\dag  \left( {{\bf{r}}'} \right)} \right] &=& \frac{{\varepsilon _0 }}{\hbar c}\frac{1}{{k_\omega  k_{\omega '} \left( {k_\omega   - k_{\omega '}  - i\eta } \right)}}\left[ { - {\cal D}^ \bot  \left( {{\bf{r}} - {\bf{r}}'} \right) - \frac{{k_\omega  }}{{k_{\omega '} }}{\cal D}^\parallel  \left( {{\bf{r}} - {\bf{r}}'} \right)} \right]\mathord{\buildrel{\lower3pt\hbox{$\scriptscriptstyle\leftarrow$}} 
\over T} _{\omega '\nu '} \left( {{\bf{r}}'} \right), \nonumber \\ 
 \left[ {{\bf{\hat F}}_\omega  \left( {\bf{r}} \right),{\bf{\hat h}}_{\omega '\nu '} \left( {{\bf{r}}'} \right)} \right] &=& \frac{{\varepsilon _0 }}{\hbar c}\frac{1}{{k_\omega  k_{\omega '} \left( {k_\omega   + k_{\omega '} } \right)}}\left[ {{\cal D}^ \bot  \left( {{\bf{r}} - {\bf{r}}'} \right) - \frac{{k_\omega  }}{{k_{\omega '} }}{\cal D}^\parallel  \left( {{\bf{r}} - {\bf{r}}'} \right)} \right]\mathord{\buildrel{\lower3pt\hbox{$\scriptscriptstyle\leftarrow$}} 
\over T} _{\omega '\nu '} \left( {{\bf{r}}'} \right).
\end{eqnarray}
Eventually, the commutators involving only the operators ${\bf{\hat h}}_{\omega \nu }$ are
\begin{eqnarray} \label{[hh]_[hhc]}
 \left[ {{\bf{\hat h}}_{\omega \nu } \left( {\bf{r}} \right),{\bf{\hat h}}_{\omega '\nu '}^\dag  \left( {{\bf{r}}'} \right)} \right] &=& \delta _{\nu \nu '} \delta \left( {\omega  - \omega '} \right)\delta \left( {{\bf{r}} - {\bf{r}}'} \right){\cal I}, \nonumber \\
\left[ {{\bf{\hat h}}_{\omega \nu } \left( {\bf{r}} \right),{\bf{\hat h}}_{\omega '\nu '} \left( {{\bf{r}}'} \right)} \right] &=& 0.
\end{eqnarray}

\section{Vanishing of the asymptotic surface integrals}
The six vanishing asymptotic surface integrals required for the derivation of the bosonic commutation relations in Sec.IV are
\begin{eqnarray} \label{Asy_Int}
 \int\limits_{S_\infty  } {dS_r } \;\frac{{\left[ {{\bf{u}}_r  \times {\cal F}_\omega  \left( {\left. {\bf{r}} \right|{\bf{n}}} \right)} \right]^{T*}  \cdot \left[ {\nabla _{\bf{r}}  \times {\cal F}_{\omega '} \left( {\left. {\bf{r}} \right|{\bf{n}}'} \right)} \right] - \left[ {\nabla _{\bf{r}}  \times {\cal F}_\omega  \left( {\left. {\bf{r}} \right|{\bf{n}}} \right)} \right]^{T*}  \cdot \left[ {{\bf{u}}_r  \times {\cal F}_{\omega '} \left( {\left. {\bf{r}} \right|{\bf{n}}'} \right)} \right]}}{{k_\omega   - k_{\omega '}  - i\eta }} &=& 0 ,\nonumber \\ 
 \int\limits_{S_\infty  } {dS_r } \left\{ {\left[ {{\bf{u}}_r  \times {\cal F}_\omega  \left( {\left. {\bf{r}} \right|{\bf{n}}} \right)} \right]^T  \cdot \left[ {\nabla _{\bf{r}}  \times {\cal F}_{\omega '} \left( {\left. {\bf{r}} \right|{\bf{n}}'} \right)} \right] - \left[ {\nabla _{\bf{r}}  \times {\cal F}_\omega  \left( {\left. {\bf{r}} \right|{\bf{n}}} \right)} \right]^T  \cdot \left[ {{\bf{u}}_r  \times {\cal F}_{\omega '} \left( {\left. {\bf{r}} \right|{\bf{n}}'} \right)} \right]} \right\} &=& 0 ,\nonumber \\ 
 \int\limits_{S_\infty  } {dS_s } \;\frac{{\left[ {{\bf{u}}_s  \times {\cal G}_\omega  \left( {\left. {\bf{s}} \right|{\bf{r}}} \right)} \right]^{T*}  \cdot \left[ {\nabla _{\bf{s}}  \times {\cal G}_{\omega '} \left( {\left. {\bf{s}} \right|{\bf{r}}'} \right)} \right] - \left[ {\nabla _{\bf{s}}  \times {\cal G}_\omega  \left( {\left. {\bf{s}} \right|{\bf{r}}} \right)} \right]^{T*}  \cdot \left[ {{\bf{u}}_s  \times {\cal G}_{\omega '} \left( {\left. {\bf{s}} \right|{\bf{r}}'} \right)} \right]}}{{k_\omega   - k_{\omega '}  - i\eta }} &=& 0 ,\nonumber \\ 
 \int\limits_{S_\infty  } {dS_s } \left\{ {\left[ {{\bf{u}}_s  \times {\cal G}_\omega  \left( {\left. {\bf{s}} \right|{\bf{r}}} \right)} \right]^T  \cdot \left[ {\nabla _{\bf{s}}  \times {\cal G}_{\omega '} \left( {\left. {\bf{s}} \right|{\bf{r}}'} \right)} \right] - \left[ {\nabla _{\bf{s}}  \times {\cal G}_\omega  \left( {\left. {\bf{s}} \right|{\bf{r}}} \right)} \right]^T  \cdot \left[ {{\bf{u}}_s  \times {\cal G}_{\omega '} \left( {\left. {\bf{s}} \right|{\bf{r}}'} \right)} \right]} \right\} &=& 0 ,\nonumber \\ 
 \int\limits_{S_\infty  } {dS_s } \;\frac{{\left[ {{\bf{u}}_s  \times {\cal F}_\omega  \left( {\left. {\bf{s}} \right|{\bf{n}}} \right)} \right]^{T*}  \cdot \left[ {\nabla _{\bf{s}}  \times {\cal G}_{\omega '} \left( {\left. {\bf{s}} \right|{\bf{r}}} \right)} \right] - \left[ {\nabla _{\bf{s}}  \times {\cal F}_\omega  \left( {\left. {\bf{s}} \right|{\bf{n}}} \right)} \right]^{T*}  \cdot \left[ {{\bf{u}}_s  \times {\cal G}_{\omega '} \left( {\left. {\bf{s}} \right|{\bf{r}}} \right)} \right]}}{{k_\omega   - k_{\omega '}  - i\eta }} &=& 0 ,\nonumber \\ 
 \int\limits_{S_\infty  } {dS_s } \left\{ {\left[ {{\bf{u}}_s  \times {\cal F}_\omega  \left( {\left. {\bf{s}} \right|{\bf{n}}} \right)} \right]^T  \cdot \left[ {\nabla _{\bf{s}}  \times {\cal G}_{\omega '} \left( {\left. {\bf{s}} \right|{\bf{r}}} \right)} \right] - \left[ {\nabla _{\bf{s}}  \times {\cal F}_\omega  \left( {\left. {\bf{s}} \right|{\bf{n}}} \right)} \right]^T  \cdot \left[ {{\bf{u}}_s  \times {\cal G}_{\omega '} \left( {\left. {\bf{s}} \right|{\bf{r}}} \right)} \right]} \right\} &=& 0.
\end{eqnarray}
Given their similar mathematical structure, these integrals can be evaluated using a consistent methodology. Accordingly, we provide a detailed proof for the first of Eqs.(\ref{Asy_Int}) and briefly outline the derivations for the remaining integrals. Using the prescription of Eq.(\ref{asym_sur_int}) together with the asymptotic behaviors of the modal dyadic (in the second of Eqs.(\ref{GS})) and of its curl, i.e.
\begin{eqnarray}
 \left[ {{\cal F}_\omega  \left( {\left. {\bf{r}} \right|{\bf{n}}} \right)} \right]_{{\bf{r}} = r{\bf{m}}} & \mathop  \approx \limits_{r \to  + \infty } & e^{i\left( {k_\omega  {\bf{n}}} \right) \cdot \left( {r{\bf{m}}} \right)} {\cal I}_{\bf{n}}  + \frac{{e^{ik_\omega  r} }}{r}{\cal S}_\omega  \left( {{\bf{m}}\left| {\bf{n}} \right.} \right), \nonumber  \\ 
 \left[ {\nabla _{\bf{r}}  \times {\cal F}_\omega  \left( {\left. {\bf{r}} \right|{\bf{n}}} \right)} \right]_{{\bf{r}} = r{\bf{m}}} & \mathop  \approx \limits_{r \to  + \infty } & ik_\omega  \left[ {e^{i\left( {k_\omega  {\bf{n}}} \right) \cdot \left( {r{\bf{m}}} \right)} {\bf{n}} \times {\cal I}_{\bf{n}}  + \frac{{e^{ik_\omega  r} }}{r}{\bf{m}} \times {\cal S}_\omega  \left( {{\bf{m}}\left| {\bf{n}} \right.} \right)} \right],
\end{eqnarray}
the first contribution on the LHS of the first of Eqs.(\ref{Asy_Int}) becomes
\begin{eqnarray} \label{I1_01}
&& \frac{1}{{ik_{\omega '} }} \int\limits_{S_\infty  } {dS_r } \frac{{\left[ {{\bf{u}}_r  \times {\cal F}_\omega  \left( {\left. {\bf{r}} \right|{\bf{n}}} \right)} \right]^{T*}  \cdot \left[ {\nabla _{\bf{r}}  \times {\cal F}_{\omega '} \left( {\left. {\bf{r}} \right|{\bf{n}}'} \right)} \right]}}{{k_\omega   - k_{\omega '}  - i\eta }} =  \nonumber \\ 
&& \quad \quad   = \left\{ {\left\{ {\mathop {\lim }\limits_{r \to \infty } \left[ {\frac{1}{{k_\omega   - k_{\omega '}  - i\eta }}r^2 \int {do_{\bf{m}} } e^{i{\bf{K}} \cdot \left( {r{\bf{m}}} \right)} {\bf{m}}} \right]} \right\}_{{\bf{K}} = k_\omega  {\bf{n}} - k_{\omega '} {\bf{n}}'}  \times {\cal I}_{\bf{n}} } \right\}^{T*}  \cdot \left( {{\bf{n}}' \times {\cal I}_{{\bf{n}}'} } \right) \nonumber \\ 
&& \quad \quad  + \mathop {\lim }\limits_{r \to \infty } \left\{ {\frac{{e^{ik_{\omega '} r} }}{{k_\omega   - k_{\omega '}  - i\eta }}r\int {do_{\bf{m}} } e^{ - i\left( {k_\omega  {\bf{n}}} \right) \cdot \left( {r{\bf{m}}} \right)} \left( {{\bf{m}} \times {\cal I}_{\bf{n}} } \right)^{T*}  \cdot \left[ {{\bf{m}} \times {\cal S}_{\omega '} \left( {{\bf{m}}\left| {{\bf{n}}'} \right.} \right)} \right]} \right\} \nonumber \\ 
&& \quad \quad  + \mathop {\lim }\limits_{r \to \infty } \left\{ {\frac{{e^{ - ik_\omega  r} }}{{k_\omega   - k_{\omega '}  - i\eta }}r\int {do_{\bf{m}} } e^{i\left( {k_{\omega '} {\bf{n}}'} \right) \cdot \left( {r{\bf{m}}} \right)} \left[ {{\bf{m}} \times {\cal S}_\omega  \left( {{\bf{m}}\left| {\bf{n}} \right.} \right)} \right]^{T*}  \cdot \left( {{\bf{n}}' \times {\cal I}_{{\bf{n}}'} } \right)} \right\} \nonumber \\ 
&& \quad \quad  + \mathop {\lim }\limits_{r \to \infty } \left\{ {\frac{{e^{i\left( {k_{\omega '}  - k_\omega  } \right)r} }}{{k_\omega   - k_{\omega '}  - i\eta }}\int {do_{\bf{m}} } \left[ {{\bf{m}} \times {\cal S}_\omega  \left( {{\bf{m}}\left| {\bf{n}} \right.} \right)} \right]^{T*}  \cdot \left[ {{\bf{m}} \times {\cal S}_{\omega '} \left( {{\bf{m}}\left| {{\bf{n}}'} \right.} \right)} \right]} \right\}.
\end{eqnarray}
By applying the divergence theorem, it is straightforward to demonstrate that
\begin{equation} \label{diverg}
\mathop {\lim }\limits_{r \to  + \infty } \left[ {r^2 \int {do_{\bf{m}} e^{i{\bf{K}} \cdot \left( {r{\bf{m}}} \right)} } {\bf{m}}} \right] = i\left( {2\pi } \right)^3 {\bf{K}}\delta \left( {\bf{K}} \right) = 0,
\end{equation}
consequently, the first term on the RHS of Eq.(\ref{I1_01}) also vanishes, as the distribution $(k_\omega - k_{\omega'} - i\eta)^{-1}$ can be factored out of the limit (refer to the discussion following Eq.(\ref{F_F})). To evaluate the second and third contributions on the RHS of Eq.(\ref{I1_01}), we apply Jones' lemma as expressed in Eq.(\ref{JonLem}), which transforms Eq.(\ref{I1_01}) into
\begin{eqnarray} \label{I1_02}
&& \frac{1}{{ik_{\omega '} }}\int\limits_{S_\infty  } {dS_r } \frac{{\left[ {{\bf{u}}_r  \times {\cal F}_\omega  \left( {\left. {\bf{r}} \right|{\bf{n}}} \right)} \right]^{T*}  \cdot \left[ {\nabla _{\bf{r}}  \times {\cal F}_{\omega '} \left( {\left. {\bf{r}} \right|{\bf{n}}'} \right)} \right]}}{{k_\omega   - k_{\omega '}  - i\eta }}  \nonumber \\ 
&& \quad \quad  = \mathop {\lim }\limits_{r \to  + \infty } \left\{ {\frac{{e^{ - i\left( {k_\omega   - k_{\omega '} } \right)r} }}{{\left( {k_\omega   - k_{\omega '} } \right) - i\eta }}\left[ {e^{i2k_\omega  r} {\cal B}_{\omega ,\omega '}^{\left( 1 \right)} \left( {{\bf{n}}\left| {{\bf{n}}'} \right.} \right) + e^{ - i2k_{\omega '} r} {\cal B}_{\omega ,\omega '}^{\left( 2 \right)} \left( {{\bf{n}}\left| {{\bf{n}}'} \right.} \right) + {\cal B}_{\omega ,\omega '}^{\left( 3 \right)} \left( {{\bf{n}}\left| {{\bf{n}}'} \right.} \right)} \right]} \right\}  
\end{eqnarray}
where the finite dyadics $\mathcal{B}_{\omega,\omega'}^{(j)}$ contain the angular integrals, and their specific algebraic form is not required for the present discussion. Indeed, as a consequence of the distribution relations in Eqs.(\ref{Dist_FUND}), the following limits hold independently
\begin{align}
 \mathop {\lim }\limits_{r \to  + \infty } e^{ \pm i2k_\omega  r}  =& 0,  &
 \mathop {\lim }\limits_{r \to  + \infty } \frac{{e^{ - i\left( {k_\omega   - k_{\omega '} } \right)r} }}{{\left( {k_\omega   - k_{\omega '} } \right) - i\eta }} =& 0
\end{align}
so that the entire RHS of Eq.(\ref{I1_02}) vanishes, along with the first contribution in the first of Eqs.(\ref{Asy_Int}). A parallel analysis demonstrates that the second contribution vanishes as well, thereby completing the proof of the first of Eqs.(\ref{Asy_Int}). By generalizing the previous methodology, the third and fifth integrals in Eqs.(\ref{Asy_Int}) are shown to vanish. These terms share the same frequency configuration and Plemelj factor, $(k_\omega - k_{\omega'} - i\eta)^{-1}$, with the first integral; thus, their vanishing is a direct consequence of the second relation in Eqs.(\ref{Dist_FUND}). Similarly, the second, fourth, and sixth integrals in Eqs.(\ref{Asy_Int}) are easily seen to vanish identically due to the first relation in Eqs.(\ref{Dist_FUND}).

\section{Evaluation of the Hamiltonian density contributions} 
In this Appendix, we provide the explicit evaluation of the terms composing the density of the Hamiltonian in Eq.(\ref{CQME_H}) upon substitution of the canonical field operators of Eqs.(\ref{Direct_Can_Fie_Ope}). The total electric Hamiltonian density, using the first of Eqs.(\ref{CQME_EB}), directly recovers the standard macroscopic form
\begin{equation}
\frac{1}{{2\varepsilon _0 }}\left( {{\bf{\hat \Pi }}_A  + \int {d\Omega } \;\alpha ^\Omega  {\bf{\hat X}}^\Omega  } \right)^2  = \frac{{\varepsilon _0 }}{2}{\bf{\hat E}}^2 .
\end{equation}
For the magnetic Hamiltonian density, we first evaluate the integral $\int {d\Omega } \;\beta ^\Omega  {\bf{\hat Y}}^\Omega$ which, expanding the definition of ${\bf{\hat Y}}^\Omega$ in Eqs.(\ref{Direct_Can_Fie_Ope}), becomes
\begin{equation}
\int {d\Omega } \;\beta ^\Omega  {\bf{\hat Y}}^\Omega   = \int {d\omega } \int{d\Omega } \frac{{\left( {\beta ^\Omega  } \right)^2 }}{{\Omega ^2  - \left( {\omega  + i\eta } \right)^2 }}\frac{{\nabla  \times {\bf{\hat E}}_\omega  }}{{i\omega }} + \int {d\Omega } \;\beta ^\Omega  \sqrt {\frac{\hbar }{{2\Omega }}} {\bf{\hat f}}_{\Omega m}  + {\rm H.c.}
\end{equation}
in turn, by using the dispersion relation in the second of Eqs.(\ref{Int_1}) and the spectral identity $\nabla \times {\bf{\hat A}} = \int d\omega (i\omega)^{-1} \nabla \times {\bf{\hat E}}_\omega + {\rm H.c.}$ (see the first of Eqs.(\ref{Direct_Can_Fie_Ope})), yielding
\begin{equation} \label{I_be_Y}
\int {d\Omega } \;\beta ^\Omega  {\bf{\hat Y}}^\Omega  = \frac{1}{{\mu _0 }}\nabla  \times {\bf{\hat A}} + \left( \int {d\omega } i{\bf{\hat Q}}_{\omega m}  + {\rm H.c.} \right)
\end{equation}
where the auxiliary operator ${\bf{\hat Q}}_{\omega m}$ is defined in Eq.(\ref{Qw_def}). Now the suitably symmetrized Hamiltonian density 
\begin{equation}
\left( {\nabla  \times {\bf{\hat A}}} \right) \cdot \left( {\frac{1}{{2\mu _0 }}\nabla  \times {\bf{\hat A}} - \int {d\Omega } \;\beta ^\Omega  {\bf{\hat Y}}^\Omega  } \right) = \frac{1}{{2\mu _0 }}\left( {\nabla  \times {\bf{\hat A}}} \right)^2 - \frac{1}{2}\left[ {\left( {\nabla  \times {\bf{\hat A}}} \right) \cdot \int {d\Omega } \;\beta ^\Omega  {\bf{\hat Y}}^\Omega   + \int {d\Omega } \;\beta ^\Omega  {\bf{\hat Y}}^\Omega   \cdot \left( {\nabla  \times {\bf{\hat A}}} \right)} \right],
\end{equation}
using Eq.(\ref{I_be_Y}), becomes 
\begin{eqnarray}
&& \left( {\nabla  \times {\bf{\hat A}}} \right) \cdot \left( {\frac{1}{{2\mu _0 }}\nabla  \times {\bf{\hat A}} - \int {d\Omega } \;\beta ^\Omega  {\bf{\hat Y}}^\Omega  } \right) =  - \frac{1}{{2\mu _0 }}\left( {\nabla  \times {\bf{\hat A}}} \right)^2 \nonumber  \\ 
&& \quad \quad  - \frac{1}{2}\int {d\omega } \int {d\omega '} \frac{1}{{\omega '}}\left[ {\left( {\nabla  \times {\bf{\hat E}}_{\omega '}  - \nabla  \times {\bf{\hat E}}_{\omega '}^\dag  } \right) \cdot {\bf{\hat Q}}_{\omega m}  + {\bf{\hat Q}}_{\omega m}  \cdot \left( {\nabla  \times {\bf{\hat E}}_{\omega '}  - \nabla  \times {\bf{\hat E}}_{\omega '}^\dag  } \right) + {\rm H.c.}} \right]
\end{eqnarray}
where the vector potential in the second contribution has been expressed in terms of its spectral components. Next, the electric reservoir Hamiltonian density, after some lengthy but direct algebra, yields
\begin{eqnarray}
&& \frac{1}{2}\int {d\Omega } \left( {{\bf{\hat \Pi }}_X^{\Omega 2}  + \Omega ^2 {\bf{\hat X}}^{\Omega 2} } \right) =  \frac{1}{2}\int {d\omega \int {d\omega '} } \left\{ \int {d\Omega } \frac{{\left( {\Omega ^2  - \omega \omega '} \right)\left( {\alpha ^\Omega  } \right)^2 }}{{\left[ {\Omega ^2  - \left( {\omega  + i\eta } \right)^2 } \right]\left[ {\Omega ^2  - \left( {\omega ' + i\eta } \right)^2 } \right]}}{\bf{\hat E}}_\omega   \cdot {\bf{\hat E}}_{\omega '}  + {\rm H.c.} \right\} \nonumber \\
&& \quad \quad  + \frac{1}{2}\int {d\omega \int {d\omega '} } \int {d\Omega } \frac{{\left( {\Omega ^2  + \omega \omega '} \right)\left( {\alpha ^\Omega  } \right)^2 }}{{\left[ {\Omega ^2  - \left( {\omega  + i\eta } \right)^2 } \right]\left[ {\Omega ^2  - \left( {\omega ' - i\eta } \right)^2 } \right]}}\left( {{\bf{\hat E}}_\omega   \cdot {\bf{\hat E}}_{\omega '}^\dag   + {\bf{\hat E}}_{\omega '}^\dag   \cdot {\bf{\hat E}}_\omega  } \right) \nonumber \\
&& \quad \quad + \frac{1}{2}\left[ \int {d\Omega } \;i\alpha ^\Omega  \sqrt {\frac{{\hbar \Omega }}{2}} \int {d\omega } \left( \frac{{{\bf{\hat E}}_\omega   \cdot {\bf{\hat f}}_{\Omega e}  + {\bf{\hat f}}_{\Omega e}  \cdot {\bf{\hat E}}_\omega  }}{{\Omega  + \omega }} + \frac{{{\bf{\hat E}}_\omega ^\dag   \cdot {\bf{\hat f}}_{\Omega e}  + {\bf{\hat f}}_{\Omega e}  \cdot {\bf{\hat E}}_\omega ^\dag  }}{{\Omega  - \omega  + i\eta }} \right) + {\rm H.c.} \right] \nonumber \\
&& \quad \quad + \frac{1}{2}\int {d\Omega } \hbar \Omega \left( {{\bf{\hat f}}_{\Omega e}  \cdot {\bf{\hat f}}_{\Omega e}^\dag   + {\bf{\hat f}}_{\Omega e}^\dag   \cdot {\bf{\hat f}}_{\Omega e} } \right) .
\end{eqnarray}
Performing the integrations over $\Omega$ via the generalized dispersion relations in the first of Eqs.(\ref{Int_3}), the reservoir density correctly reconstructs the free-evolution energy and the macroscopic auxiliary operators,
\begin{eqnarray}
&& \frac{1}{2}\int {d\Omega } \left( {{\bf{\hat \Pi }}_X^{\Omega 2}  + \Omega ^2 {\bf{\hat X}}^{\Omega 2} } \right) = \frac{1}{2}\int {d\omega } \hbar \omega \left( {{\bf{\hat f}}_{\omega e}^\dag   \cdot {\bf{\hat f}}_{\omega e}  + {\bf{\hat f}}_{\omega e}  \cdot {\bf{\hat f}}_{\omega e}^\dag  } \right) - \frac{1}{2}\varepsilon _0 \left( {\int {d\omega } {\bf{\hat E}}_\omega   + H.c.} \right)^2  \nonumber \\
&& \quad \quad + \frac{1}{2}\left[ \int {d\omega } \;\int {d\omega '} \left( \frac{{{\bf{\hat Q}}_{\omega e}  \cdot {\bf{\hat E}}_{\omega '}  + {\bf{\hat E}}_{\omega '}  \cdot {\bf{\hat Q}}_{\omega e} }}{{\omega  + \omega '}} + \frac{{{\bf{\hat Q}}_{\omega e}  \cdot {\bf{\hat E}}_{\omega '}^\dag   + {\bf{\hat E}}_{\omega '}^\dag   \cdot {\bf{\hat Q}}_{\omega e} }}{{\omega  - \omega ' + i\eta }} \right) + {\rm H.c.} \right] .
\end{eqnarray}
where the auxiliary operator ${\bf{\hat Q}}_{\omega e}$ is defined in Eq.(\ref{Qwe_def}). The evaluation of the magnetic reservoir energy density follows symmetrically from the electric counterpart. Applying the duality transformations $\varepsilon_0 \to -1/\mu_0$, $\varepsilon_\omega \to 1/\mu_\omega$, ${\bf{\hat E}}_\omega \to (i\omega)^{-1}\nabla \times {\bf{\hat E}}_\omega$, and ${\bf{\hat f}}_{\Omega e} \to -i{\bf{\hat f}}_{\Omega m}$, the density maps to
\begin{eqnarray}
&& \frac{1}{2}\int {d\Omega } \left( {{\bf{\hat \Pi }}_Y^{\Omega 2}  + \Omega ^2 {\bf{\hat Y}}^{\Omega 2} } \right) = \frac{1}{2}\int {d\omega } \hbar \omega \left( {{\bf{\hat f}}_{\omega m}^\dag   \cdot {\bf{\hat f}}_{\omega m}  + {\bf{\hat f}}_{\omega m}  \cdot {\bf{\hat f}}_{\omega m}^\dag  } \right) + \frac{1}{{2\mu _0 }}\left( {\int {d\omega } \frac{{\nabla  \times {\bf{\hat E}}_\omega  }}{{i\omega }} + H.c.} \right)^2   \nonumber \\
&& \quad \quad + \frac{1}{2}\left[ \int {d\omega } \int {d\omega '} \frac{\omega }{{\omega '}}\left( \frac{{{\bf{\hat Q}}_{\omega m}  \cdot \nabla  \times {\bf{\hat E}}_{\omega '}  + \nabla  \times {\bf{\hat E}}_{\omega '}  \cdot {\bf{\hat Q}}_{\omega m} }}{{\omega  + \omega '}}  - \frac{{{\bf{\hat Q}}_{\omega m}  \cdot \nabla  \times {\bf{\hat E}}_{\omega '}^\dag   + \nabla  \times {\bf{\hat E}}_{\omega '}^\dag   \cdot {\bf{\hat Q}}_{\omega m} }}{{\omega  - \omega ' + i\eta }} \right) + {\rm H.c.} \right] .
\end{eqnarray}
Finally, we evaluate the curl of the magnetic auxiliary operator in Eq.(\ref{Qw_def}), i.e.
\begin{equation} \label{rot_Qm}
\nabla  \times {\bf{\hat Q}}_{\omega m} = \frac{1}{{\omega \mu _0 }}\left( \nabla  \times \frac{1}{{\mu _\omega  }}\nabla  \times    \right) {\bf{\hat E}}_\omega - i\sqrt {\frac{\hbar }{{2\omega }}} \nabla  \times \left( {\beta ^\omega  {\bf{\hat f}}_{\omega m} } \right) .
\end{equation}
After noting that the electric field spectrum ${{\bf{\hat E}}_\omega  }$ in Eq.(\ref{Eom2}) is found to satisfy the inhomogeneous Helmholtz equation in Eq.(\ref{CQME_InhHel}), Eq.(\ref{rot_Qm}) can be cast as
\begin{equation}
\nabla  \times {\bf{\hat Q}}_{\omega m} = \omega \varepsilon _0 \varepsilon _\omega  {\bf{\hat E}}_\omega   + i\sqrt {\frac{{\hbar \omega }}{2}} \alpha ^\omega  {\bf{\hat f}}_{\omega e},
\end{equation}
which, in view of the definition of the electric auxiliary operator in Eq.(\ref{Qwe_def}), directly yields
\begin{equation}
\nabla \times {\bf{\hat Q}}_{\omega m} = {\bf{\hat Q}}_{\omega e}.
\end{equation}


\begin{thebibliography}{10} 
\bibitem{Glaub1} R. J. Glauber and M. Lewenstein,    
                 Quantum optics of dielectric media,
                 Phys. Rev. A {\bf 43}, 467-491 (1991).                
\bibitem{Dalto1} B. J. Dalton, E. S. Guerra and P. L. Knight,
                 Field quantization in dielectric media and the generalized multipolar Hamiltonian,
                 Phys. Rev. A {\bf 54}, 2292-2313 (1996).
\bibitem{Fanooo} U. Fano,
                 Atomic Theory of Electromagnetic Interactions in Dense Materials,
                 Phys. Rev. {\bf 103}, 1202-1218 (1956).
\bibitem{Hopfi1} J. J. Hopfield,
                 Theory of the Contribution of Excitons to the Complex Dielectric Constant of Crystals,
                 Phys. Rev. {\bf 112}, 1555-1567 (1958).
\bibitem{Huttn1} B. Huttner and S. M. Barnett,             
                 Dispersion and Loss in a Hopfield Dielectric,
                 Europhys. Lett. {\bf 18}, 487-492 (1992).                 
\bibitem{Huttn2} B. Huttner and S. M. Barnett,
                 Quantization of the electromagnetic field in dielectrics,
                 Phys. Rev. A {\bf 46}, 4306-4322 (1992).                               
\bibitem{Sutto1} L. G. Suttorp and M. Wubs,
                 Field quantization in inhomogeneous absorptive dielectrics,
                 Phys. Rev. A {\bf 70}, 013816 (2004)
\bibitem{Sutto2} L. G. Suttorp and A. J. van Wonderen,
                 Fano diagonalization of a polariton model for an inhomogeneous absorptive dielectric,
                 Europhys. Lett. {\bf 67}, 766–772 (2004).
\bibitem{Kheir1} F. Kheirandish and M. Soltani,
                 Extension of the Huttner-Barnett model to a magnetodielectric medium,
                 Phys. Rev. A {\bf 78}, 012102 (2008).
\bibitem{Matloo} R. Matloob,
                 Electromagnetic field quantization in a linear isotropic dielectric,
                 Phys. Rev. A {\bf 69}, 052110 (2004).
\bibitem{Kheir2} F. Kheirandish and M. Amooshahi,
                 Electromagnetic field quantization in a linear polarizable and magnetizable medium,
                 Phys. Rev. A {\bf 74}, 042102 (2006).
\bibitem{Bhattt} N. A. R. Bhat and J. E. Sipe,
                 Hamiltonian treatment of the electromagnetic field in dispersive and absorptive structured media,
                 Phys. Rev. A {\bf 73}, 063808 (2006).
\bibitem{Amoos1} M. Amooshahi and F. Kheirandish,
                 Electromagnetic field quantization in a magnetodielectric medium with external charges,
                 Phys. Rev. A {\bf 76}, 062103 (2007).              
\bibitem{Sutto3} L. G. Suttorp,                
                 Field quantization in inhomogeneous anisotropic dielectrics with spatio-temporal dispersion,
                 J. Phys. A: Math. Theor. {\bf 40}, 3697–3719 (2007).             
\bibitem{Amoos2} M. Amooshahia,
                 Canonical quantization of electromagnetic field in an anisotropic polarizable and magnetizable medium,
                 J. Math. Phys. {\bf 50}, 062301 (2009).
\bibitem{Philbi} T. G. Philbin,               
                 Canonical quantization of macroscopic electromagnetism,
                 New J. Phys. {\bf 12}, 123008 (2010).  
\bibitem{Grune2} T. Gruner and D. G. Welsch,
                 Green-function approach to the radiation-field quantization for homogeneous and inhomogeneous Kramers-Kronig dielectrics,
                 Phys. Rev. A {\bf 53}, 1818-1829 (1996).
\bibitem{Schee3} S. Scheel, L. Kn{\"o}ll, and D. G. Welsch
                 QED commutation relations for inhomogeneous Kramers-Kronig dielectrics,
                 Phys. Rev. A {\bf 58}, 700-706 (1998).
\bibitem{Dungg1} H. T. Dung, L. Kn{\"o}ll, and D. G. Welsch, 
                 Three-dimensional quantization of the electromagnetic field in dispersive and absorbing inhomogeneous dielectrics,
                 Phys. Rev. A {\bf 57}, 3931-3942 (1998). 
\bibitem{Schee4} S. Scheel and S. Y. Buhmann,
                 Macroscopic quantum electrodynamics - concepts and applications,
                 Acta Phys. Slovaca {\bf 58}, 675-809 (2008)
\bibitem{Buhma1} S. Y. Buhmann, D. T. Butcher and S. Scheel,
                 Macroscopic quantum electrodynamics in nonlocal and nonreciprocal media,
                 New J. Phys. {\bf 14},  083034 (2012).
\bibitem{Hanso1} G. W. Hanson, F. Lindel, S. Y. Buhmann,
                 The Langevin Noise Approach for Lossy Media and the Lossless Limit,
                 J. Opt. Soc. Am. B {\bf 38}, 758-768 (2021).
\bibitem{DiStef} O. Di Stefano, S. Savasta and R. Girlanda,
                 Mode expansion and photon operators in dispersive and absorbing dielectrics,
                 J. Mod. Opt. {\bf 48}, 67-84 (2001)
\bibitem{Dreze1} A. Drezet,
                 Quantizing polaritons in inhomogeneous dissipative systems,
                 Phys. Rev. A {\bf 95}, 023831 (2017).  
\bibitem{Dorie1} V. Dorier, J. Lampart, S. Guérin, and H. R. Jauslin,               
                 Canonical quantization for quantum plasmonics with finite nanostructures,
                 Phys. Rev. A {\bf 100}, 042111 (2019).
\bibitem{Naaaa2} D. Y. Na, T. E. Roth , J. Zhu,  W. C. Chew and C. J. Ryu,            
                 Numerical framework for modeling quantum electromagnetic systems involving finite-sized lossy dielectric objects in free space,
                 Phys. Rev. A {\bf 107}, 063702 (2023).
\bibitem{Ciatt1} A. Ciattoni,                
                 Quantum electrodynamics of lossy magnetodielectric samples in vacuum: Modified Langevin noise formalism,
                 Phys. Rev. A {\bf 110}, 013707 (2024). 
\bibitem{Miano1} G. Miano, L. M. Cangemi and C. Forestiere,
                 Quantum emitter interacting with a dispersive dielectric object: a model based on the modified Langevin noise formalism,
                 Nanophotonics, 2025. https://doi.org/10.1515/nanoph-2024-0703.  
\bibitem{Miano2} G. Miano, L. M. Cangemi and C. Forestiere,
                 Spectral densities of a dispersive dielectric sphere in the modified Langevin noise formalism,              
                 Phys. Rev. A {\bf 112}, 033712 (2025).
\bibitem{Miano3} G. Miano, L. M. Cangemi and C. Forestiere,
                 Modified Langevin noise formalism for multiple quantum emitters in dispersive electromagnetic environments out of equilibrium,
                 Phys. Rev. A {\bf 113}, 023720 (2026).
\bibitem{Ciatt2} A. Ciattoni,
                 Quantum-optical scattering by macroscopic lossy objects: A general approach,
                 Phys. Rev. A {\bf 112}, 013704 (2025). 
\bibitem{Ciatt3} A. Ciattoni,                 
                 Quantum interference effects in two-photon scattering by a macroscopic lossy sphere,
                 submitted for publication on Physical Review A.
\bibitem{Dreze2} A. Drezet,
                 Equivalence between the Hamiltonian and Langevin noise descriptions of plasmon polaritons in a dispersive and lossy inhomogeneous medium,
                 Phys. Rev. A {\bf 96}, 033849 (2017).  
\bibitem{Frank1} S. Franke, J. Ren, S. Hughes and M. Richter,
                 Fluctuation-dissipation theorem and fundamental photon commutation relations in lossy nanostructures using quasinormal modes,
                 Phys. Rev. Res. {\bf 2}, 033332 (2020).   
\bibitem{Kanwal} R. P. Kanwal,
                 Generalized Functions: Theory and Applications 3rd ed.,
                 Birkhäuser, Boston (2004). 
\bibitem{Krist1} G. Kristensson,
                 Scattering of Electromagnetic Waves by Obstacles,
                 Scitech Publishing, New York (2016).
\bibitem{Bornn1} M. Born and E. Wolf,
                 Principle of Optics,
                 Cambridge University Press (2019).                            
\end{thebibliography}
\end{document}